\documentclass[rmp,aps,twocolumn,floatfix]{revtex4-1}

\usepackage{graphics}
\usepackage{epsfig}
\usepackage{amsmath}
\usepackage{slashbox}

\usepackage{lineno}

\def\inbar{\,\vrule height1.5ex width.4pt depth0pt}
\def\IR{\relax{\rm I\kern-.18em R}}
\def\IC{\relax\hbox{$\inbar\kern-.3em{\rm C}$}}

\newcommand{\be}{\begin{linenomath}\begin{equation}}
\newcommand{\ee}{\end{equation}\end{linenomath}}
\newcommand{\beqa}{\begin{linenomath}\begin{eqnarray}}
\newcommand{\eeqa}{\end{eqnarray}\end{linenomath}}

\newcommand{\footfrac}[2]%

\newcommand{\open}{{<\kern -0.3em{\scriptscriptstyle )}}}

\newcommand{\Compass}{{\sc Compass}}

\newcommand{\GeV}{\mathrm{GeV}}

\begin{document}


\title{The Spin Structure of the Nucleon}

\author{Christine A. Aidala}
\email{caidala@umich.edu}
\affiliation{Physics Department,
University of Michigan,
450 Church Street,
Ann Arbor, MI 48109-1040, U.S.A.}
\author{Steven D. Bass}
\email{Steven.Bass@uibk.ac.at}
\affiliation{Institute for Theoretical Physics,
University of Innsbruck, 
Technikerstrasse 25,
A-6020 Innsbruck, Austria}
\author{Delia Hasch}
\email{Delia.Hasch@lnf.infn.it}
\affiliation{INFN-Frascati,
via E. Fermi 40, 00044 Frascati (Rm), Italy}
\author{Gerhard K. Mallot}
\email{Gerhard.Mallot@cern.ch}
\affiliation{CERN, CH-1211 Gen\`eve 23, Switzerland}

\begin{abstract}
This article reviews our present understanding of QCD spin physics:
the proton spin puzzle and new developments aimed at
understanding the transverse structure of the nucleon.
We discuss present experimental investigations of
the nucleon's
internal spin structure,
the theoretical interpretation of the different measurements
and the open questions and challenges for future investigation.
\end{abstract}

\date{9 January 2013}
\maketitle
\tableofcontents

\section{Introduction}

There has been a vigorous and global program of experiments and
theoretical developments in the last 25 years aimed at understanding
the internal spin structure of the proton.
How is the proton's spin built up from the spin and
orbital angular momentum of the quarks and gluons inside?
Tremendous progress has been made with unraveling the proton's
spin structure with advances in experimental techniques,
theoretical models,
perturbative QCD, non-perturbative QCD and lattice calculations.

This activity was inspired by the initial
European Muon Collaboration (EMC) data 
which suggested the puzzling result that quark intrinsic spin contributes little to the proton's spin \cite{Ashman:1987hv}.
Today
there is good convergence of the theoretical and experimental 
understanding the proton's longitudinal spin structure.
Further puzzling data in measurements of transverse single-spin asymmetries 
revealed up to 40\% asymmetries in proton-proton collisions
(and 5--10\% in lepton-nucleon collisions with unpolarized 
 leptons and
 transversely polarized nucleons)
which persist to high energies.
These single-spin asymmetries indicate significant spin-orbit coupling in the nucleon associated with quark transverse momentum and the bound state structure of the nucleon.
The study of transverse momentum and associated orbital
angular-momentum processes has spawned new programs
to map out the three-dimensional structure of the nucleon.
In this article we review these developments
highlighting the considerable and exciting developments in QCD
spin physics in recent years,
together with an outlook to the future:
What are the main open questions and
the planned experiments to help answer them?

In 1988 EMC published their polarized deep inelastic measurement 
of the proton's $g_1$ spin dependent structure function and
the flavor-singlet axial-charge $g_A^{(0)}$
(the nucleon's ``quark spin content") 
suggesting that quark spins
summed over up, down and strange quark flavors
contribute only a small fraction of the proton's spin.
This result inspired considerable theoretical activity and
new experiments at 
CERN,
SLAC, DESY,
Jefferson Laboratory (JLab) and the
Relativistic Heavy Ion Collider (RHIC) at 
Brookhaven National Laboratory (BNL)
to understand the spin structure of the nucleon.
The first task was to check the initial curious result
from EMC and
second to resolve the spin-flavor structure of the proton.
How is the spin content of the proton 
distributed among the valence and sea quarks and gluons?
What about orbital angular momentum in the nucleon?

We now know that the nucleon's flavor-singlet axial-charge 
measured in polarized deep inelastic scattering 
is $g_A^{(0)} \sim 0.35$.
This value was surprising from the viewpoint of early quark models.
In the static quark model 
-- the eightfold way picture of Gell-Mann --
before inclusion of quark motion, quark spin contributes 100\% of 
the proton's spin.
Relativistic quark models without gluonic or pion cloud degrees of
freedom generally predict about 60\% of the proton's spin should
be carried by the quarks, 
with the remaining 40\% in quark orbital angular momentum.
Today data and theory point to a consistent picture where the
proton spin puzzle is a valence quark effect.
Valence quark contributions to $g_A^{(0)}$
approximately saturate the measured value.
While polarized glue may contribute a significant fraction of
the proton's spin (perhaps up to 50\% at the scale of present
experiments),
sea quark and QCD gluon corrections to the singlet 
axial-charge are small and within the expectations of quark models.
The pion cloud of the nucleon acts
to shift angular momentum from spin to orbital angular momentum
and
induces SU(3) breaking in the nucleon's axial-charges.
There is also a fascinating theoretical possibility that
the valence quarks may polarize the QCD vacuum in a
nucleon through gluon topological effects so that some
fraction of the proton's singlet axial-charge resides at
zero parton momentum (or Bjorken $x$).
Non-zero orbital angular momentum of the valence quarks is expected,
induced also by confinement which introduces a transverse scale in
the physics.
This orbital angular momentum through spin-orbit coupling
is a prime candidate to explain the large single spin asymmetries
observed in proton-proton collisions.
Information about quark total angular momentum in the proton
can be extracted from deeply virtual Compton scattering and
high-energy single spin asymmetry data in model-dependent analyses.
The results are consistent with QCD lattice calculations.

This Review is organized as follows.
In the first part (Sections II--III)
we give a brief introduction to nucleon spin physics and the experiments that have been performed to investigate it.
Then, in Section IV, we discuss the proton spin puzzle and the small value of $g_A^{(0)}$
extracted from polarized deep inelastic scattering.
In Section V
we give an overview of the present global
program aimed at disentangling the spin-flavor structure of
the proton.
Section VI covers the theoretical interpretation of
longitudinal spin data and understanding of the proton spin puzzle.
We next turn our attention to the transverse structure of the
nucleon and manifestations of orbital angular momentum in the
nucleon in Section VII.
This discussion introduces generalized parton distributions 
(GPDs),
which describe hard exclusive reaction processes, 
and
transverse momentum dependent distributions (TMDs),
which describe spin-momentum correlations and spin-orbit couplings
in the nucleon.
The TMDs are manifest in high-energy single-spin and 
azimuthal asymmetries.
A summary of key issues and challenging questions for the next 
generation of experiments is given in Section VIII and IX.

Earlier review articles on the spin structure of the proton as well 
as complementary more recent reviews,
each with a different emphasis,
are given in
\textcite{Anselmino:1994gn},
\textcite{Ellis:1995de},
\textcite{Cheng:1996jr},
\textcite{Altarelli:1998nb},
\textcite{Shore:1998dn},
\textcite{Lampe:1998eu},
\textcite{Filippone:2001ux},
\textcite{Jaffe:2001dm},
\textcite{Bass:2004xa},
\textcite{Kuhn:2008sy},
\textcite{Barone:2010zz},
\textcite{Burkardt:2008jw},
\textcite{Myhrer:2009uq}
and the monograph \textcite{Bass:2007zzb}.

\section{Spin Structure Functions and Parton Distributions}

Our knowledge about the high-energy spin structure of the nucleon comes from both polarized deep inelastic scattering (DIS)
experiments and high-energy polarized proton-proton collisions.
Polarized deep inelastic scattering (pDIS) experiments
involve scattering a longitudinally polarized high-energy
lepton beam from a longitudinally or transversely polarized
nucleon at large momentum transfer. Inclusive measurements,
where only the scattered lepton is observed in the final
state, and semi-inclusive measurements,
where one tags on at least one high-energy final state hadron
in coincidence with the scattered lepton, have been performed.
The experiments were performed
with an electron beam at SLAC and JLAB, with electron and
positron beams at DESY and with muon beams at CERN. In
proton-proton scattering the protons are either longitudinally
or transversely polarized.
Polarized deep inelastic scattering experiments have so far all
been performed using a fixed target.
A future polarized electron-ion collider is in planning.
Details of the experiments are given in Section III.
Historically,
information about
the proton's internal spin structure came first from
measuring the proton's $g_1$ and $g_2$ spin structure functions
in inclusive deep inelastic scattering and,
more recently, from semi-inclusive reactions
in both lepton-nucleon and proton-proton collisions and
hard exclusive processes in lepton-nucleon scattering.

Measurements with longitudinally polarized targets and beams tell us about the helicity distributions of quarks and gluons in the nucleon, which at leading order can be thought of as the difference in probability of finding a parton with longitudinal polarization parallel or anti-parallel to that of the nucleon.  Measurements with transversely polarized targets are particularly sensitive to quark and gluon transverse and orbital angular momentum.  Studies of transverse degrees of freedom in the nucleon and in fragmentation processes are a current subject of experimental investigation with sensitivity to spin-orbit couplings in QCD.

For polarized lepton-proton scattering, specialize to the target rest frame and let $E$ denote the energy of the incident lepton which is scattered through an angle $\theta$ to emerge in the final state with energy $E'$.
Let $\uparrow \downarrow$ denote the longitudinal polarization of
the lepton beam.
In photon-nucleon scattering the spin dependent structure functions
$g_1$ and $g_2$ are defined through the imaginary part of the forward Compton scattering amplitude.
The structure functions contain all of
the target-dependent information in the deep inelastic process.
Consider the amplitude for forward scattering of a photon carrying momentum $q_{\mu}$ ($q^2 = -Q^2 \leq 0$) from a polarized nucleon
with momentum $p_{\mu}$, mass $M$ and spin $s_{\mu}$.
We work with the kinematic
Bjorken variable $x = Q^2/2p \cdot q = Q^2/ 2 M\nu$ 
where
$\nu = p \cdot q/M = E - E'$,
and
let $y = p \cdot q / p \cdot k = \nu / E$.
For a longitudinally polarized proton target 
(with spin denoted $\Uparrow \Downarrow$)
the unpolarized and polarized differential cross-sections are
\begin{eqnarray}
& &
{d^2 \sigma \uparrow \Downarrow \over d x d y }
+
{d^2 \sigma \uparrow \Uparrow \over d x d y }
=
\nonumber \\
& &
{2 \pi \alpha^2 \over M E x^2 y^2} \
\biggl[ \biggl( 1 - y - {Mxy \over 2 E} \biggr) F_2 (x, Q^2)
 +
 x y^2 F_1 (x, Q^2 ) \biggr]
\nonumber \\
\label{eqb21}
\end{eqnarray}
and
\begin{eqnarray}
& &
{d^2 \sigma \uparrow \Downarrow \over d x d y }
-
{d^2 \sigma \uparrow \Uparrow \over d x d y }
=
\nonumber \\
& &
{4 \alpha^2 \over M E x y} \
\biggl[ \biggl( 2-y-{Mxy \over E} \biggr) g_1 (x,Q^2)
- {2 Mx \over E} g_2 (x, Q^2)
\biggr]
\nonumber \\
\label{eqb22}
\end{eqnarray}
where the mass of the lepton is neglected.
The relation between the structure functions in deep inelastic 
lepton-nucleon scattering and the virtual-photon nucleon 
cross-sections is discussed and derived in various textbooks, 
{\it e.g.}\ \textcite{Roberts:1990ww}.
One finds
\begin{eqnarray}
A_1 =
{\sigma_{1 \over 2} - \sigma_{3 \over 2}
\over
 \sigma_{1 \over 2} + \sigma_{3 \over 2}}
 =
{g_1 - {Q^2 \over \nu^2} \ g_2 \over F_1}
\rightarrow
{g_1 \over F_1}
\label{eqb25}
\end{eqnarray}
where $\sigma_{3 \over 2}$ and $\sigma_{1 \over 2}$ are 
the cross-sections for the absorption of a transversely
polarized photon with spin polarized parallel and anti-parallel 
to the spin of the longitudinally polarized nucleon.
For a longitudinal polarized target the $g_2$ contribution
to the differential cross-section and the longitudinal
spin asymmetry is suppressed relative to the $g_1$ contribution
by the kinematic factor ${M / E} \ll 1$.
For a transverse polarized target this kinematic suppression
factor for $g_2$ is missing
implying
that transverse polarization is vital to measure $g_2$.
We refer to \textcite{Roberts:1990ww} and
\textcite{Windmolders:2002na}
for the procedure how the spin dependent structure functions
are extracted from the spin asymmetries measured in polarized
deep inelastic scattering.

In high-$Q^2$ deep inelastic scattering the structure functions
$F_1$, $F_2$, $g_1$ and $g_2$
exhibit approximate scaling.
They are to a very good approximation independent of $Q^2$ and 
depend
{\it only} on Bjorken $x$.
(The small $Q^2$ dependence which is present in these structure
 functions is
 logarithmic and determined by perturbative QCD evolution.)

In the (pre-QCD) parton model the deep inelastic structure functions
$F_1$ and $F_2$ are written as
\begin{eqnarray}
F_1 (x) = {1 \over 2x} F_2 (x) = {1 \over 2} \sum_q e_q^2 \{q + {\bar q} \}(x)
\label{eqb26}
\end{eqnarray}
and the polarized structure function $g_1$ is
\begin{eqnarray}
g_1 (x) = {1 \over 2} \sum_q e_q^2 \Delta q(x)
. 
\label{eqb27}
\end{eqnarray}
Here $e_q$ denotes the electric charge of the struck quark and
\begin{eqnarray}
\{q + {\bar q} \}(x)
&=& (q^{\uparrow} + {\overline q}^{\uparrow})(x) +
                        (q^{\downarrow} + {\overline q}^{\downarrow})(x)
\nonumber \\
\Delta q(x)
&=& (q^{\uparrow} + {\overline q}^{\uparrow})(x) -
                        (q^{\downarrow} + {\overline q}^{\downarrow})(x)
\label{eqb28}
\end{eqnarray}
denote the spin-independent (unpolarized) and spin-dependent
quark parton distributions
which measure the distribution of quark momentum and spin in
the proton.
For example,
${\overline q}^{\uparrow}(x)$
is interpreted as the probability to find an anti-quark of
flavor $q$
with plus component of momentum $x p_+$
($p_+= p_0+p_3$ 
 is the plus component of the target proton's momentum)
and spin polarized in the same direction as the spin of the
target proton.
When we integrate out the momentum fraction $x$
the quantity
$
\Delta q = \int_0^1 dx \ \Delta q(x)
$
is interpreted as the fraction of the proton's spin which
is carried by quarks (and anti-quarks) of flavor $q$.
Hence summing over the up, down and strange quark $\Delta q$
contributions gives the total fraction of the proton's spin 
carried by the spins of these quarks.

What values should we expect for the $\Delta q$?
First, consider the static quark model.
The simple SU(6) proton wavefunction
\begin{eqnarray}
|p\uparrow \rangle &=&
{1 \over \sqrt{2}} | u\uparrow (ud)_{S=0} \rangle
+
{1 \over \sqrt{18}} | u\uparrow (ud)_{S=1} \rangle
\nonumber \\
& & -
{1 \over 3} | u\downarrow (ud)_{S=1} \rangle
-
{1 \over 3} | d\uparrow (uu)_{S=1} \rangle
\nonumber \\
& & +
{\sqrt{2} \over 3} | d\downarrow (uu)_{S=1} \rangle
\label{eqa9}
\end{eqnarray}
yields the values
$\Delta u - \Delta d = {5 \over 3}$ and 
$\Delta u + \Delta d =1$.
In relativistic quark models one has to take into account
the four-component Dirac spinor
$\psi \sim
\biggl({ f \atop i \sigma \cdot {\hat{r}} g }\biggr)$.
The lower component of the Dirac spinor is p-wave with intrinsic spin
primarily pointing in the opposite direction to the 
spin of the proton \cite{Jaffe:1989jz}.
Relativistic effects renormalize the axial charges by the depolarization factor 0.65 with a net transfer of angular
momentum from intrinsic spin to orbital angular momentum.
In QCD and in more sophisticated models further depolarization is induced by gluonic and pion-cloud degrees of freedom -- see Section VI.

In QCD the flavor-singlet combination of the $\Delta q (x)$ quark parton distributions mixes with the spin-dependent gluon distribution
under $Q^2$ evolution \cite{Altarelli:1977zs}.
This spin dependent gluon distribution measures the momentum and 
spin dependence of glue in the proton.
The second spin structure function $g_2$ vanishes without the effect 
of quark transverse momentum and has a non-trivial parton interpretation \cite{Jaffe:1989xx,Roberts:1990ww}.

The parton model description of polarized deep inelastic scattering
involves writing the deep inelastic structure functions as the sum
over the convolution of ``soft'' quark and gluon parton distributions
with ``hard'' photon-parton scattering coefficients
\begin{eqnarray}
g_1^p (x)
&=&
\Biggl\{
         {1 \over 12} (\Delta u - \Delta d) 
       + {1 \over 36} (\Delta u + \Delta d - 2 \Delta s) 
\Biggr\}
\otimes 
C^q_{ns} 
\nonumber \\
& &        
+ {1 \over 9}  
\Biggl\{
(\Delta u + \Delta d + \Delta s) 
\otimes 
C^q_s 
+
f 
\Delta g 
\otimes 
C^g 
\Biggr\}
.
\nonumber \\
\label{eqd80}
\end{eqnarray}
Here
$\Delta q(x)$ and $\Delta g(x)$ 
denote the polarized quark and gluon
parton distributions, $C^q$ and $C^g$ 
denote the corresponding hard-scattering coefficients, 
and $f$ is the number of quark flavors liberated into 
the final state
($f=3$ below the charm production threshold).
The parton distributions contain all the target-dependent information and describe a flux of quark and gluon partons into the (target independent) interaction between the hard photon and the parton
which is described by the coefficients $C^q$ and $C^g$.
These coefficients are calculated using perturbative QCD via the
cross-section for the hard photon scattering from a quark or gluon parton ``target".
They are independent of infra-red mass singularities
(terms involving the quark mass or virtuality of the
 parton in the photon-parton collision)
which are absorbed into the parton distributions
(and softened by confinement related physics).
If the same recipe (``factorization scheme'') for separating hard and soft parts of the parton phase space is applied consistently to all hard processes then the factorization theorem asserts that
the parton distributions
that one extracts from different experiments are process independent.
In other words, the same polarized quark and gluon distributions
should be obtained from experiments involving polarized
hard QCD processes in polarized proton-proton collisions
and polarized deep inelastic scattering experiments.
For example, colliding longitudinally polarized proton beams
provides sensitivity to the gluon-helicity distribution function at leading order.
For hadron production with transverse momentum $p_T$,
the helicity-dependent difference in hadron production is defined as
\begin{eqnarray}
{d \Delta \sigma \over d p_T}
\equiv
{1 \over 2}
\biggl[
{d \sigma^{++} \over d p_T} -
{d \sigma^{+-} \over d p_T} \biggr]
\end{eqnarray}
where the superscripts $++$ and $+-$ refer to same and opposite
helicity combinations of the colliding protons.
Factorization allows this to be written as a convolution of
the long- and short-distance terms summed over all possible
flavors for the partonic interaction $a + b \rightarrow jet + X$
\begin{eqnarray}
{d \Delta \sigma \over d p_T}
&=&
\sum_{ab} \int dx_a dx_b
\Delta f_a (x_a, \mu ) \Delta f_b (x_b, \mu )
\nonumber \\
& &
\ \ \ \ \ \
\times
{d \Delta {\hat \sigma}^{a b \rightarrow jet + X}
\over d p_T}
(x_a P_a, x_b P_b, \mu ) .
\nonumber \\
\end{eqnarray}
Here $P_a$ and $P_b$ denote the momenta of the incident protons;
$\Delta f_a(x_a, \mu)$ are the polarized parton distributions of the colliding partons carrying light-cone momentum fraction $x$ evaluated at factorization and renormalization scale $\mu$.
The helicity-dependent difference in the cross-section of the hard partonic scattering $a+b \rightarrow jet +X$ is denoted
by $d \Delta {\hat \sigma}$ and is calculable in perturbative QCD.
Partonic cross-section calculations are carried out 
to finite order in $\alpha_s$ and have a dependence 
on factorization and renormalization scales, denoted $\mu$.
The final hadronic cross-section
is independent of 
the factorization and renormalization scales and the scheme used.
The QCD parton model treatment readily generalizes to the production
of high-energy hadrons in the final state, with the produced ``fast"
hadron carrying a significant fraction of the momentum of a ``parent"
parton. The parton-to-hadron process is parametrized by fragmentation functions which also obey process-independent factorization in perturbative QCD calculations.

Analogous to the helicity distributions measured with longitudinal
polarization, transversity distributions 
describe the density of transversely polarized quarks inside a 
transversely polarized proton,  see e.g.\ \textcite{Barone:2001sp}.
The transversity distributions, 
which were introduced in
\textcite{Ralston:1979ys}, \textcite{Artru:1989zv}, 
\textcite{Jaffe:1991ra} and \textcite{Cortes:1991ja}, 
are interpreted in parton language as follows.
Consider a nucleon moving with (infinite) momentum in the
$\hat e_{3}$-direction, but polarized 
transverse to $\hat e_{3}$.
Then $\delta q (x)$ 
(also denoted $\Delta_T q (x)$ and $h_1^q(x)$ in the literature)
counts the quarks with flavor $q$,
momentum fraction $x$ and their spin parallel to the spin of a nucleon minus the number anti-parallel.
That is, in analogy with Eq.~(\ref{eqb28}), $\delta q(x)$ measures the distribution of partons with transverse polarization in a transversely polarized nucleon, {\it viz.}
\begin{eqnarray}
\delta q (x) = 
q^{\uparrow}(x) + {\bar q}^{\uparrow}(x)
- 
q^{\downarrow}(x) - {\bar q}^{\downarrow}(x).
\label{eqj145}
\end{eqnarray}
In a helicity basis transversity corresponds to helicity-flip 
making it
a probe of chiral symmetry breaking \cite{Collins:1992kk}.
There is no gluon analogue of transversity in the nucleon so
$\delta q$ evolves in $Q^2$ like a valence or non-singlet quark
distribution, without mixing with glue.
If quarks moved non-relativistically in the nucleon $\delta q$ and
$\Delta q$ would be identical
since rotations and Euclidean boosts commute and a series of boosts
and rotations can convert a longitudinally polarized
nucleon into a transversely polarized nucleon at infinite momentum.
The difference between the transversity and helicity distributions
reflects the relativistic character of quark motion in the nucleon.

Following the discovery that the quark spin contribution to the proton's spin is small, there has been a vigorous program to 
measure the separate contributions of up, down and strange quark flavors 
as well as the gluon spin and the orbital contributions.
This has inspired dedicated spin programs in semi-inclusive
deep inelastic scattering (SIDIS) and
polarized proton-proton collisions to measure the separate
valence and sea quark as well as gluon polarization.
As efforts to investigate nucleon spin in more detail intensified and
various experimental programs were being developed in the 1990s,
new theoretical ideas arose as well.
TMD distributions,
describing spin-momentum correlations in the nucleon, were
initially proposed~\cite{Sivers:1989cc}
to explain the very large transverse single spin asymmetries
involved in polarized hadronic scattering that were 
first observed in the 1970s
by \textcite{Klem:1976ui} and \textcite{Dragoset:1978gg}.
The GPDs
introduced in
\textcite{Mueller:1998fv}, \textcite{Ji:1996ek} and 
\textcite{Radyushkin:1997ki}
to describe hard exclusive reactions
provided for the first time a means of describing the radial 
position distributions of partons 
at a specific longitudinal momentum 
within the nucleon.
Both TMD distributions and GPDs offer links
to the 
orbital angular momentum contributions 
to the nucleon's spin.
These processes and the present status of experimental and
theoretical investigation are described in Section VII.

\section{Experiments}

Experiments that have probed the nucleon spin structure 
are outlined in Table I.
This includes both polarized deep inelastic lepton-nucleon scattering 
and proton-proton collision experiments.
Considerable effort 
was invested in developing polarized beam and
target technology, yielding physics results with ever increasing precision.
The first experiments focused on 
inclusive deep inelastic measurements of nucleon spin structure.
More recent experiments, described in detail below,
were able to detect and identify hadrons in the final state leading
to new probes of the nucleon in
semi-inclusive and hard exclusive reactions.
Future experimental programs 
(COMPASS-II, the 12 GeV upgrade of JLab and experiments at
 Fermilab and RHIC)
 with high luminosity and acceptance are planned to explore the
three-dimensional structure of the nucleon in spatial and transverse
momentum degrees of freedom. We discuss these future programs 
in Section VIII.

\begin{widetext}

\begin{table}[t!]
\label{tab:exp-overview}
\caption{\small
High energy spin experiments:
the kinematic ranges in $x$ and $Q^2$ correspond to the average kinematic values of the highest statistics measurement of each experiment, which is typically the inclusive spin asymmetry;
$x$ denotes Bjorken $x$ unless specified.
}
\begin{center}
\begin{tabular}{|c|c|c|c|c|c|c|}
\hline
Experiment  & Year & Beam   & Target  &  Energy (GeV) & $Q^2$ (GeV$^2$) & $x$ \\
\hline
\multicolumn{7}{|c|}{Completed experiments}\\
\hline
SLAC -- E80, E130 & 1976--1983 & $e^-$ & H-butanol  & $\lesssim$23 & 1--10 & 0.1--0.6 \\
SLAC -- E142/3 & 1992--1993 &  $e^-$ & NH$_3$, ND$_3$ & $ \lesssim 30$ & 1--10 & 0.03--0.8  \\
SLAC -- E154/5 & 1995--1999 &  $e^-$ & NH$_3$, $^6$LiD, $^3$He  & $\lesssim 50$& 1--35 & 0.01--0.8  \\
CERN -- EMC  & 1985 &  $\mu^+$ & NH$_3$ & 100, 190 & 1--30 & 
0.01--0.5 \\
CERN -- SMC  & 1992--1996 &  $\mu^+$ & H/D-butanol, NH$_3$ & 100, 190& 1--60 & 0.004--0.5  \\
FNAL E581/E704 & 1988--1997 & $p$ & $p$ & 200 & $\sim 1$ & $0.1 < x_F < 0.8$ \\
\hline
\multicolumn{7}{|c|}{Analyzing and/or Running}\\
\hline
DESY -- HERMES & 1995--2007 & $e^+$, $e^-$ & H, D, $^3$He & $\sim 30$ & 1--15 & 0.02--0.7 \\
CERN -- COMPASS & 2002--2012 & $\mu^+$  & NH$_3$, $^6$LiD & 160, 200 & 1--70 & 0.003--0.6   \\ 
JLab6 -- Hall A & 1999--2012 & $e^-$ & $^3$He & $\lesssim 6$ & 1--2.5 &  0.1--0.6 \\
JLab6 -- Hall B & 1999--2012 & $e^-$ & NH$_3$, ND$_3$ & $\lesssim 6$ & 1.-5 & 0.05--0.6 \\
RHIC -- BRAHMS & 2002--2006 & $p$ & $p$ (beam) & $2 \times$~(31--100) & 
$\sim$~1--6 & $-0.6 < x_F < 0.6$ \\
RHIC -- PHENIX, STAR & 2002+ & $p$ & $p$ (beam)& $2 \times$~(31--250)
& $\sim$~1--400 &
$\sim$~0.02--0.4 \\
\hline
\multicolumn{7}{|c|}{Approved future experiments (in preparation) }\\
\hline
CERN -- COMPASS--II  & 2014+ &  $\mu^+$, $\mu^-$  &  unpolarized H$_2$ & 160 & $\sim$~1--15 & $\sim$~0.005--0.2 \\
&  &  $\pi^-$  & NH$_3$ & 190 &    & 
$-0.2 < x_F < 0.8$ \\
JLab12 -- HallA/B/C  & 2014+ & $e^-$ & HD, NH$_3$, ND$_3$, $^3$He & $\lesssim$12  & $\sim$~1--10 & $\sim$~0.05--0.8 \\
\hline
\end{tabular}
\end{center}
\end{table}

\end{widetext}

\subsection{SLAC experiments}

SLAC experiments pioneered polarized DIS measurements and set many standards
in polarized beam and target technologies.
Their spin program focused on high statistics measurements of the inclusive
asymmetries.
The first measurements of the proton spin structure were performed
by the experiments 
E80 \cite{Alguard:1976bm,Alguard:1978gf}
and E130 \cite{Baum:1980mh,Baum:1983ha},
followed by a series of
high precision experiments
E142 \cite{Anthony:1996mw},
E143 \cite{Abe:1998wq},
E154 \cite{Abe:1997cx}
and
E155 \cite{Anthony:1999rm,Anthony:2000fn} a decade later.
These experiments utilized polarized electrons
which were produced by laser photoemission and
subsequently accelerated.
The longitudinal polarization of the beam was frequently inverted 
and the
polarization measured using M$\o$ller scattering.
A rapid cycling of the beam and/or target polarization reduces systematic uncertainties in the measured spin 
asymmetries related to the stability of the experimental setup.
Polarized target materials involved solid-state butanol and 
ammonia (NH$_3$) for the proton and 
D-butanol, ND$_3$ as well as $^6$LiD 
for the deuteron \cite{Crabb:1997cy,Meyer:2004dy}.
For the most recent E154 and E155 experiments the target
polarization was typically 
38\% for $^3$He, 90\% for NH$_3$ and 22\% for LiD 
with beam polarization about 80\%.
The target material, 
doped with a paramagnetic substance or
irradiated with electron beams, was
polarized using dynamic nuclear polarization, which requires
temperatures of about 1 K and strong magnetic holding fields.
Such targets contain a considerable amount of non-polarizable nucleons, which is parametrized by the so-called dilution factor.
This factor depends on all kinematic variables relevant for the process under study and needs,
in principle, to be determined for each type of measurement.
Typical values for polarized solid state targets 
range between 0.1 and 0.2 with the exception of 
$^6$LiD (0.4-0.5)
and represent an important factor 
in the extraction of physical observables from the measured ones.
Information on the neutron structure was obtained either from 
the combination of measurements with proton and deuteron 
targets or by using a polarized $^3$He target
which is dominated by the neutron 
since the two proton spins in $^3$He are anti-aligned.
Here, polarization was obtained from optical pumping and adiabatic spin exchange.
The target polarization was measured using the NMR technique.
Scattered electrons were detected with magnetic spectrometers optimized for 
high-momentum-resolution and good electron identification.

\subsection{CERN experiments}

\subsubsection{The EMC and SMC experiments}

Following the early measurements at SLAC, the European Muon Collaboration (EMC) experiment performed at CERN in 1985 
the first polarized DIS measurements at $x < 0.1$ down to $x=0.01$
after a series of measurements of unpolarized nucleon and nuclear structure functions.
The experiment used the polarized CERN muon beam up to momenta of 
200~GeV 
and a solid-state irradiated ammonia target.
Their low-$x$ measurements, accessible due to the high energy of the muons, suggested the breakdown of the naive parton picture that quarks provide essentially all of the spin of 
the nucleon \cite{Ashman:1987hv, Ashman:1989ig}.

This triggered more detailed and precise measurements 
by the Spin Muon Collaboration (SMC) in 1992--1996, and 
by COMPASS 
(since 2002). 
The beam line and the principal ideas of the CERN 
muon experiments are described 
in the COMPASS Section~\ref{sssec:compass}.
The EMC Spectrometer is described in \textcite{Aubert:1980rm}.
The polarization of the CERN muon beam was measured by
SMC \textcite{Adeva:1993kp}. 
A detailed description of the SMC deuteron
target polarization is given in \textcite{Adeva:1994ic}.
The COMPASS experiment used 
the SMC target in the initial period of data taking
up to 2005 as reported in \textcite{Ball:2003vb}. 
A new target is used since 2006 \cite{Gautheron:2007zz}.

After 1987 the focus was on the region $x<0.1$ and 
the flavor-singlet axial-charge (Ellis--Jaffe sum-rule) 
for the neutron. 
The latter must deviate from the naive prediction in 
a similar way as 
for the proton in order to preserve the fundamental 
isovector Bjorken sum-rule for $g_1^p-g_1^n$. 
(These sum-rules are discussed below.)
The SMC experiment could extend the measured 
$x$-range down to $x=0.004$ (for $Q^2>1~\GeV^2$) and 
established the validity of the Bjorken sum-rule with measurements using polarized proton (butanol and ammonia) and deuteron (D-butanol) targets \cite{Adeva:1993km, Adeva:1998vv}. 
The large acceptance of the SMC spectrometer in the forward direction allowed them to present the first determination of individual quark distributions for different flavors \cite{Adeva:1995yi,Adeva:1997qz} from semi-inclusive DIS.
A dedicated polarimeter confirmed the validity of the beam polarization obtained from Monte Carlo simulations \cite{Adeva:1993kp,Adams:1999af} 
used in the EMC, SMC and COMPASS analyses.

\subsubsection{The COMPASS experiment}

\label{sssec:compass}

The COMPASS spectrometer 
(\textcite{Abbon:2007pq}, Fig.~1) 
is installed at the muon beam line of the CERN SPS accelerator.
A polarized muon beam of energy 160--200~GeV and with a polarization of about 80\% impinges
on a solid-state polarized target consisting of two or three cells with proton
or deuteron target material polarized in opposite directions. 
The usable beam intensity is typically $2\times10^7$/s during a 
9.6~s long spill. The repetition rate varies and is typically
about 1/40~s. 
The muon polarization arises naturally from the weak decay of the parent pions produced by the primary proton beam of 400~GeV. 
The momentum of each beam muon is measured in the beam momentum station.
Downstream of the target, the scattered muon and produced hadrons 
are detected in a two-stage magnetic spectrometer with the two dipole magnets (SM1, SM2).

\begin{figure}[b]
  \begin{center}
    \includegraphics[width=\hsize]{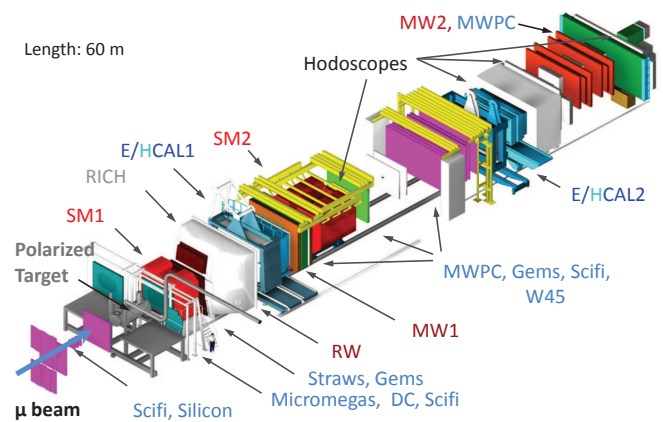}
    \end{center}
  \caption{\label{fig:spectrometer}
    The \Compass\ spectrometer, for a description see text.}
  \end{figure}

Charged particles are tracked in the beam regions by scintillating fiber stations (SciFi) and by silicon detectors. 
In the inner region close to the beam, gaseous detectors of the micromegas and gas-electron-multiplier (Gem) types with high rate capabilities are deployed.
The backbone of tracking in the intermediate region is multiwire proportional chambers (MWPCs). 
Finally, the large area tracking away from the beam region is 
covered by drift chambers (DC, W45) and drift tubes (Straws, RW, MW).

The velocity of charged particles is measured in a ring-imaging Cherenkov detector (RICH), which can separate pions and kaons from 
9~GeV up to 50~GeV. 
The inner quarter of the photon detector is made of
multianode-photomultiplier tubes, while the outer part relies on 
MWPCs with a photosensitive CsI cathode.

The energy of charged particles is measured in sampling
hadron calorimeters (HCAL), while neutral particles, in particular
high-energy photons, are detected in electromagnetic
calorimeters (ECAL).
They comprise lead glass modules as well as scintillator/lead
``shashlik" modules in the inner high-radiation region.

Event recording is triggered by the scattered muon, which is ``identified" by its ability to traverse thick hadron 
absorbers, located just upstream of the Muon Wall detectors (MW). 
The event selection is based on  various systems of scintillator hodoscopes and logic modules applying selection criteria like 
target pointing and energy loss in the scattering. The patterns
causing a trigger were optimized by Monte Carlo simulations. The spectrometer has about 250k read-out channels, which can be recorded
with a frequency of 20~kHz for an event size of the order of 40~kByte.

The heart of the experiment is the polarized target system.
While the muon beam comes naturally polarized due to the parity violation in the decay of the parent pions, 
polarizing protons and deuterons is very difficult. 
Gas targets can not be used with the muon beam due to the low beam intensity compared to electron beams. 
An advantage of muon beams is the high muon energy, which presently can not be reached by electron beams. 
The polarized target system comprises a 2.5~T solenoid magnet, 
a 0.6~T dipole magnet, a $^3$He/$^4$He dilution refrigerator, 
a 70~GHz microwave system and an NMR system to measure the target polarization.
The target material is cooled down to about 60~mK in frozen spin mode. 
The nucleons/nuclei are
polarized by dynamic nuclear polarization which only is applicable for particular materials.
In COMPASS irradiated ammonia (NH$_3$) and lithium-6 deuteride ($^6$LiD) were
selected as proton and deuteron targets, respectively. Lithium-6 is very close to a system
of a free deuteron and a helium-4 core and has essentially the same magnetic moment as
the deuteron. Thus $^6$LiD corresponds to two deuterons plus
a helium nucleus.
Typically, polarizations of 85\% for protons and 50\% for deuterons were reached.
A key feature of COMPASS is that both target polarizations are present simultaneously in separate target cells along the beam, 
e.g.\ ``$\rightarrow, \leftarrow$" for the two-cell
configuration until 2004 and 
``$\rightarrow, \leftarrow, \rightarrow$" 
for the three-cell configuration from 2006 onward. 
In the former configuration the length of the cells was twice
60~cm while in the latter it was 30~cm, 60~cm, 30~cm, respectively.
Thus in an asymmetry measurement most systematic uncertainties cancel.
Using the dipole and solenoid magnet, the magnetic field can be rotated from {\it e.g.}\ pointing 
downstream to transverse to upstream. 
The spin follows the magnetic field adiabatically and
thus the spin orientations can be changed within 30~min. 
Such a field rotation is performed typically once per day for the longitudinal polarization in order to cancel potentially remaining systematic effects. 
The field can also be kept transverse for measurements with 
transverse target polarization.
Here the polarization is inverted by repolarizing typically 
once per week.
In the shutdown year 2005 the superconducting target magnet was replaced by a new one, increasing
the angular acceptance from $\pm70~\mathrm{mrad}$ 
to $\pm180~\mathrm{mrad}$.

The experiment is taking data since 2002.
The main focus has been on inclusive and semi-inclusive polarized
deep inelastic scattering.
As schematically depicted in Fig.~2, the detection of a hadron in the final state provides information about the flavor of the struck quark, 
while the kinematics of the DIS event is fixed by the incoming and scattered lepton.
The years 2008--2009 were dedicated to the hadron spectroscopy 
program of COMPASS with pion, kaon and proton beams.
In 2012 the pion polarizability is being measured using a 
negative pion beam and a thin nickel target.
A pilot run for deeply virtual Compton scattering and 
hard exclusive meson production has been successfully completed in 2012.

\subsection{The HERMES experiment at DESY}

\begin{figure}[tbp]
  \begin{center}
    \includegraphics[width=\hsize]{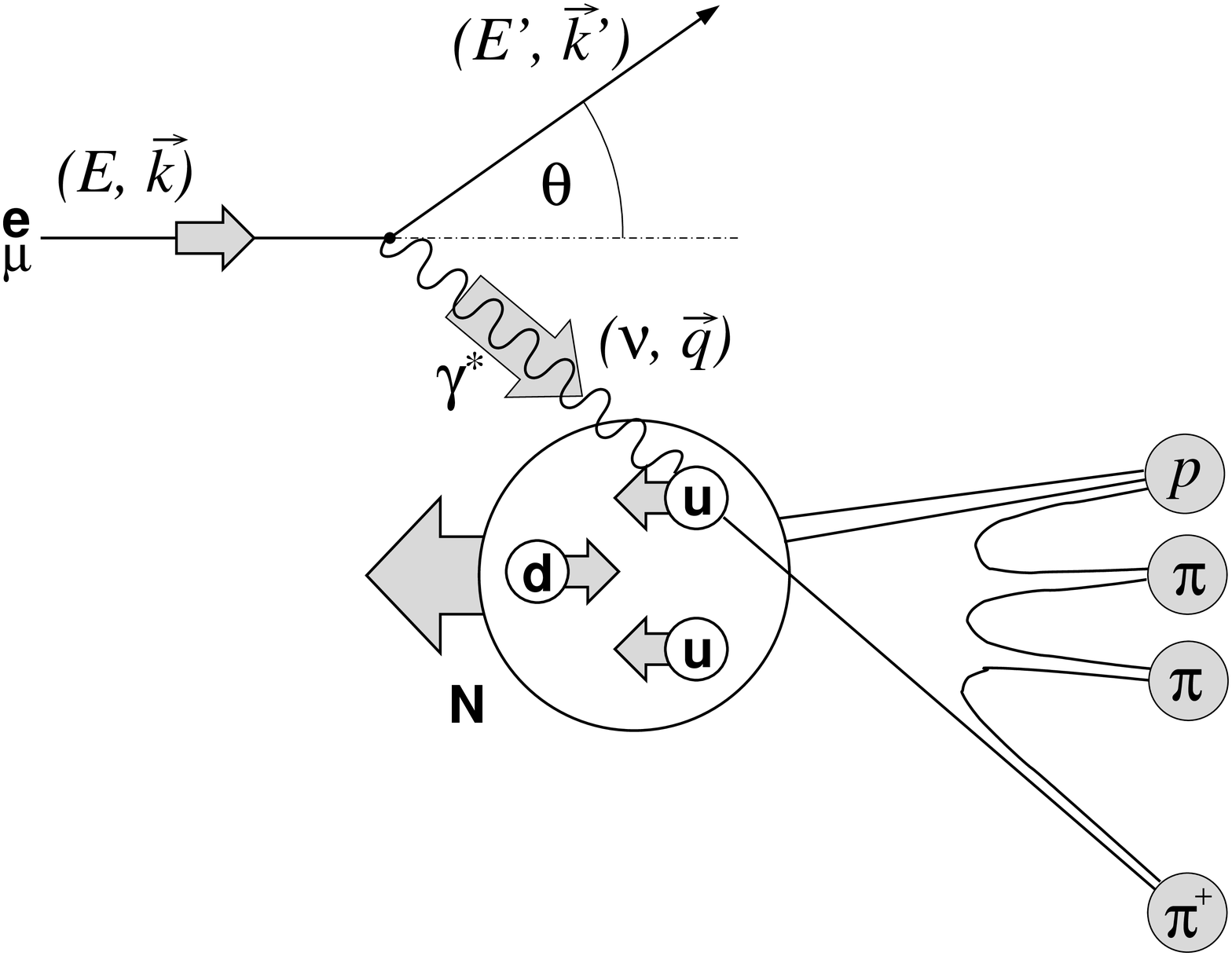}
    \end{center}
  \caption{\label{fig:sidis-pict}
    Semi-inclusive DIS studied at COMPASS, HERMES and JLab.}
  \end{figure}

The HERMES experiment employed an innovative technique for the polarized target, which is
very different from all other polarized DIS experiments.
Gas targets of pure nuclear-polarized atoms of hydrogen or deuterium were used, which
permit essentially background-free measurements from highly polarized nucleons with little
or no dilution of the signal from unpolarized nucleons in the target.
This choice eliminates one of the main systematic sources in polarized DIS, the uncertainty
in the determination of the dilution factor.

The HERMES gas targets were highly longitudinally ($\sim 85\%$) or transversely (75\%) polarized
with the ability to invert the direction of the spin of the nucleons within milliseconds.
Due to the low densities, however, such targets are only practicable in the high currents of storage rings.
HERMES was operating from 1995 to 2007 at the HERA lepton storage ring, which provided
electron or positron beams of typically 40 mA and with an energy of 27.5 GeV.
In order to enhance the target density, the novel technique of a storage cell was used \cite{Baumgarten:2001ym,Baumgarten:2003fh,Baumgarten:2003fg,Airapetian:2003aj}.
Here, the gas was fed into a T-shaped open-ended elliptical cell coaxial to the lepton beam.
The gas atoms underwent several hundred wall bounces before escaping from the ends where they were differentially pumped away by a large system of turbo-pumps.
This increased the density by a factor of about 100 compared to free gas jet targets.

The polarized atoms were injected into the cell from an atomic beam source based on
Stern-Gerlach polarization filtering and radio-frequency transitions between atomic substates
in a magnetic field~\cite{Airapetian:2004yf}.
A small sample gas diffused from the middle of the cell into a 
Breit-Rabi polarimeter
which measured the atomic polarization \cite{Baumgarten:2001ym}, 
or into a target gas analyzer which measured the atomic and the molecular content of the 
sample \cite{Baumgarten:2003rp}.
A magnet surrounding the storage cell provided a holding field defining the polarization axis and
prevented spin relaxation via spin exchange or wall collisions.
The cell temperature was kept at about 100~K, the value for which atomic recombination and spin relaxation during wall collisions are minimal.

Stored high energy electron beams may become spontaneously transversely polarized via a small polarization asymmetry in the emission of synchrotron radiation by the beam particles as they
are deflected by the magnetic fields of the ring (Sokolov-Ternov effect)~\cite{Sokolov:1963zn}.
The beam polarization grows and approaches asymptotically an equilibrium value with a time constant depending on the 
characteristics of the ring, for HERA typically 1/2 hour.
Polarizations as large as 60\% were achieved.
Spin rotators and polarimeters were essential components of the HERA lepton beam
~\cite{Buon:1985de,Barber:1993ui,Barber:1992fc,Beckmann:2000cg}.
Spin rotators in front of and behind the experiment provided longitudinal polarization at the interaction point and at one of the two beam polarimeters. The two beam polarimeters were based on Compton back-scattering of circularly polarized laser light. They continuously monitored the transverse and longitudinal polarization of the lepton beam.

The HERMES spectrometer was designed to detect the scattered lepton and produced hadrons within
a wide angular acceptance and with good momentum resolution.
Particular emphasis was given to the particle identification capabilities which allowed for pion, kaon
and proton separation over almost the whole 
momentum range \cite{Akopov:2000qi}.
The HERA beam lines passed through the non-instrumented horizontal mid-plane of the spectrometer.
A horizontal iron plate shielded the beam lines from the 1.5 Tm dipole field of the spectrometer magnet,
thus dividing the spectrometer in two identical halves.
The geometrical acceptance of $\pm 170$ mrad 
horizontally and $\pm(40-140)$ mrad vertically resulted in
detected scattering angles ranging from 40 to 220 mrad.
Tracking was provided by several stages of drift chambers before and after the spectrometer magnet.
The combination of signals from a lead-glass calorimeter, a preshower detector, a transition radiation detector and a ring-imaging Cherenkov provided lepton identification with very high efficiency and purity better than 99\% as well as pion, kaon and proton separation over almost the whole momentum range of 2--15~GeV.
All components are described in detail in 
~\textcite{Ackerstaff:1998xb}.

A recoil detector 
was installed in the target region for the last 1.5 years of HERMES data taking with
unpolarized hydrogen and deuterium targets in order to enhance access to hard exclusive processes, in particular to deeply virtual Compton
scattering.

\subsection{JLab experiments}

Experiments at Jefferson Lab utilized the highest polarization electron beams (85\%) with energy ranging from 0.8 GeV close to 6 GeV.
The technologies of polarizing beam and target follow those pioneered and further developed at SLAC.

The beam was provided by the Continuous Electron Beam Accelerator Facility (CEBAF) \cite{Leemann:2001dg}, 
which used polarized electron guns based on a ``superlattice" of a thin gallium arsenide (GaAs) layer on top of
GaAs-phosphide bulk matter illuminated by circularly polarized 
photons from high intensity lasers 
\cite{Sinclair:2007ez,Stutzman:2007ny}.
Subsequently, the polarized electrons passed up to five times the 
two linear accelerators based on superconducting radio frequency technology and connected by two recirculation arcs.
The spin direction of the electrons was manipulated using the crossed electric and magnetic fields of Wien filters, which allow for rapid spin rotation.
Their direction was inverted every about 30 ms.
Beam polarimetry was employed at several stages of the acceleration process.
CEBAF delivered polarized beams simultaneously to the three experimental halls (Hall A, B and C)
with the option to independently dial the energy and intensity.
Typical beam intensities ranged from a few nA in Hall B to over 100 $\mu$A in the other two halls \cite{Kazimi:2004zv}.

Longitudinal polarized solid state ammonia (NH$_3$) targets 
for the
proton and ND$_3$ for the deuteron were 
employed at Hall B \cite{Keith:2003ca}.
These targets are based on similar techniques as discussed before for the SLAC and CERN experiments for both polarization and polarimetry.
Hall A used a polarized $^3$He target.
The target polarization was measured by both the NMR technique of 
adiabatic fast passage and a technique based on electron 
paramagnetic resonance~\cite{Romalis:1998ik}.
Average target polarizations of about 55\% were obtained.

Hall A and C were both instrumented with  small acceptance but high resolution spectrometers that could cope with the highest beam intensities but measured at fixed scattering angles.
These spectrometers are equipped for high resolution tracking, precise time-of-flight measurements and 
lepton/hadron separation~\cite{Alcorn:2004sb}.

Hall B was instrumented with the CEBAF Large Acceptance Spectrometer (CLAS)~\cite{Mecking:2003zu}.
The CLAS design was based on a toroidal field, generated by six superconducting coils arranged around the beam line.
The six coils naturally divided the detector into six independent spectrometers, each of them containing a set of drift chambers for tracking, a gas Cerenkov counter for electron/pion separation, an 
array of scintillator counters for particle identification using 
time of flight measurements, and electromagnetic calorimeters for neutral particle identification. For charged particles, CLAS covered polar angles between 8$^\circ$ and 142$^\circ$ in the laboratory frame and between 60\% and 80\% of the azimuthal angles.

\subsection{Hadronic scattering experiments}

While deep inelastic lepton-nucleon scattering has long been a standard tool of the trade in the study of unpolarized and polarized nucleon  structure, much has been learned from polarized hadronic scattering as well. 
The first high-energy primary polarized proton beams were achieved at the Zero-Gradient Synchrotron at Argonne National Laboratory in 1973.  Proton beams there were initially accelerated to 6~GeV with a polarization of about $60$\%, and shortly thereafter polarized beams up to 12~GeV were achieved.  In the 1990s at Fermilab, secondary beams of polarized protons or antiprotons from lambda or antilambda decays opened up new kinematic regions for polarized hadronic scattering, with polarized beams of up to 200~GeV ($\sqrt{s}=19$~GeV).  Polarized hadronic scattering experiments at center-of-mass energies more than an order of magnitude higher were achieved with the inauguration of the Relativistic Heavy Ion Collider for polarized protons in 2001.

\subsubsection{The Relativistic Heavy Ion Collider}

RHIC is located at Brookhaven National Laboratory in New York.  
RHIC was built to collide heavy ions at center-of-mass energies of up to 200~GeV per colliding nucleon pair and polarized protons at center-of-mass energies ranging from 50 to 500~GeV. Collision of asymmetric species, 
{\it i.e.}~different species in the two beams, is also possible due to independent rings with independent steering magnets.  The first polarized proton collisions were achieved at a center-of-mass energy of 200~GeV in December 2001.

The RHIC storage ring is 3.83~km in circumference and is designed with six interaction points (IPs) at which beam collisions are possible. Up to 112 particle bunches per ring can be injected, in which case the time between bunch crossings at the IPs is 106~ns.  Polarizations of up to 65\% for 100~GeV proton beams and about
$60$\% for 255~GeV beams have been achieved.  The maximum luminosities achieved thus far are $5 \times 10^{31}$~$\textrm{cm}^{-2}$~$\textrm{s}^{-1}$ at $\sqrt{s}=200$~GeV and $2 \times 10^{32}$~$\textrm{cm}^{-2}$~$\textrm{s}^{-1}$ at $\sqrt{s}=510$~GeV.

Three experiments have studied polarized proton collisions at RHIC.  There are two ongoing large experiments, STAR~\cite{Ackermann:2002ad} and PHENIX~\cite{Adcox:2003zm}, each of which have more than 500 collaborators total working on both the heavy ion and polarized proton programs, and the smaller BRAHMS~\cite{Adamczyk:2003sq} experiment, with fewer than 100 collaborators, which took data through 2006.
In additional to the program of proton spin structure measurements,
the transverse single-spin asymmetry in elastic proton-proton scattering has also been measured to constrain the hadronic 
spin-flip amplitude in this reaction~\cite{Adamczyk:2012kn}.

\subsubsection{RHIC as a polarized $p+p$ collider} \label{section:collider}
RHIC is the first and only high-energy polarized proton-proton collider in the world.
A number of technological developments and advances over the past several decades have made it possible to create a high-current polarized proton source, maintain the beam polarization throughout acceleration and storage, and obtain accurate measurements of the degree of beam polarization at various stages from the source to full-energy beams in RHIC.  
For an overview of RHIC as a polarized-proton collider 
see \textcite{Alekseev:2003sk}.  
In the case of polarized-proton running at RHIC, a pulse of polarized H$^-$ ions from the source is accelerated to 200~MeV in the linac, then stripped of its electrons as it is injected and captured as a single bunch of polarized protons in the Booster, which accelerates the protons to 1.5~GeV. The bunch of polarized protons is then transferred to the Alternating Gradient Synchrotron (AGS) and accelerated to 24~GeV before injection into RHIC.  Each bunch is accelerated in the AGS and injected into RHIC independently, with the two RHIC rings being filled one bunch at a time. The direction of the spin vector is selected for each bunch separately.  The nominal fill duration is eight hours, after which the beams are dumped and fresh beams are injected into RHIC.  The bunch-by-bunch spin patterns in consecutive fills are varied in order to reduce potential systematic effects.

Polarized proton injection uses an optically-pumped polarized H$^-$ ion source (OPPIS)~\cite{Zelenski:2002gb}.  
H$^-$ polarization at the source
of 85\% has been achieved.

Siberian snakes \cite{Derbenev:1978hv}, a series of spin-rotating dipoles, so named because of the beam trajectory through the magnets and the fact that they were developed at Novosibirsk in Russia, are used to overcome both imperfection and intrinsic depolarizing
resonances in RHIC. There are two snakes installed in each RHIC ring at diametrically opposite points along the rings. The two snakes in each ring rotate the spin vector $180^\circ$ about perpendicular horizontal axes, without perturbing the stable spin direction and with only local distortion of the beam orbit. In this way, all additive depolarizing effects from resonances are avoided.

For RHIC to provide full-energy polarized beams, the polarization must be measurable at various stages of acceleration in order to identify and address possible origins of depolarization at each step. Only RHIC polarimetry will be discussed here. There are two types of polarimeters installed in RHIC. The fast proton-carbon ($p$C) polarimeter~\cite{Nakagawa:2008zzb} takes advantage of a known analyzing power, $A_N^{p\textrm{C}} \approx 0.01$, in the elastic scattering of polarized protons with carbon atoms ($p^{\uparrow } + \textrm{C}\rightarrow p^{\uparrow } + \textrm{C}$), which originates from interference between electromagnetic and hadronic elastic scattering amplitudes.  The $p$C polarimeter can make measurements in less than ten seconds and provide immediate information on the stability or decay of the beam polarization from a few data points taken over the several hours of a fill.  Calibration of the $p$C polarimeter to within an absolute beam polarization of less than 5\% can then be provided by measuring polarized elastic $p+p$ scattering with a polarized hydrogen-jet-target polarimeter~\cite{Zelenski:2005mz}.  With the hydrogen-jet-target polarization of greater than 90\% known to better than 2\% in absolute polarization ~\cite{Okada:2005gu}, the absolute beam polarization can be determined by exploiting the symmetry of the process.

The naturally stable spin direction through acceleration and storage in RHIC is transverse to the proton's momentum, in the vertical direction.  Spin rotator dipole magnets have been used to achieve both radial and longitudinal spin \cite{MacKay:2003}.  
The rotators are located outside the interaction regions of the PHENIX and STAR experiments, giving both experiments the ability to choose independently whether they want longitudinally or transversely polarized collisions.  The BRAHMS experiment, having no spin rotators available, focused on transverse spin measurements.  The local nature of the spin rotator magnets means that the STAR and PHENIX experiments must each have their own way of checking the direction of the spin vector at their respective interaction regions.

Observed azimuthal transverse single spin asymmetries in the production of forward neutrons~\cite{Bazilevsky:2003bm} and forward charged particles can be used to provide local polarimetry.  These asymmetries are exploited by the experiments to measure the degree to which the beam polarization is vertically transverse, radially transverse, or longitudinal. 
More information on local polarimetry at PHENIX and STAR can be found in \textcite{Adare:2007dg} and \textcite{Kiryluk:2005gg}.

\begin{figure}[tbp]
  \begin{center}
    \includegraphics[width=\hsize]{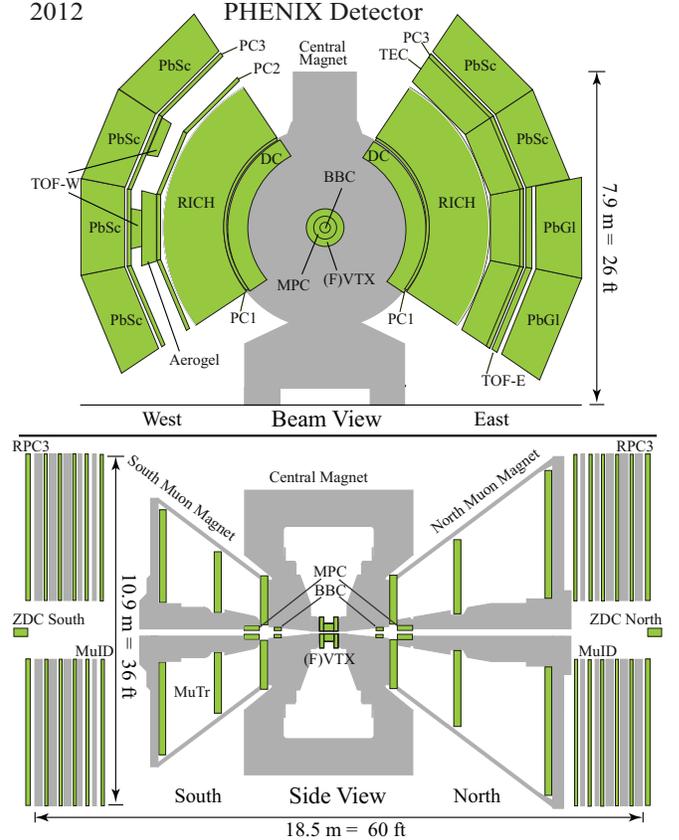}
    \end{center}
  \caption{\label{fig:PHENIXDet}
    The PHENIX detector at RHIC as configured for data taking in 2012.}
  \end{figure}

\subsubsection{RHIC experiments}
\paragraph{The PHENIX detector}
PHENIX was designed as a large, multi-purpose experiment with fast data acquisition and high granularity over a limited acceptance.  See Fig.~\ref{fig:PHENIXDet} for beam and side views of the PHENIX detector as configured for data taking in 2012. 
There are two central arms with an acceptance covering a pseudorapidity range $|\eta|<0.35$ and $\Delta\phi=\frac{\pi}{2}$ each in azimuth.  
The central arms include drift and pad
chambers (DC, PC1, PC2, PC3) for momentum and position measurements, a
ring-imaging Cherenkov detector (RICH) primarily for electron
identification, small-acceptance time-of-flight and aerogel
counters (TOF-E, TOF-W, Aerogel) for charged hadron identification, and electromagnetic
calorimetry (PbSc, PbGl). Electronics-level triggering in
the central arms uses information from the calorimetry
and ring-imaging Cherenkov detector. There are
two muon spectrometers covering a pseudorapidity of
$1.2 < |\eta| < 2.4$, consisting of tracking chambers and
muon identifier panels (MuTr, MuID). Resistive plate chambers (RPC3) were added in 2011 and 2012 to improve triggering on high-momentum
muons for $W$ boson measurements. 
Forward electromagnetic calorimetry (MPC) covering 
$3.0 < |\eta| < 3.9$ was added in 2006 and 2007, and silicon vertex detectors ((F)VTX) for heavy flavor measurements over $|\eta| < 2.4$ were added in 2011 and 2012.

For luminosity measurements, identical zero-degree hadronic calorimeters (ZDC) are located in the RHIC tunnel at $\pm 18$~m from the nominal IP for all RHIC experiments.
PHENIX also uses quartz Cherenkov beam-beam counters (BBC) positioned around
the beam pipe at $\pm 1.44$~m from the nominal interaction point
as a minimum-bias trigger detector and for polarization-averaged as well as spin-dependent luminosity measurements.  Collision rates for 500~GeV $p+p$ running reach $\sim$3~MHz, and the electronics-level triggers select events to reduce this rate to approximately 7~kHz of recorded data.

\paragraph{The STAR detector}
The Solenoidal Tracker at RHIC (STAR)
was designed as a large, multi-purpose detector with wide acceptance, making it well suited for correlation measurements.  The core of STAR is a time-projection chamber, which covers $2\pi$ in azimuth and has tracking capabilities over $|\eta|<1.3$ and good particle identification for $|\eta|<1$.  There is electromagnetic calorimetry for $-1 < \eta < 2$.  In the forward direction, there is additional electromagnetic calorimetry for $2.5<\eta<4.0$.  Recent upgrades include a time-of-flight detector with 100~ps resolution for additional particle identification, and tracking based on 
Gem detectors for $1<\eta<2$ was partially installed for 2012 data-taking to enable charge-sign discrimination of forward electrons from $W$ boson decays.

In addition to the zero-degree hadronic calorimeters identical among the RHIC experiments, STAR has scintillator beam-beam counters positioned around the beam pipe covering $3.4 < |\eta| < 5.0$, which provide a minimum-bias trigger as well as spin-averaged and spin-dependent luminosity measurements along with the ZDCs.

\paragraph{The BRAHMS detector}
The BRAHMS detector was a smaller experiment at RHIC designed for excellent momentum measurement and charged particle identification over a very broad range of rapidities.  It consisted of two movable spectrometer arms covering small solid angles, the Forward Spectrometer, which could be positioned as close as 2.3$^\circ$ from the beam pipe, and the Midrapidity Spectrometer, which could be moved to cover an angular range from $30^\circ < \theta < 95^\circ$.  The spectrometer arms included five dipole magnets, time-projection chambers, multi-wire drift chambers, time-of-flight hodoscopes, and threshold as well as ring-imaging Cherenkov detectors.  Global detectors consisted of a silicon array for multiplicity measurements, threshold Cherenkov beam-beam counters for event vertex and timing determination as well as luminosity measurements, and ZDCs identical to those used by PHENIX and STAR.

\section{The Proton Spin Puzzle}

We begin our discussion of physics results by first describing 
how the small value of the ``quark spin content" 
$g_A^{(0)}$ is obtained 
from polarized deep inelastic scattering and the
first moment of the $g_1$ spin structure function.

In QCD the first moment of $g_1$ is determined from 
the dispersion relation for polarized photon-nucleon scattering and
the light-cone operator product expansion.
One finds that the first moment of $g_1$ 
is related to the scale-invariant axial charges of the target
nucleon by
\begin{eqnarray}
& &
\int_0^1 dx \ g_1^p (x,Q^2)
\nonumber \\
&=&
\Biggl( {1 \over 12} g_A^{(3)} + {1 \over 36} g_A^{(8)} \Biggr)
\Bigl\{1 + \sum_{\ell\geq 1} c_{{\rm NS} \ell\,}
\alpha_s^{\ell}(Q)\Bigr\}
\nonumber \\
& &
+ {1 \over 9} g_A^{(0)}|_{\rm inv}
\Bigl\{1 + \sum_{\ell\geq 1} c_{{\rm S} \ell\,}
\alpha_s^{\ell}(Q)\Bigr\}  +  {\cal O}({1 \over Q^2})
 + \ \beta_{\infty}
.
\nonumber \\
\label{eqc50}
\end{eqnarray}
Here $g_A^{(3)}$, $g_A^{(8)}$ and $g_A^{(0)}|_{\rm inv}$ are the
isovector, SU(3) octet and scale-invariant  flavor-singlet
axial-charges respectively.
The flavor non-singlet $c_{{\rm NS} \ell}$
and singlet $c_{{\rm S} \ell}$ Wilson coefficients
are calculable in $\ell$-loop perturbative QCD.
These perturbative QCD coefficients
have been calculated
to $O(\alpha_s^3)$ precision \cite{Larin:1997qq}.
For $\alpha_s = 0.3$ typical of
the deep inelastic experiments one finds
$\Bigl\{1 + \sum_{\ell= 1}^3 c_{{\rm NS} \ell\,}
\alpha_s^{\ell}(Q)\Bigr\}
=
0.85
$
and
$\Bigl\{1  + \sum_{\ell = 1}^3 c_{{\rm S} \ell\,}
\alpha_s^{\ell}(Q)\Bigr\}
=
0.96
$.
The term $\beta_{\infty}$
represents a possible leading-twist subtraction constant from
the circle at infinity when one closes the contour in the
complex plane in the dispersion relation
\cite{Bass:2004xa}.
The subtraction constant affects just the first moment and corresponds to a contribution at Bjorken $x$ equal to zero.

In terms of the flavor-dependent axial-charges
\begin{equation}
2M s_{\mu} \Delta q =
\langle p,s |
{\overline q} \gamma_{\mu} \gamma_5 q
| p,s \rangle
\label{eqc55}
\end{equation}
the isovector, octet and singlet axial charges are
\begin{eqnarray}
g_A^{(3)} &=& \Delta u - \Delta d
\nonumber \\
g_A^{(8)} &=& \Delta u + \Delta d - 2 \Delta s
\nonumber \\
g_A^{(0)}|_{\rm inv}/E(\alpha_s)
\equiv
g_A^{(0)}
&=& \Delta u + \Delta d + \Delta s
.
\label{eqc56}
\end{eqnarray}
Here
\begin{equation}
E(\alpha_s) = \exp \int^{\alpha_s}_0 \! d{\tilde \alpha_s}\,
\gamma({\tilde \alpha_s})/\beta({\tilde \alpha_s})
\label{eqc54}
\end{equation}
is a renormalization group factor
which corrects
for the (two loop) non-zero anomalous dimension
$\gamma(\alpha_s)$
of the singlet axial-vector current
\begin{equation}
J_{\mu5} = 
\bar{u}\gamma_\mu\gamma_5u
                  + \bar{d}\gamma_\mu\gamma_5d
                  + \bar{s}\gamma_\mu\gamma_5s  
\label{eqc53}
\end{equation}
which 
is close to one and which 
goes to one in the limit
$Q^2 \rightarrow \infty$.
The symbol $\beta$ denotes the QCD beta function
$\beta (\alpha_s)
= - \biggl(11 - \frac{2}{3} f\biggr) (\alpha_s^2 / 2 \pi) + ...$
and
$\gamma$ is given by
$\gamma(\alpha_s) = f (\alpha_s / \pi)^2 + ...$
where $f$ (=3) is the number of active flavors \cite{Kodaira:1979pa}.
The singlet axial charge, $g_A^{(0)}|_{\rm inv}$,
is independent of the renormalization scale $\mu$
and corresponds
to
$g_A^{(0)}(Q^2)$ evaluated in the limit $Q^2 \rightarrow \infty$.
The flavor non-singlet axial-charges
$g_A^{(3)}$ and $g_A^{(8)}$ are renormalization group invariants.
We are free to choose
the QCD coupling $\alpha_s(\mu)$ at either a hard or a soft scale
$\mu$.
The perturbative QCD expansion of $E(\alpha_s)$
remains close to one -- even for large values of $\alpha_s$.
If we take $\alpha_s \sim 0.6$ as typical of the infra-red
then
$
E(\alpha_s) \simeq
1 - 0.13 - 0.03 + ... = 0.84 + ...
$
where $-0.13$ and $-0.03$ are the ${\cal O}(\alpha_s)$
and ${\cal O}(\alpha_s^2)$ corrections respectively.

In the naive parton model $g_A^{(0)}$ is interpreted
as the fraction of the proton's spin which is carried by the intrinsic spin of its quark and anti-quark constituents.
The experimental value of $g_A^{(0)}$ is obtained
through measuring $g_1$ and combining the first moment integral
in Eq.(12) with knowledge of $g_A^{(3)}$ and $g_A^{(8)}$
from other processes plus theoretical calculation of
the perturbative QCD Wilson coefficients.

The isovector axial-charge is measured independently in neutron
$\beta$-decays
($g_A^{(3)} = 1.270 \pm 0.003$
\cite{PhysRevD.86.010001})
and the octet axial charge is commonly taken to be the value 
extracted from hyperon $\beta$-decays assuming a 2-parameter 
SU(3) fit ($g_A^{(8)} = 0.58 \pm 0.03$ \cite{Close:1993mv}).
However, it should be noted
the uncertainty quoted for $g_A^{(8)}$ has been a matter of
some debate \cite{Jaffe:1989jz,Ratcliffe:2004jt}.
SU(3) symmetry may be badly broken and some have suggested
that the error on $g_A^{(8)}$ should
be as large as 25\%~\cite{Jaffe:1989jz}.
A recent re-evaluation of the nucleon's axial-charges
in the
Cloudy Bag model taking into account the effect of the
one-gluon-exchange hyperfine interaction and the pion cloud
plus kaon loops led to the value
$g_A^{(8)} = 0.46 \pm 0.05$~\cite{Bass:2009ed}.
The model reduction of $g_A^{(8)}$ from the SU(3) value comes
primarily from the pion cloud with $g_A^{(3)}$ taking its physical
value.

\begin{figure}[tbp]
\begin{center}
\includegraphics[width=\hsize]{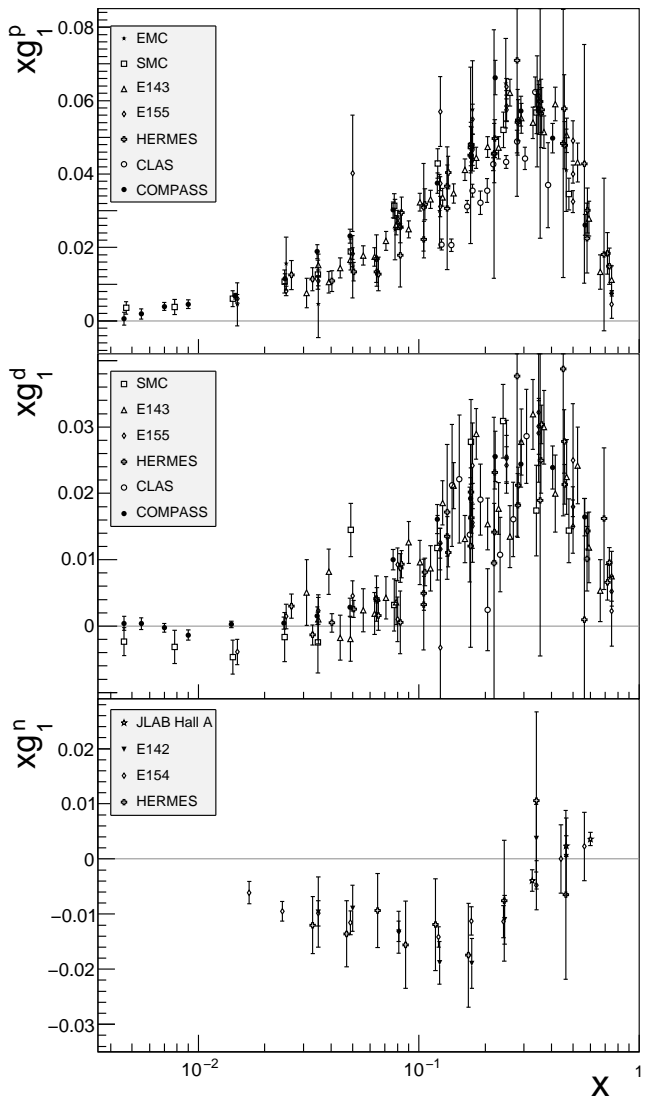}
{\caption[Delta]
{World data on $x g_1$ as a function of $x$ for the proton (top),
 the deuteron (middle) and the neutron (bottom) at the $Q^2$ of
 the measurement. 
 Only data points for $Q^2> 1~\GeV^2$ and $W>2.5~\GeV$ are shown.
 Error bars are statistical errors only.}
\label{fig:fig1}}
\end{center}
\end{figure}

Deep inelastic measurements of $g_1$ have been performed in experiments at CERN, DESY, JLab and SLAC.
An overview of the world data on the nucleon's $g_1$ spin structure
function is shown in Fig.~4.
This data is published in
EMC \cite{Ashman:1989ig},
SMC \cite{Adeva:1998vv},
E142 \cite{Anthony:1996mw},
E143 \cite{Abe:1998wq},
E154 \cite{Abe:1997cx},
E155 \cite{Anthony:2000fn},
E155 \cite{Anthony:1999rm},
HERMES \cite{Airapetian:2007mh},
JLab \cite{Zheng:2003un,Dharmawardane:2006zd},
and COMPASS \cite{Alekseev:2010hc,Alexakhin:2006vx}.
There is a general consistency among all data sets.
The kinematic reach of the different experiments is visible in
Fig.~5.
COMPASS have the smallest-$x$ data, down to $x \sim 0.004$.

\begin{figure}[tbp]
  \begin{center}
    \includegraphics[width=\hsize]{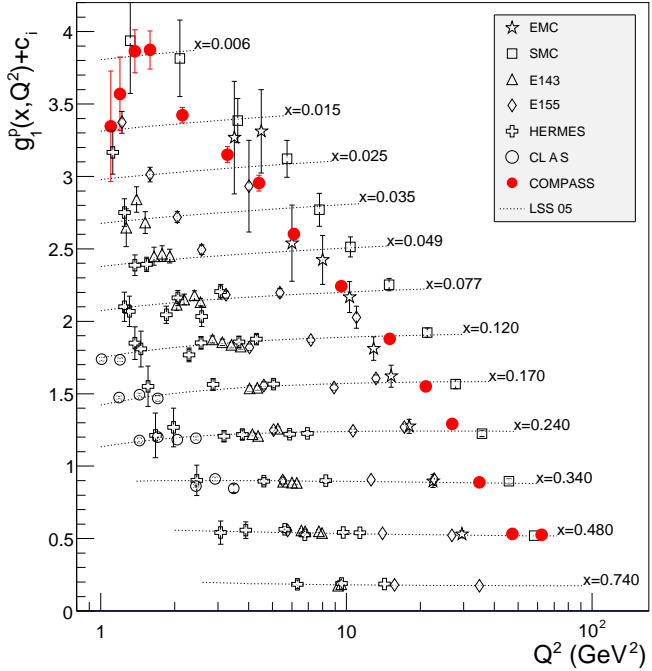}
    \caption{\label{fig:kinematics}
World data for $g_1(x,Q^2)$ for the proton with $Q^2> 1~\GeV^2$
and $W>2.5~\GeV$. For clarity a constant $c_i = 0.28(11.6-i)$ has
been added to the $g_1$ values within a particular $x$ bin starting
with $i=0$ for $x=0.006$.
Error bars are statistical errors only.
(Also shown is the QCD fit of \textcite{Leader:2005ci}.)}
    \end{center}
  \end{figure}

There are several striking features in the data.
COMPASS measurements of the deuteron spin structure function $g_1^d$
show the remarkable feature that $g_1^d$ is consistent with zero in the small-$x$ region between 0.004 and 0.02 \cite{Alexakhin:2006vx}.
In contrast, the isovector part of $g_1$ 
is observed to rise at small $x$ as 
$g_1^{p-n} \sim x^{-0.22 \pm 0.07}$~\cite{Alekseev:2010hc}
and 
is much bigger than the isoscalar $g_1^{d}$.
This compares to the situation in the unpolarized
structure function $F_2$ where 
the small-$x$ region is dominated by isoscalar gluonic exchanges.

\subsection{Spin sum-rules}

To test deep inelastic sum-rules it is necessary to have all data points at the same value of $Q^2$.
Next-to-leading order (NLO) QCD-motivated fits taking into account
the scaling violations associated with perturbative QCD
are used to evolve all the data points to the same $Q^2$.
First moment sum-rules are then evaluated
by extrapolating these fits to $x=0$ and to $x=1$,
or using a Regge-motivated extrapolation of the data.
Next-to-leading order (NLO) QCD-motivated fits discussed in Section V.C are used
to extract from these scaling violations the parton distributions
and in particular the gluon polarization.

Polarized deep inelastic scattering experiments are interpreted
in terms of a small value for the flavor-singlet axial-charge.
For example, COMPASS found 
using the SU(3) value for $g_A^{(8)}$ 
\cite{Alexakhin:2006vx}
and no leading twist subtraction constant
\begin{equation}
g_A^{(0)}|_{\rm pDIS, Q^2 \rightarrow \infty}
=
0.33 \pm 0.03 ({\rm stat.}) \pm 0.05 ({\rm syst.}) .
\end{equation}
(This deep inelastic quantity misses 
 any contribution to $g_A^{(0)}|_{\rm inv}$ 
 from a possible delta function at $x=0$).
When combined with $g_A^{(8)} = 0.58 \pm 0.03$,
the value of $g_A^{(0)}|_{\rm pDIS}$ in Eq.(17)
corresponds to a negative strange-quark polarization
\begin{eqnarray}
\Delta s_{Q^2 \rightarrow \infty}
&=&
\frac{1}{3}
(g_A^{(0)}|_{\rm pDIS, Q^2 \rightarrow \infty} - g_A^{(8)})
\nonumber \\
&=&
- 0.08 \pm 0.01 ({\rm stat.}) \pm 0.02 ({\rm syst.})
\end{eqnarray}
-- that is,
polarized in the opposite direction to the spin of the proton.
With this $\Delta s$, the following values for the
up and down quark polarizations are obtained
\begin{eqnarray}
\Delta u_{Q^2 \rightarrow \infty}
=
0.84 \pm 0.01 ({\rm stat.}) \pm 0.02 ({\rm syst.})
\nonumber \\
\Delta d_{Q^2 \rightarrow \infty}
=
-0.43 \pm 0.01 ({\rm stat.}) \pm 0.02 ({\rm syst.})
\end{eqnarray}
The non-zero value of $\Delta s_{Q^2 \rightarrow \infty}$
in Eq.(18) is known as
the violation of the Ellis-Jaffe sum-rule \cite{Ellis:1973kp}.

The extracted value of $g_A^{(0)}|_{\rm pDIS}$ required 
to be understood by theory,
and the corresponding polarized strangeness, 
depend on
the value of $g_A^{(8)}$.
If we instead use the value $g_A^{(8)} = 0.46 \pm 0.05$
the corresponding experimental value of
$g_A^{(0)}|_{\rm pDIS}$
would increase to $g_A^{(0)}|_{\rm pDIS} = 0.36 \pm 0.03 \pm 0.05$
with
\begin{equation}
\Delta s 
\sim -0.03 \pm 0.03 .
\end{equation}
We shall discuss the value of $\Delta s$ in more detail in
Sections V and VI in connection with more direct measurements
from semi-inclusive deep inelastic scattering plus global
fits to spin data, models and recent lattice calculations 
with disconnected diagrams (quark sea contributions) included.

The Bjorken sum-rule \cite{Bjorken:1966jh,Bjorken:1969mm}
for the isovector part of $g_1$ follows from 
current algebra and is a fundamental prediction of QCD.
The first moment of the isovector part of $g_1$ is
determined 
by the nucleon's isovector axial-charge
\begin{equation}
\int_0^1 dx g_1^{p-n}
=
{1 \over 6} g_A^{(3)}
\Bigl\{1 + \sum_{\ell\geq 1} c_{{\rm NS} \ell\,}
\alpha_s^{\ell}(Q)\Bigr\} .
\end{equation}
up to a 1\% correction from charge symmetry violation
suggested by a recent lattice calculation \cite{Cloet:2012db}.
It has been confirmed in polarized deep inelastic scattering at
the level of 5\%.
The value of $g_A^{(3)}$ extracted from the most recent
COMPASS data is
$1.28 \pm 0.07 ({\rm stat.}) \pm 0.010 ({\rm syst.})$
\cite{Alekseev:2010hc}
and compares well 
with the Particle Data Group value
$1.270 \pm 0.003$ deduced from neutron beta-decays
\cite{PhysRevD.86.010001}.

\begin{figure}[tbp]
\centerline{
\epsfig{file=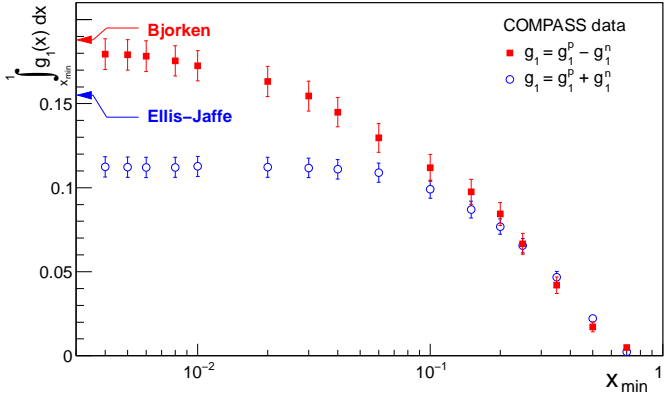,width=0.49\textwidth,clip=}}
\caption{\small
Convergence of the first moment integral of $g_1$ as a
function of the lower integration limit $x_{\rm min}$ 
for the Bjorken integral 
(isospin non-singlet) and 
the Ellis--Jaffe integral (iso-singlet)
from the COMPASS proton and deuteron data at 
$Q^2=3~\GeV^2$.
The arrows indicate the theoretical expectations.
Error bars are statistical errors only.}
\label{fig:g1NS}
\end{figure}

The evolution of the Bjorken integral 
$\int_{x_{\rm min}}^1 dx g_1^{p - n}$ 
as a function of $x_{min}$
as well as the isosinglet integral 
$\int_{x_{\rm min}}^1 dx g_1^{p + n}$ are shown in Fig.~6.
The Bjorken sum-rule and isosinglet integral converges
in the measured $x$ region.
Note that a large contribution, about $50\%$,
of the Bjorken sum-rule comes from $x \lesssim 0.15$.
The integral for the first moment of 
$g_1^{p + n}$ saturates at $x \sim 0.05$:
the isosinglet part of $g_1$ is close to 
zero in this measured range of small Bjorken $x$.

The nucleon's second spin structure function $g_2$ is believed to satisfy the Burkhardt-Cottingham sum-rule
$\int_0^1 dx g_2 =0$ \cite{Burkhardt:1970ti}.
The most precise measurements to date in polarized deep inelastic scattering come from
the SLAC E155 and E143 experiments, which report
$\int_{0.02}^{0.8} dx  \ g_2^p = -0.042 \pm 0.008$
for the proton and
$\int_{0.02}^{0.8} dx  \ g_2^d = -0.006 \pm 0.011$
for the deuteron at $Q^2=5$~GeV$^2$ \cite{Anthony:2002hy}.
E155 estimate a contribution about $0.02$ 
 to the first moment of the proton $g_2$ 
 come from the $x$ range between 0 and 0.02 
 from the twist-two (Wandzura-Wilczek) part of $g_2$:
 $g_2^{\rm WW}(x) = \int_x^1 {dy \over y} g_1(y) - g_1(x)$.

\subsection{Proton spin puzzles}

The results from polarized deep inelastic scattering pose the following questions:
\begin{itemize}
\item
How is the spin ${1 \over 2}$ of the proton built up from the spin and
orbital angular momentum of the quarks and gluons inside?
\item
Why is the quark spin content $g_A^{(0)}|_{\rm pDIS}$ so small?
\item
How about $g_A^{(0)} \neq g_A^{(8)}$~?
What separates the values of the octet and singlet axial-charges?
How reliable is the SU(3) value of $g_A^{(8)}$~?
\item
Is the proton spin puzzle a valence quark or sea/glue effect?
\item
Can we extract information about the quark and gluon
orbital angular momentum contributions from experiments,
and with minimal model dependence?
\end{itemize}
We next discuss the theoretical development and experimental 
work that has been performed to address these questions and
the physics interpretation of present measurements.

\subsection{Spin and the singlet axial-charge}

There are two key issues involved in understanding the small
value of $g_A^{(0)}|_{\rm pDIS}$:
the physics interpretation of the flavor-singlet axial-charge
$g_A^{(0)}$
and possible SU(3) breaking in the extraction of $g_A^{(8)}$
from hyperon $\beta$-decays.
How big really is the OZI violation
$\Delta s = \frac{1}{3} (g_A^{(0)}|_{\rm pDIS} - g_A^{(8)})$?

Theoretical QCD analysis based on the axial anomaly
leads to the formula
\begin{equation}
g_A^{(0)}
=
\biggl(
\sum_q \Delta q - 3 {\alpha_s \over 2 \pi} \Delta g \biggr)_{\rm partons}
+ {\cal C}_{\infty}
\label{eqa10}
\end{equation}
-- see
~\textcite{Altarelli:1988nr}, \textcite{Efremov:1988zh},
\textcite{Carlitz:1988ab}, \textcite{Bass:1991yx} and
\textcite{Bass:2004xa}.
Here $\Delta g_{\rm partons}$ 
is the amount of spin carried
by polarized
gluon partons in the polarized proton
with
$\alpha_s \Delta g \sim {\tt constant}$ 
as $Q^2 \rightarrow \infty$
\cite{Altarelli:1988nr,Efremov:1988zh});
$\Delta q_{\rm partons}$ measures the spin carried by quarks and
anti-quarks carrying ``soft'' transverse momentum
$k_t^2 \sim {\cal O}(P^2, m^2)$ 
where $P^2$ is a typical gluon virtuality in the nucleon
and $m$ is the light quark mass.
The polarized gluon term is associated with events in polarized
deep inelastic scattering where the hard photon strikes a
quark or anti-quark generated from photon-gluon fusion and
carrying $k_t^2 \sim Q^2$ \cite{Carlitz:1988ab}.
It is associated with the QCD axial anomaly in perturbative QCD.
${\cal C}_{\infty}$ denotes a potential non-perturbative gluon
topological contribution
with support only at Bjorken $x=0$ \cite{Bass:2004xa}.
This term is discussed in Section VI on theoretical understanding.
It is associated with the possible subtraction constant in
the
dispersion relation for $g_1$.
If non-zero it would mean that
$\lim_{\epsilon \rightarrow 0} \int_{\epsilon}^1 dx g_1$
will measure
the difference of
the singlet axial-charge and the subtraction constant contribution;
that is, polarized deep inelastic scattering measures the combination
$g_A^{(0)}|_{\rm pDIS} = g_A^{(0)} - {\cal C}_{\infty}$.

Possible explanations for the small value of $g_A^{(0)}|_{\rm pDIS}$
extracted from polarized deep inelastic experiments that
have been suggested in the theoretical literature include
screening from positive gluon polarization,
possible SU(3) breaking in the isosinglet axial-charges $g_A^{(8)}$ and $g_A^{(0)}$,
negative strangeness polarization in the nucleon,
a possible topological contribution at $x=0$
plus connections to axial U(1) dynamics
discussed in 
\textcite{Fritzsch:1989ty}, 
\textcite{Narison:1994hv}, 
\textcite{Shore:2007yn} and \textcite{Bass:1999is}.

The two-loop QCD evolution factor $E(\alpha_s)$ in Eq.(15)
is associated with the polarized gluon term which carries all
the scale dependence.
The quark spin contribution $\Delta q_{\rm partons}$ and
the subtraction constant in Eq.(22) are QCD scale invariant.
The quark spin term $\Delta q_{\rm partons}$
 is also known as the
 JET and chiral scheme 
\cite{Leader:1998qv,Cheng:1996jr}
 and AB scheme 
\cite{Ball:1995td}
 version of quark polarization -- see Section V.C.
In an alternative approach, 
called
the $\overline{\rm MS}$ scheme \cite{Bodwin:1989nz},
$\sum_q \Delta q_{\overline{\rm MS}}$,
is defined as the total matrix element of 
the flavor-singlet 
axial-current (including the gluonic terms in Eq.(22)).
We return to this issue in Section V.C 
with discussion of QCD fits to experimental data.
The growth in the gluon polarization $\Delta g \sim 1 / \alpha_s$
at large $Q^2$ is compensated by growth with opposite sign in the
gluon orbital angular momentum.

One would like to understand the dynamics which yield a small
value of the singlet axial-charge extracted from polarized
deep inelastic scattering
and also the sum-rule for the longitudinal spin structure
of the nucleon
\begin{equation}
\frac{1}{2} = \frac{1}{2} \sum_q \Delta q + \Delta g + L_q + L_g
\end{equation}
where $L_q$ and $L_g$ denote the orbital angular momentum contributions.
Operator definitions of the different terms or combinations of
terms in this equation are discussed in
\textcite{Jaffe:1989jz}, \textcite{Ji:1996ek}, \textcite{Shore:1999be},
\textcite{Bakker:2004ib}, \textcite{Bass:2004xa}, 
\textcite{Chen:2008ag}, 
\textcite{Wakamatsu:2010qj}, \textcite{Leader:2011za} 
and most recently in \textcite{Hatta:2011ku}, \textcite{Ji:2012ba} 
and \textcite{Lorce:2012rr}.
We discuss orbital angular momentum and attempts to measure it
in Sections VI--VII.

There is presently a vigorous global program to disentangle the different contributions.
Key experiments include semi-inclusive polarized deep inelastic scattering (COMPASS and HERMES)
and polarized proton-proton collisions (PHENIX and STAR),
as well as deeply virtual Compton scattering 
and hard exclusive meson production
to learn about total angular momentum (COMPASS, HERMES and JLab).
Single spin observables in semi-inclusive scattering
from transversely polarized targets
is sensitive to orbital angular momentum in the proton.

\section{Quark and Gluon Polarization from Data}

Key observables needed to understand the small value of the singlet
axial-charge $g_A^{(0)}|_{\rm pDIS}$ are the
polarized strangeness and polarized glue in the nucleon.
The search for polarized strangeness has inspired a dedicated 
experimental
program with semi-inclusive deep inelastic scattering.
Further, much activity was motivated by the discovery of 
\textcite{Altarelli:1988nr} and \textcite{Efremov:1988zh} 
that polarized glue makes a scaling contribution to the 
first moment of $g_1$,
$\alpha_s \Delta g \sim {\tt constant}$.
If there were a large negative contribution
$-3 {\alpha_s \over 2 \pi} \Delta g$
with {\it e.g.}\ 
gluon polarization of the order of $\Delta g \simeq 2.5$ at 
$Q^2=10~\GeV^2$, then this could reconcile the small measured 
value of 
$g_A^{(0)}|_{\rm pDIS}$
with the 
naive parton model expectation of 
about 0.6 through Eq.~(22).
This suggestion
sparked a vigorous and ambitious program to measure
$\Delta g$.
Interesting channels include gluon mediated processes in
semi-inclusive polarized deep inelastic scattering
(COMPASS and HERMES) and hard QCD processes
in high-energy polarized proton-proton collisions at RHIC.

\subsection{Valence and sea polarization}

Semi-inclusive measurements of fast pions and kaons in the current
fragmentation region with final state particle identification can be
used to reconstruct the individual up, down and strange quark
contributions to the proton's spin.
In contrast to inclusive polarized deep inelastic scattering
where the $g_1$
structure function is deduced by detecting only the scattered
lepton, the detected particles in the semi-inclusive experiments are
high-energy (greater than 20\% of the energy of the incident photon)
charged pions and kaons in coincidence with the scattered lepton.
For large energy fraction $z=E_h/E_{\gamma} \rightarrow 1$ the most
probable occurrence is that the detected $\pi^{\pm}$ and $K^{\pm}$
contain the struck quark or anti-quark in their valence Fock state.
They therefore act as a tag of the flavor of the struck quark
\cite{Close:1979bt}.

\begin{widetext}

\begin{figure}[t!]
\includegraphics[width=\hsize]{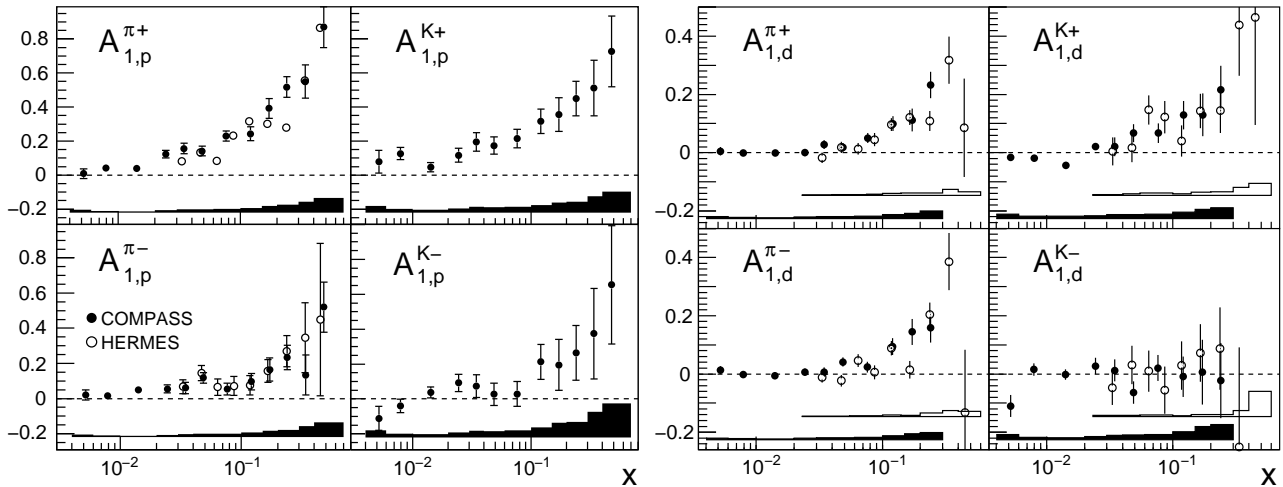}
\caption
{Semi-inclusive longitudinal double-spin asymmetries for identified pions and kaons from COMPASS \cite{Alekseev:2009ac,Alekseev:2010ub} and HERMES \cite{Airapetian:2004zf} for the proton (left) and the deuteron (right) as function of $x$ at the $Q^2$ of the measurements. 
The error bars and bands indicate the statistical and
systematic uncertainties, respectively.
Figure adapted from \textcite{Alekseev:2009ac} (left, proton target)
and \textcite{Alekseev:2010ub} (right, deuteron target).
}
\label{fig:fig9}
\end{figure}

\end{widetext}

In leading order the virtual-photon--proton double-spin 
(cross-section) asymmetry is
\begin{equation}
A_{1p}^{h} (x,Q^2) \simeq
{
\sum_{q,h} e_q^2 
\Delta f_q(x,Q^2) \int_{z_{\rm min}}^1 dz D_f^h (z, Q^2)
\over
\sum_{q,h} e_q^2 
f_q(x,Q^2) \int_{z_{\rm min}}^1 dz D_f^h (z, Q^2)
}
\label{eqi142}
\end{equation}
where $z_{\rm min} \sim 0.2$.
Here
$\Delta f_q(x,Q^2)$
is the quark (or anti-quark) polarized parton distribution, 
$f_q(x,Q^2)$ the unpolarized distribution
and
$e_q$ is the quark charge;
$
D_f^h (z, Q^2) = \int d p_t^2 D_q^h (z, p_t^2, Q^2)
$
is the fragmentation function for the struck quark or anti-quark
to produce a hadron $h$ ($=\pi^{\pm}, K^{\pm}$) carrying energy
fraction $z=E_h/E_{\gamma}$ in the target rest frame;
Note the integration over the transverse momentum 
$p_t$ \cite{Close:1991vp}.
Since pions and kaons have spin zero, the fragmentation functions
are the same for both polarized and unpolarized leptoproduction.
The fragmentation functions 
for $u \rightarrow \pi^+$ and $d \rightarrow \pi^-$
are known as ``favored" 
(where the fragmenting quark has the same 
 flavor as a valence quark in the final state hadron);
the fragmentation functions for 
$u \rightarrow \pi^-$ and $d \rightarrow \pi^+$
are known as ``unfavored".

This program for polarized deep inelastic scattering was 
pioneered by the SMC 
\cite{Adeva:1995yi,Adeva:1997qz} 
and
the HERMES 
\cite{Ackerstaff:1999ey,Airapetian:2003ct,Airapetian:2004zf}
experiments.
The most recent measurements 
from HERMES are reported in \textcite{Airapetian:2008qf} and
from COMPASS in \textcite{Alekseev:2010ub}.

The experimental strategy has been to measure the asymmetries
$A_{1}^h$ 
for charged hadron production and separated charged pion and
kaon production from 
proton and deuteron targets.
There is good agreement between the COMPASS and HERMES data
in the kinematic region of overlap -- see Fig.~7.
Flavor-separated polarized quark distributions for valence and
sea quarks are then extracted from the data 
using fragmentation functions 
that have been fitted to previous hadron production data, 
with the most accurate taken to be those from the DSS group
\cite{deFlorian:2007aj}
from a global fit to 
single-hadron production in $e^+e^-$, $ep$ and $pp$ collisions.

The polarizations of the up and down quarks are positive and
negative respectively, while the extracted sea polarization
data are consistent with zero 
-- see Table II which 
includes measurements from COMPASS \cite{Alekseev:2010ub},
HERMES \cite{Airapetian:2004zf} and SMC \cite{Adeva:1997qz}.

\begin{widetext}

\begin{table}[t!]
\caption{First moments for valence quark and light-sea polarization
from 
SMC, 
HERMES, 
and COMPASS. 
For each experiment the integrated sea is evaluated from data up to $x=0.3$ and, for SMC, assuming an isospin symmetric polarized sea.
}
{\begin{tabular}{lcccccr}
Experiment  &   $x$-range & $Q^2$ (GeV$^2$) &
$\Delta u_v$ & $\Delta d_v$ & $\Delta\bar{u}$ & $\Delta\bar{d}$ \\
\hline
SMC    & 0.003--0.7 & 10  & 
$0.73 \pm 0.10 \pm 0.07$ & $-0.47 \pm 0.14 \pm 0.08$ &
$0.01 \pm 0.04 \pm 0.03$ &  $0.01 \pm 0.04 \pm 0.03$ 
\\
HERMES & 0.023--0.6 & 2.5 & 
$0.60 \pm 0.07 \pm 0.04$ & $-0.17 \pm 0.07 \pm 0.05$ &
$0.00 \pm 0.04 \pm 0.02$ & $-0.05 \pm 0.03 \pm 0.01$
\\
COMPASS  & 0.006--0.7 & 10  & 
$0.67 \pm 0.03 \pm 0.03$  & $-0.28 \pm 0.06 \pm 0.03$ &
$0.02 ´\pm 0.02 \pm 0.01$ & $-0.05 \pm 0.03 \pm 0.02$ 
\\
\end{tabular}}
\vspace{3ex}
\end{table}

\end{widetext}

The COMPASS and HERMES determinations of 
the sum of strange and anti-strange polarisation $\Delta s(x)$
are shown together in Fig.~8, 
plotted in the combination $x \Delta s (x)$.
There is no evidence in the semi-inclusive data for 
large 
negative strange quark polarization in the nucleon.
The HERMES data covers the region $0.02 < x < 0.6$,
where
the extracted $\Delta s$
is consistent with a zero or small positive value.
This data integrates to 
$\int_{0.02}^{0.6} dx \Delta s = 0.037 \pm 0.019 \pm 0.027$
\cite{Airapetian:2008qf}
in contrast with the negative value for polarized strangeness,
Eq.(18), extracted from inclusive measurements of $g_1$.
COMPASS measurements \cite{Alekseev:2009ac,Alekseev:2010ub}
show no evidence of strangeness polarization
in the region $x > 0.004$
with the integrated $\Delta s = -0.02 \pm 0.02 \pm 0.02$.

The precise value of $\Delta s$ extracted from semi-inclusive scattering may be affected by any possible future improvement 
in the accuracy of the kaon fragmentation functions $D_q^{K}(z)$.
However a drastic change in the ratio
$\int dz D_{\bar s}^{K^+} / \int dz D_{u}^{K^+}$
would be needed to bring the first moment of $\Delta s$
extracted from semi-inclusive scattering
in agreement with the inclusive value, Eq.(18), obtained
using the SU(3) value of $g_A^{(8)}$ \cite{Alekseev:2010ub}.
More experimental data, especially on kaon fragmentation 
processes, are needed for improved precision on strangeness polarization in the nucleon.

\begin{figure}[tbp]
\includegraphics[width=0.45\textwidth]{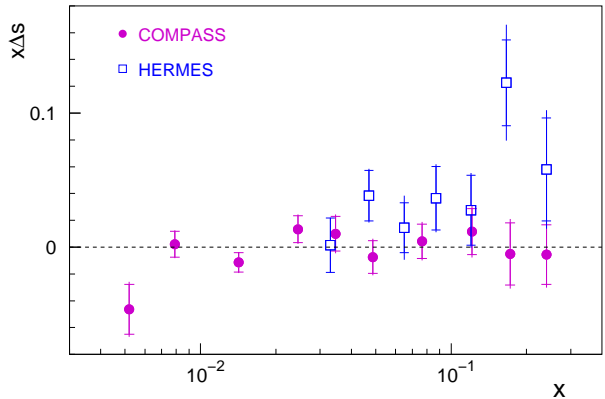}
\caption
{COMPASS \cite{Alekseev:2010ub} and 
HERMES \cite{Airapetian:2008qf}
 results for the strangeness polarization 
 $x\Delta s(x)$ 
 as function of $x$.
 The data are obtained in a leading-order analysis of SIDIS 
 asymmetries
(including those for charged kaons) and using the DSS fragmentation
        functions \cite{deFlorian:2007aj}.
The inner error bar represents the statistical uncertainty; the full bar the quadratic sum of statistical and systematic uncertainties.}
\label{fig:fig9}
\end{figure}

Semi-inclusive data are consistent with a small positive or zero
isospin asymmetry in the polarized sea 
$\Delta {\bar u} - \Delta {\bar d}$.
For the COMPASS data at 3 GeV$^2$
one finds
$
\int_{0.004}^{0.3} dx (\Delta {\bar u} - \Delta {\bar d})
 =
0.06 \pm 0.04 ({\rm stat.}) \pm 0.02 ({\rm syst.})$.
For HERMES data at 2.5 GeV$^2$
$
\int_{0.023}^{0.3} dx (\Delta {\bar u} - \Delta {\bar d})
 =
0.048 \pm 0.057 ({\rm stat.}) \pm 0.028 ({\rm syst.})$
\cite{Airapetian:2004zf}.
These values compare with the unpolarized sea measurement
$
\int_0^1 dx ({\bar u} - {\bar d})
 =
-0.118 \pm 0.012
$
from the E866 experiment at FNAL \cite{Towell:2001nh}.
A compilation of theoretical predictions is given in
\textcite{Peng:2003zm}.
Meson cloud models predict small negative isospin asymmetries
in the polarized sea
\cite{Cao:2001nu,Kumano:2001cu}
whereas statistical \cite{Bourrely:2001du}
and chiral quark soliton models \cite{Wakamatsu:2003wg}
predict positive values.
The COMPASS and HERMES results are consistent 
with these predictions within uncertainties.

The $W$-boson production program at RHIC~\cite{Bunce:2000uv} 
will provide additional flavor-separated measurements of 
polarized up and down quarks and anti-quarks.  
At RHIC the polarization of the $u, \bar{u}, d$, and $\bar{d}$ quarks in the proton is being measured directly using $W$ boson production in $u\bar{d} \rightarrow W^+$ and $d\bar{u} \rightarrow W^-$.  The charged weak boson is produced through a pure V$-$A coupling and the chirality of the quark and anti-quark in the reaction is fixed.  The $W$ is observed through its leptonic decay $W\rightarrow l\nu$, and the charged lepton is measured.  Measurement of the flavor-separated anti-quark helicity distributions via $W$ production in $p+p$ collisions is complementary to measurements via SIDIS in that there is no dependence on details of the fragmentation process, and the process scale, $Q^2 \approx M_W^2$ is significantly higher than any data from existing fixed-target polarized DIS experiments.  A parity-violating asymmetry for $W^+$ production in $p+p$ collisions at $\sqrt{s}=500$~GeV consistent with predictions based on anti-quark helicity distributions extracted from SIDIS has already been observed by both
PHENIX~\cite{Adare:2010xa} and
STAR~\cite{Aggarwal:2010vc}
based on data collected in 2009.  
Considerably improved results are expected from data taken in 2011 and 2012 with higher luminosities and polarization.
Preliminary results for both $W^+$ and $W^-$ asymmetries 
from STAR, 
based on data taken at the beginning of 2012,
are consistent with results from 
SIDIS and
suggest the possible asymmetry
$\Delta \bar{u} > \Delta \bar{d}$ 
for $x$ from 0.05--1~\cite{RHICWP:2012}.

An independent measurement of the strange-quark axial-charge
could be made through neutrino-proton elastic scattering.
This process measures the combination
$\frac{1}{2}
( \Delta u - \Delta d - \Delta s )_{\rm inv} 
- 0.01 g_A^{(0)}|_{\rm inv}$,
where the small last term is a correction from heavy-quarks
which has been calculated to 
LO \cite{Kaplan:1988ku} and NLO \cite{Bass:2002mv} accuracy.
The axial-charge measured in $\nu p$ elastic scattering
is independent of any assumptions about possible SU(3) breaking,
the presence or absence of a subtraction at
infinity in the dispersion relation for $g_1$ and the $x \sim 0$
behavior of $g_1$.
A recent suggestion for an experiment using low-energy neutrinos 
produced from pion decay at rest is discussed in 
\textcite{Pagliaroli:2012hq}.

In a recent analysis \cite{Pate:2008va} of parity violating
quasi-elastic electron and neutrino scattering data between
0.45 and 1 GeV$^2$
(from the JLab experiments G0 and HAPPEx and the Brookhaven experiment E734), 
the axial form-factor was extrapolated to $Q^2=0$ and favored negative or zero values of $\Delta s$ with large uncertainty.

\subsection{Gluon polarization}

\begin{figure}[tbp]
\includegraphics{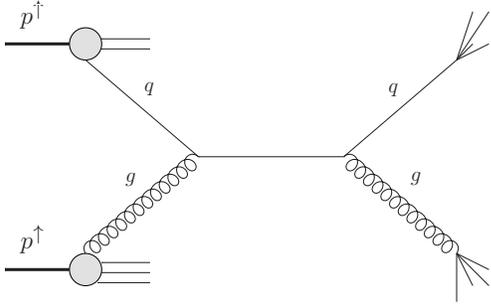}
\vspace{4.8cm}
\parbox{8.0cm}
{\caption[Delta]
{Jet production from quark-gluon scattering in polarized 
proton-proton collisions.} 
\label{figure:qgFeynman}}
\end{figure}

Polarized proton-proton scattering is sensitive to the ratio of
polarized to unpolarized glue, $\Delta g /g$, via leading-order interactions of gluons, 
as illustrated in Fig.~\ref{figure:qgFeynman}.
The first experimental attempt to look at gluon polarization was 
made by the FNAL E581/704 Collaboration using a 200~GeV polarized proton beam and a polarized proton target. 
They measured a 
longitudinal double-spin asymmetry $A_{LL}$ for inclusive 
multi-$\gamma$ and $\pi^0 \pi^0$ production consistent with zero within their sensitivities, suggesting that $\Delta g/g$ is not 
so large in the region of 
$0.05 \lesssim x_g \lesssim 0.35$~\cite{Adams:1994bg}.

COMPASS was conceived to measure $\Delta g$ via the study of the photon-gluon fusion process, as shown in Fig.~10.
The cross-section for this process is directly related to the 
(polarized) gluon distribution at the Born level.
The experimental technique consists of the reconstruction of 
charmed mesons~\cite{Alekseev:2009ad,Adolph:2012ca}
or high-$p_T$ hadrons~\cite{Ageev:2005pq}
in the final state to access $\Delta g$.  
For the charmed meson case COMPASS also performed a
NLO analysis which shifts probed $x_g$ to larger values.
The high-$p_T$ particle method leads to samples with larger statistics, 
but these have higher background contributions from QCD Compton processes and fragmentation. 
High-$p_T$ hadron 
production was also used in early attempts to access gluon polarization 
by HERMES~\cite{Airapetian:1999ib} and SMC~\cite{Adeva:2004dh}
and the most recent HERMES
determination \cite{Airapetian:2010ac} 
and COMPASS measurement \cite{Adolph:2012vj}.

\begin{figure}[htp]
\includegraphics[width=0.28\textwidth]{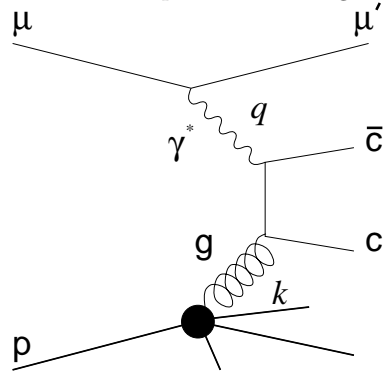}
{\caption[Delta]
{Production of a $c \bar{c}$ pair in polarized 
photon gluon fusion is being
used to measure gluon polarization in the polarized proton.
 }
\label{fig:fig16}}
\end{figure}

\begin{figure}[tbp]
\begin{center}
\includegraphics[width=0.45\textwidth]{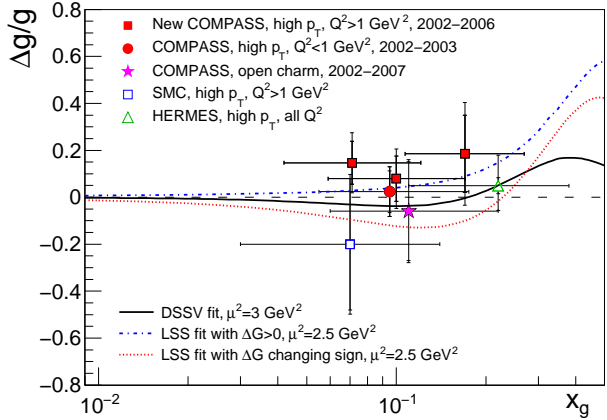}
\end{center}
\caption{
Gluon polarization $\Delta g/g$ from leading-order analyses of hadron
or hadron-pair production as function of the probed gluon momentum
fraction $x_g$. 
Also shown are NLO fits from 
\textcite{deFlorian:2009vb} and \textcite{Leader:2010rb}. 
(Figure adapted from~\textcite{Adolph:2012vj}.)
The inner error bar represents the statistical uncertainty; 
the full bar the quadratic sum of statistical and systematic uncertainties.
The horizontal bar indicates the $x_g$ range of the measurement.}
\label{fig:Delta_GbyG} \end{figure}

These measurements in lepton-nucleon scattering are listed in 
Table~\ref{tab:DIS_DG} for the ratio of the polarized to unpolarized 
glue $\Delta g/g$ and shown in 
Fig.~\ref{fig:Delta_GbyG} 
for leading-order (LO) analyses of the data.  
The data cluster around $x_g \sim 0.1$
with the exception of the COMPASS NLO point from open charm.
There is no evidence in the data 
for non-zero gluon polarization at this value of $x_g$.

The chance to measure $\Delta g$
was a main physics drive for polarized RHIC.
Experiments using the PHENIX and STAR detectors are investigating polarized glue in the proton.
Measurements of $\Delta g/g$ from RHIC
are sensitive to gluon polarization in the range
$0.02 \lesssim x_g \lesssim 0.3$
($\sqrt{s} = 200$~GeV) and
$0.06 \lesssim x_g \lesssim 0.4$
($\sqrt{s} = 62.4$~GeV)
for the neutral pion $A_{LL}$ measured
by PHENIX~\cite{Adare:2008aa, Adare:2008qb}
and inclusive jet production measured by STAR at 200~GeV
center-of-mass energy~\cite{Abelev:2007vt, Adamczyk:2012qj}.
Additional channels sensitive to $\Delta g$ at RHIC 
have been published as well~\cite{Abelev:2009pb, Adare:2010cy, Adare:2012nq}.

\begin{widetext}

\begin{table}[t!]
\caption{Polarized gluon measurements from deep inelastic experiments.}
\vspace{3ex}
{\begin{tabular}{@{}lllrr@{}}
  Experiment  &  process            &  $\langle x_g \rangle$
& $\langle \mu^2 \rangle$ (GeV$^2$) &  $\Delta g / g$   \\
\hline
HERMES \cite{Airapetian:1999ib}     &  hadron pairs        & 0.17 &  $ \sim 2 $  &
$ 0.41 \pm 0.18 \pm 0.03$
\\
HERMES \cite{Airapetian:2010ac}
     &  inclusive hadrons   & 0.22 &  $ 1.35 $    &
$ 0.049 \pm 0.034 \pm 0.010^{+0.125}_{-0.099}$
\\
SMC \cite{Adeva:2004dh}       &  hadron pairs        & 0.07 &              &
$ -0.20 \pm 0.28 \pm 0.10$
\\
COMPASS \cite{Ageev:2005pq,Procureur:2006sg}    & hadron pairs, $Q^2 < 1$  &  $ 0.085 $  &  $ 3 $ &
$0.016 \pm 0.058 \pm 0.054$
\\
COMPASS \cite{Adolph:2012vj} & hadron pairs, $Q^2 > 1$  & $0.09 $ &  $3$          &
$0.125 \pm 0.060 \pm 0.063$
\\
%
COMPASS \cite{Adolph:2012ca} 
& open charm (LO)  & $0.11$ & 13 & 
$-0.06 \pm 0.21\pm 0.08$ \\
COMPASS \cite{Adolph:2012ca} 
& open charm (NLO) & $0.20$ & 13 & 
$-0.13 \pm 0.15 \pm 0.15$ \\
\end{tabular}
}
\label{tab:DIS_DG}
\end{table}

\end{widetext}

\begin{figure}[tbp]
\centering
\includegraphics[width=\hsize]
{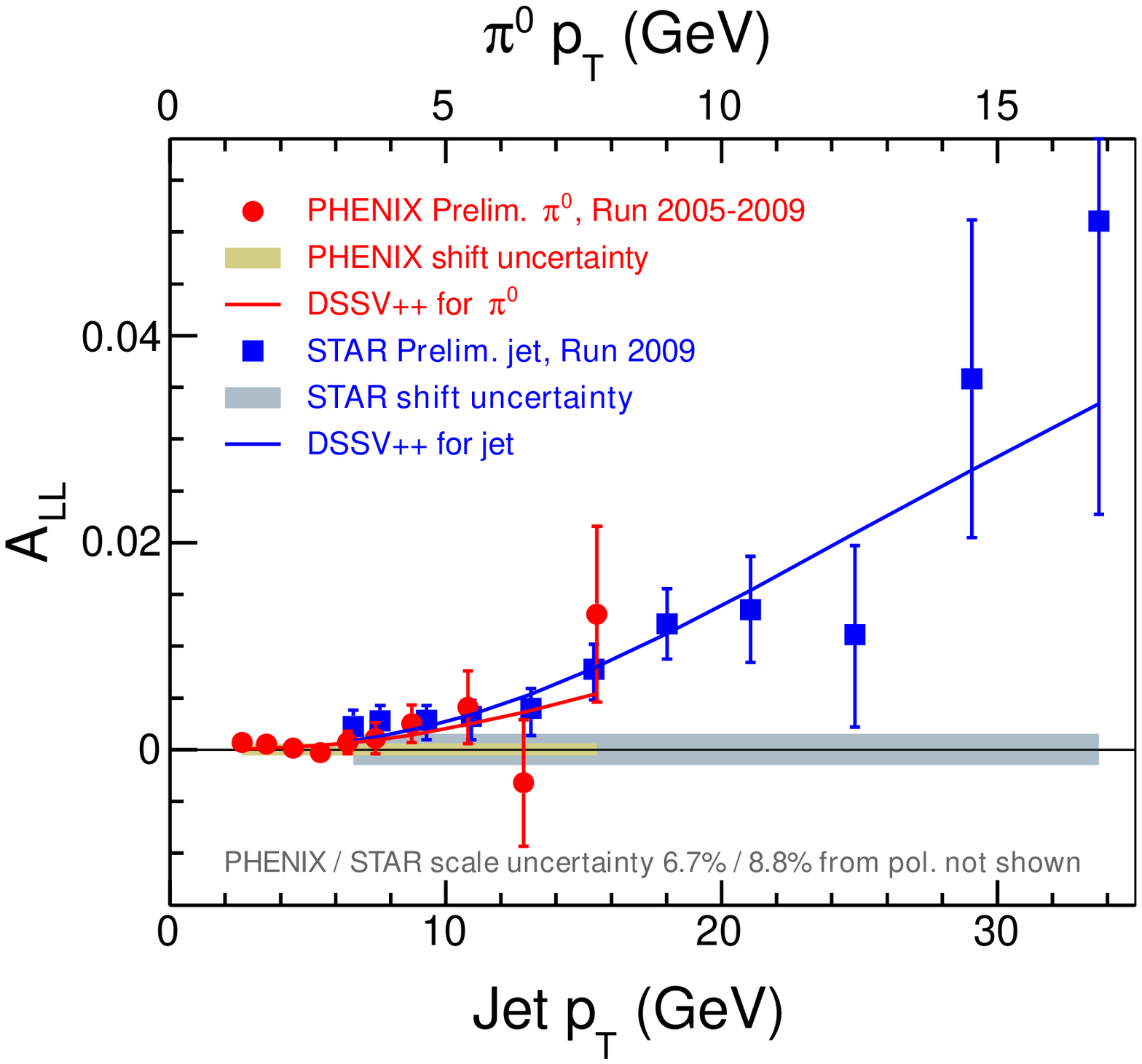}
%
\caption[Longitudinal double-spin asymmetry in $\pi^0$ production from PHENIX.]{The longitudinal double-spin asymmetry in $\pi^0$ production measured by PHENIX~\cite{Manion:2011zz} and in jet production measured by STAR~\cite{Djawotho:2011fga}, shown with calculations based on the DSSV polarized parton distributions that were updated to include these results; 
Figure from \textcite{RHICWP:2012}.  
The relationship between the pion and jet $p_T$ scales is given by the mean $z$ value of 
$\sim 0.5$~\cite{Adler:2006sc}.
Error bars represent the statistical uncertainty.
\label{fig:STARPHENIXDeltaG}}
\end{figure}

Combined preliminary results from PHENIX and STAR using more recent 200 GeV data than those published 
in~\textcite{Adare:2008aa} and~\textcite{Abelev:2007vt} 
are shown in Fig.~\ref{fig:STARPHENIXDeltaG}. The longitudinal double spin asymmetry in neutral pion production measured by PHENIX based on combined data from 2005, 2006, and 2009 is shown as a function of pion $p_T$ (upper scale) ~\cite{Manion:2011zz}.  Figure~\ref{fig:STARPHENIXDeltaG} also shows the asymmetry in single-inclusive jet production as a function of jet $p_T$ (lower scale) measured by STAR based on data taken in 2009 \cite{Djawotho:2011fga}, providing the first evidence for non-zero gluon polarization in the proton.  The relationship between the pion and jet $p_T$ scales is given by the mean $z$ value of $\sim 0.5$~\cite{Adler:2006sc}.  The data are shown with a calculation using helicity distributions extracted from a global fit to polarized world data from DIS, semi-inclusive DIS, and proton-proton collisions ("DSSV") ~\cite{deFlorian:2008mr,deFlorian:2009vb} that was updated to include these results~\cite{RHICWP:2012}.  See the following section for more details about fits to helicity distributions. A given $p_T$ bin for single inclusive jet or hadron production generally samples a wide range of $x_g$ values. However, dijet measurements in $p+p$ collisions provide better constraints on the $x_g$ values probed. Preliminary STAR results for dijet production have also been released~\cite{Walker:2011vs} and confirm the non-zero double-spin asymmetry seen in single jet production.

While there is now evidence in the RHIC data that gluon polarization in the proton is non-zero, 
the measurements indicate that polarized glue, by itself, is not sufficient to resolve the difference between the small value of 
$g_A^{(0)}|_{\rm pDIS}$ and 
the naive constituent quark model prediction, $\sim 0.6$, 
through the polarized glue term
$-3 {\alpha_s \over 2 \pi} \Delta g$.
Note however that gluon polarization $\Delta g \sim 0.2 - 0.3$ 
would still make a significant contribution to the spin of the 
proton in Eq.(23).

\subsection{NLO QCD motivated fits to spin data}

Global NLO perturbative QCD analysis are performed on polarization
data sets including both lepton-nucleon and proton-proton collision
data.
The aim is to extract the polarized quark and gluon parton distributions.
These analysis, starting from 
\textcite{Ball:1995td} and
\textcite{Altarelli:1996nm},
frequently use DGLAP evolution and are performed in a given factorization scheme.
This QCD fit approach has more recently been extended to a 
global analysis of data from polarized DIS, 
semi-inclusive polarized DIS and 
high-energy polarized proton-proton collisions
\cite{deFlorian:2008mr,deFlorian:2009vb}.

The separation of $g_1$ into ``hard'' and ``soft'' 
contributions is not unique and depends on the choice of ``factorization scheme''. 
For example, one might use a kinematic cut-off on the partons'
transverse momentum squared ($k_t^2 > \mu^2$) to define
the factorization scheme and thus separate the hard and soft
parts of the phase space for the photon-parton collision.
The cut-off $\mu^2$ is called the factorization scale.
The coefficients in Eq.(8) have the perturbative expansion
$C^q = \delta(1-x) +
         {\alpha_s \over 2\pi} f^q(x, {Q^2 / \mu^2})$
and
  $C^g = {\alpha_s \over 2\pi} f^g(x, {Q^2 / \mu^2})$
where
the strongest singularities in the functions $f^q$ and $f^g$
as $x \rightarrow 1$ are $\ln (1-x)/(1-x)_+$ and $\ln (1-x)$
respectively,
see {\it e.g.}\ \textcite{Lampe:1998eu}.
The deep inelastic structure functions are dependent on 
$Q^2$ and independent of the factorization scale $\mu^2$ and
the ``scheme'' used to separate
the $\gamma^{*}$-parton cross-section into ``hard'' and ``soft''
contributions.

Examples of different ``schemes'' used in the literature are
the modified minimal subtraction (${\rm \overline{MS}}$)
\cite{'tHooft:1972fi,Bodwin:1989nz}
to regulate the mass singularities which arise in scattering
from massless partons,
the ``AB''
\cite{Ball:1995td} and 
``CI'' (chiral invariant) \cite{Cheng:1996jr} 
or ``JET''
\cite{Leader:1998qv} schemes.
In the $\overline{\rm MS}$ scheme the polarized gluon distribution
does not contribute explicitly to the first moment of $g_1$.
In the AB, CI and JET schemes on the other hand the polarized gluon
(axial anomaly contribution) $\alpha_s \Delta g$ 
does contribute explicitly to the first moment since 
$\int_0^1 dx \ C^{(g)} = - {\alpha_s \over 2 \pi}$
-- see the spin decomposition in Eq.(22).

The $\mu^2$ dependence of the parton distributions is given
by the DGLAP equations \cite{Altarelli:1977zs}
\begin{eqnarray}
{d \over dt} \Delta \Sigma(x,t) &=& 
\biggl[
\int_x^1 {dy \over y} 
\Delta P_{qq}({x \over y}, \alpha_s(t)) \Delta \Sigma (y,t)
\nonumber \\
& & \ \ \ \ \ \ \ \ \ \ + \ 2 f
\int_x^1 {dy \over y} 
\Delta P_{qg}({x \over y}, \alpha_s(t)) \Delta g (y,t)
\biggr]
\nonumber \\
{d \over dt} \Delta g (x,t) &=& 
\biggl[
\int_x^1 {dy \over y}
\Delta P_{gq}({x \over y}, \alpha_s(t)) \Delta \Sigma (y,t)
\nonumber \\
& &
\ \ \ \ \ \ \ \ \ \ \ \ \
+
\int_x^1 {dy \over y} 
\Delta P_{gg}({x \over y}, \alpha_s(t)) \Delta g (y,t) \ \biggr]
\nonumber \\
\label{eqi141}
\end{eqnarray}
where 
$\Delta \Sigma(x,t) = \sum_q \Delta q(x,t)$, 
$t = \ln \mu^2$
and $f$ is the number of active flavors.
The splitting functions $P_{ij}$ in Eq.(25) have been calculated at
leading order by \textcite{Altarelli:1977zs}
and at next-to-leading order by
\textcite{Zijlstra:1993sh},
\textcite{Mertig:1995ny}
and
\textcite{Vogelsang:1995vh}.

The largest uncertainties in the QCD fits are associated with the ansatz chosen for the shape of the spin-dependent quark and gluon
distributions at a given input scale.
Further, the SU(3) value of $g_A^{(8)}$ ($=0.58 \pm 0.03$) 
is assumed in present fits though no significant change 
in the $\chi^2$ quality of the fits should be expected if 
one instead took a value of $g_A^{(8)}$ with possible 20\% 
SU(3) breaking 
included.\footnote{We thank S. Taneja and R. Windmolders for discussion on this issue.}
The values for the quark and gluon spin contents 
($\Delta \Sigma$ and $\Delta g$)
obtained in recent NLO fits are listed in Table IV
with results
quoted from
\textcite{Blumlein:2010rn} (BB10: DIS data),
\textcite{Nocera:2012hx} (NFRR12: DIS data),
\textcite{Leader:2010rb} (LLS10: DIS and SIDIS data) and
\textcite{Hirai:2008aj} (AAC08: DIS and RHIC data)
and
\textcite{deFlorian:2008mr,deFlorian:2009vb}
(DSSV08: DIS, SIDIS and proton-proton collision data).

The most complete fits in terms of maximum included data are 
from the DSSV group, which take the SU(3) value for $g_A^{(8)}$. 
One finds need for a large negative contribution to
$\Delta s$ 
from small $x$, outside the measured $x$ range when SIDIS data 
is included.
The values obtained in this approach 
for
$\int_{x_{\rm min}}^1 dx \Delta s(x)$
are
$\Delta s = -0.057$ with $x_{\rm min}=0$
and about $-0.001$ with $x_{\rm min}=0.001$.
That is, to reproduce the SU(3) value of the octet axial-charge,
the negative polarized strangeness obtained from inclusive
$g_1$ measurements gets pushed into the unmeasured small-$x$ 
range, $x < 0.004$.
It is interesting here to note that, 
historically (before COMPASS, HERMES and RHIC Spin),
the proton spin puzzle was assumed to be associated
with strangeness/sea/glue
polarization in the newly opened 
kinematics of EMC, SLAC and SMC, $x$ between 0.1 and 0.01.
We now have accurate SIDIS measurements down to $x \sim 0.004$
which show no evidence for large sea/glue polarization effects.
With the SIDIS measurements of $\Delta s$, either one needs SU(3) breaking in the octet axial-charge or strangeness/glue effects at 
very small $x$.
Without including the most recent data from 2009 or later,
\textcite{deFlorian:2008mr,deFlorian:2009vb}
find a best-fit 
full first moment 
$\int_0^1 dx \Delta g(x) = -0.084$ at $Q^2 = 10\,{\rm GeV}^2$.
With a very large 
$\Delta \chi^2 / \chi^2 =2\%$ allowed range,
the 
truncated first moment 
$\int_{0.001}^1 dx \Delta g = 0.013^{+0.702}_{-0.314}$
was obtained.
With these errors $\Delta g$ is still not precise.

\begin{widetext}

\begin{table}[t!]
\caption{\small
First moments of the polarized singlet-quark and gluon 
distributions at the scale
4 GeV$^2$ in the $\overline{\rm MS}$ scheme;
values quoted from \textcite{Nocera:2012hx}.}
\begin{center}
\begin{tabular}{|c|c|c|c|c|c|}
\hline
          & DSSV08	&  BB10	&  LSS10	&  AAC08 & NFRR12 \\
\hline
$\Delta \Sigma (Q^2)$ &
$0.25 \pm 0.02$ & $0.19 \pm 0.08$ & $0.21 \pm 0.03$ & 
$0.24 \pm 0.07$ & $0.31 \pm 0.10 $
\\ \hline
$\Delta g (Q^2)$      &
$-0.10 \pm 0.16$ & $0.46 \pm 0.43$ & $0.32 \pm 0.19$ & 
$0.63 \pm 0.19$ & $-0.2 \pm 1.4$ \\
\hline
\end{tabular}
\end{center}
\label{tab1}
\end{table}

\end{widetext}

A recent attempt to extract polarized parton distributions
from inclusive polarized deep inelastic data using
neural network techniques is reported in \textcite{Nocera:2012hx}.
In this approach no assumption is made about the functional form of the input distributions, greatly reducing the primary (and notoriously difficult to quantify) systematic uncertainty on 
parton distribution fits.  
The neural network method has already been used quite successfully in the parametrization of the unpolarized parton distributions, 
with results published at NLO in 2010~\cite{Ball:2010de} and 
now NNLO in 2012~\cite{Ball:2011uy}.  However, in the case of the application of the neural network method in the extraction of polarized parton distributions, 
thus far only inclusive polarized DIS data has 
been incorporated, similar to the BB10 fit~\cite{Blumlein:2010rn}.

\section{Theoretical Understanding}

In relativistic quark models some of the proton's spin is carried 
by quark orbital angular momentum.
One has to take into account the four-component Dirac spinor
for the quarks
$\psi =
{N \over \sqrt{4 \pi}}
\biggl({ f \atop i \sigma \cdot {\hat{r}} g }\biggr)$,
where 
$f$ and $g$ are functions of the spatial coordinates
and $N$ is a normalization factor.
The lower component of the Dirac spinor is p-wave with intrinsic spin
primarily pointing in the opposite direction to spin of the nucleon.
In the MIT Bag model, where quarks are confined in an infinite square well potential with radius $R$, 
one finds the depolarization factor
$N^2 \int_0^R dr r^2 (f^2 - {1 \over 3} g^2)
= 0.65$
for $\Delta q$ in the proton 
with all quarks in the $(1s)$ ground-state~\cite{Jaffe:1989jz}.
That is, 35\% of the proton's spin content is shifted 
into orbital angular momentum through the confinement potential.

More detailed calculations of non-singlet axial-charges 
in relativistic constituent quark models 
are sensitive to the confinement 
potential,
effective color-hyperfine interaction,
pion and kaon clouds plus additional wavefunction
corrections (associated with center of mass motion)
chosen to reproduce the physical value of $g_A^{(3)}$.

This physics was recently investigated within the Cloudy Bag model 
~\cite{Myhrer:2007cf,Thomas:2008ga,Bass:2009ed}.
The Cloudy Bag was designed to model confinement and spontaneous 
chiral symmetry breaking, taking into account pion physics and the manifest breakdown of chiral symmetry at the bag surface in the MIT 
bag.
If we wish to describe proton spin data including matrix elements of
$J_{\mu 5}^3$, $J_{\mu 5}^8$ and $J_{\mu 5}$,
then we would like to know that the model versions of these currents
satisfy the relevant Ward identities
(the divergence equations for these currents).
For the scale-invariant
non-singlet axial-charges $g_A^{(3)}$ and $g_A^{(8)}$,
corresponding to the matrix elements of partially conserved
currents, the model is well designed to make a solid prediction.

The effective color-hyperfine interaction has the quantum numbers of
one-gluon exchange (OGE).
In models of hadron spectroscopy this interaction plays an important role in the nucleon-$\Delta$ and $\Sigma-\Lambda$ mass differences,
as well as the nucleon magnetic moments~\cite{Close:1979bt} and
the spin and flavor dependence of parton distribution
functions~\cite{Close:1988br}.
It shifts total angular-momentum between spin and orbital
contributions and, therefore, also contributes to model
calculations of the octet axial-charges~\cite{Myhrer:1988ap}.
With OGE included
(together with a phenomenological wavefunction renormalization
 to ensure $g_A^{(3)}$ takes the physical value),
the model is in very good
agreement with the SU(3) fit to the nucleon 
and hyperon axial-charges extracted
from $\beta$-decays with $g_A^{(8)}$ predicted to be around 0.6.

Next, the pion cloud induces SU(3) breaking in the nucleon's
axial-charges. The pion cloud further shifts intrinsic spin into orbital angular momentum~\cite{Schreiber:1988uw,Tsushima:1988xv}.
Including 
pion and kaon cloud corrections gives the model result
$g_A^{(8)} = 0.46 \pm 0.05$ 
(with the corresponding semi-classical
 singlet axial-charge or spin fraction
being $0.42 \pm 0.07$ before inclusion of gluonic effects)
~\cite{Bass:2009ed}.
With this Cloudy Bag value for $g_A^{(8)}$
the corresponding experimental value of $g_A^{(0)}|_{\rm pDIS}$
would increase to 
$g_A^{(0)}|_{\rm pDIS} = 0.36 \pm 0.03 \pm 0.05$,
considerably reducing 
the apparent OZI violation 
$
\frac{1}{3}
(g_A^{(0)}|_{\rm pDIS} - g_A^{(8)})$
that one needs to explain.

A recent lattice calculation with disconnected diagrams included 
\cite{QCDSF:2011aa} gave $\Delta s = -0.02 \pm 0.01$
in the $\overline{\rm MS}$ scheme at 7.4 GeV$^2$.
This value compares with the Cloudy Bag prediction
$\Delta s \sim -0.01$ before gluonic degrees of freedom are included.
These numbers are in good agreement with
the values extracted
from polarized SIDIS data
by COMPASS and HERMES with the DSS fragmentation functions.

Gluon polarization has been investigated in bag and 
light-cone models and in studies of heavy-quark axial-charges.
The nucleon's charm-quark axial-charge was interpreted 
to give an estimate of gluon polarization 
$| \Delta g (m_c^2) | \lesssim 0.3$
with $\alpha_s (m_c^2) = 0.4$ \cite{Bass:2011zn}.
This upper bound corresponds to 
$|3{\alpha_s \over 2 \pi} \Delta g| \lesssim 0.06$. 
Values of $\Delta g \sim 0.3$ and $\sim 0.5$ at 1 GeV$^2$
were obtained in the MIT Bag model~\cite{Chen:2006ng}
and in a light-cone model~\cite{Brodsky:1994kg} respectively.
These theoretical values are consistent with the extractions
of gluon polarization from COMPASS, HERMES and RHIC Spin data.

To understand 
${\cal C_{\infty}}$,\footnote{In the notation of Eq.(12):
$\beta_{\infty}
= - {1 \over 9} {\cal C_{\infty}}
\Bigl\{1 + \sum_{\ell\geq 1} c_{{\rm S} \ell\,}
\alpha_s^{\ell}(Q)\Bigr\}  
$.}
deep inelastic sum-rules are derived using the operator product
expansion and the dispersion relation for deeply-virtual
photon-nucleon scattering.
Two important issues with the dispersion are the convergence of
the first moment integral at the highest energies and any contribution from closing the circle in the complex momentum plane.
The subtraction constant, if finite, corresponds to a constant real
term in the forward Compton scattering amplitude. It
affects just the first moment integral and thus behaves like a
$\delta (x)$ term with support only at $x=0$.
A subtraction constant yields a finite correction to the sum-rule obtained from integrating only over finite non-zero values of
Bjorken $x$. 
One can show \cite{Bass:2004xa} that non-local gluon topological structure requires consideration of a possible $\delta (x)$ subtraction constant. Whether it has finite value or not is sensitive to the realization of axial U(1) symmetry breaking by instantons 
\cite{Crewther:1978zz,'tHooft:1986nc} and the importance of topological structure in the proton.
The QCD vacuum is a Bloch superposition of states characterized
by non-vanishing topological winding number and non-trivial chiral
properties.
When we put a valence quark into this vacuum it can act as a
source which polarizes the QCD 
vacuum with net result that the spin ``dissolves''. 
Some fraction of the spin of the constituent quark
is shifted from moving partons into the vacuum at $x=0$.
This spin contribution becomes associated with
non-local gluon topology with support only at Bjorken $x=0$.

Valuable information about the spin puzzle also follows from looking at the $x$ dependence of $g_1$.
The small value of $g_A^{(0)}$ or ``missing spin'' is associated 
with a ``collapse'' in the isosinglet part of $g_1$ to something close to zero instead of a valence-like rise at $x$ 
less than about 0.05
-- see {\it e.g.}\ the $g_1$ data in Fig.~4 and 
the convergence of 
$\int_{x_{\rm min}}^1 dx g_1^{p+n}$ and
$\int_{x_{\rm min}}^1 dx g_1^{p-n}$ in Fig.~6. 
This isosinglet part is the sum of SU(3)-flavor singlet and octet contributions.
If there were a large positive polarized gluon contribution to the proton's spin, this would act to drive the small $x$ part of the singlet part of $g_1$ negative~\cite{Bass:1992bk}
-- that is, acting in the opposite direction to any valence-like
rise at small $x$.
However, gluon polarization measurements constrain this spin
contribution to be small in measured kinematics meaning that the sum of valence and sea quark contributions is suppressed at small $x$.
Neither the SU(3) flavor-singlet nor octet contribution breaks free
in the measured small $x$ region.
Hence, the suppression of $g_1^{p+n}$ at small $x$ should be either an isosinglet effect or a delicate cancellation between octet and
singlet contributions over an order of magnitude in small Bjorken $x$.

The $g_1^{p-n}$ data are consistent with quark model and 
perturbative QCD counting rules
predictions in the valence region $x > 0.2$ \cite{Bass:1999uj}.
The size of $g_A^{(3)}$ forces us to accept a large contribution 
from small $x$ (a non-perturbative constraint) and the rise in 
$g_1^{p -n}$ is in excellent agreement with the prediction
$g_1^{p-n} \sim  x^{-0.22}$ of hard Regge exchange
\cite{Bass:2006dq}
-- in particular a possible $a_1$ hard-pomeron cut involving 
the hard-pomeron which seems to play an important role in 
unpolarized deep inelastic scattering \cite{Cudell:1999kfa} 
and in the proton-proton total cross-section measured at LHC
\cite{Donnachie:2011aa}.
(Soft) Regge theory predicts that the singlet term should 
behave as $\sim N \ln x$ in the small $x$ limit, 
with the coefficient $N$ to be determined from experiment
\cite{Bass:1994xb,Close:1994yr}.
From the data, this normalization seems to be close to zero.

Where are we in our understanding of the spin structure of the proton and the small value of $g_A^{(0)}|_{\rm pDIS}$~?
Measurements of valence, gluon and sea polarization suggest that the polarized glue term 
$-3 \frac{\alpha_s}{2 \pi} \Delta g$
and the strange quark contribution $\Delta s_{\rm partons}$ 
in Eq.(22) are unable 
to resolve the small value of $g_A^{(0)}|_{\rm pDIS}$.
Two explanations are suggested within the theoretical and experimental uncertainties depending upon the magnitude of SU(3) breaking in the nucleon and hyperon axial-charges.
One is a value of $g_A^{(8)} \sim 0.5$
plus an axial U(1)
topological effect at $x=0$ associated with a finite subtraction constant in the $g_1$ dispersion relation.
The second is a much larger pion cloud reduction of $g_A^{(8)}$
to a value $\sim 0.4$.
Combining the theoretical error on the pion cloud chiral corrections
embraces both possibilities.
The proton spin puzzle seems to be telling us about the interplay of valence quarks with chiral dynamics and the complex vacuum structure 
of QCD.

Orbital angular momentum (OAM) in relativistic quark models
(for example the MIT and Cloudy bag models)
without explicit gluon degrees of freedom
has the usual interpretation of relativistic quantum mechanics.
For QCD dynamics the definition of OAM is more subtle because of
the gauge covariant derivative,
meaning that quark orbital momentum is in principle sensitive 
to the gluon fields in the nucleon that the quarks interact with.

Going beyond spin and helicity to consider also orbital and
total angular momentum, several operator decompositions
have been proposed.
Starting from the relation between angular momentum and
the energy-momentum tensor,
\textcite{Ji:1996ek} takes 
\begin{eqnarray}
     \vec{J}_q &=& \int d^3x ~\vec{x} \times \vec{T}_q \nonumber \\
                 &=& \int d^3x ~\left[ \psi^\dagger
     {\vec{\Sigma}\over 2}\psi + \psi^\dagger \vec{x}\times
          (-i\vec{D})\psi\right]
     \, ,  \nonumber \\
     \vec{J}_g &=& \int d^3x ~\vec{x} \times (\vec{E} \times \vec{B}) .
\label{ang}
\end{eqnarray}
The gauge covariant derivative $D_{\mu} = \partial_{\mu} +ig A_{\mu}$
with $A_{\mu}$ the gluon field
means 
that $L_q$ is {\it a priori}
sensitive to gluonic degrees of freedom.
The $J_q$ and $J_g$
quantities here are amenable to QCD lattice calculations and,
in principle, measurable through deeply virtual Compton scattering.
In an alternative approach, taking the + light cone component of
the QCD angular momentum tensor in $A_+=0$ gauge
\textcite{Jaffe:1989jz} proposed the operator decomposition
\begin{eqnarray}
M^{+12} &=&
\frac{1}{2} q_+^{\dagger} \gamma_5 q_+
+ q_+^{\dagger}
\biggl( \vec{x} \times i \vec{\partial} \biggr)^3 q_+
\nonumber \\
& &
+
2 {\rm Tr} F^{+j}
\biggl( {\vec x} \times i {\vec \partial} \biggr) A^j
+ {\rm Tr} \epsilon^{+-ij} F^{+i}A^j
\nonumber \\
\end{eqnarray}
where the gluon term in the gauge covariant
derivative is no longer present through the gauge fixing.

The connection between the quark and gluon total angular momentum
contributions $J^q$ and $J^g$ and
the QCD energy-momentum tensor allows us to write down
their LO QCD evolution equations \cite{Ji:1995cu}.
The quark and gluon total angular momenta in the infinite scaling
limit are given by 
$J_q ( \infty ) = \frac{1}{2} \{3 f / (16 + 3f)\}$
and
$J_g ( \infty ) = \frac{1}{2} \{16  / (16 + 3f)\}$,
with $f$ the number of active flavors
-- that is, the same scaling limit as the quark and gluon momentum contributions at infinite $Q^2$.
The Ji and Jaffe-Manohar definitions of orbital angular momentum
satisfy the same (LO) QCD evolution equation,
so, at LO, are equal in a model calculation if 
the glue contribution can be set equal to zero at a 
low-energy input scale.

To obtain information about the quark 
``orbital angular momentum'' $L_q$
we need to subtract the value of the ``intrinsic spin'' 
$S_q = {1 \over 2} \Delta q$
measured
in polarized deep inelastic scattering from the total quark
angular momentum $J_q$.
This means that $L_q$ is scheme dependent with different schemes
corresponding to different physics content depending on how the scheme handles information about the axial anomaly, large-$k_t$ physics and any possible ``subtraction at infinity'' in the dispersion relation for $g_1$.
The quark total angular momentum $J_q$ is anomaly free in QCD so 
that axial anomaly effects occur with equal magnitude and opposite sign in $L_q$ and $S_q$.
When looking at physical observables that are sensitive to OAM and
quark spin
(with possible axial-anomaly contribution)
it will be important to identify which OAM definition and which
scheme quantity is most relevant to the observable 
-- for example,
in SIDIS the largest $k_t$ events are included in the 
$\overline{\rm MS}$ version of $\Delta q$ whereas they are omitted in the JET scheme version \cite{Bass:2002jd}.

There are some theoretical subtleties when dealing with gluon angular momentum.
In the parton model the gluon polarization $\Delta g$ has a clean interpretation in light-cone gauge as the forward matrix element of 
the local Chern-Simons current $K_+$ 
(appearing in the QCD axial anomaly)
up to a surface term which has support only at $x=0$
\cite{Manohar:1990kr,Bass:2004xa}.
In light-cone gauge $K_+$ coincides 
with the gluon spin operator \cite{Jaffe:1995an}.
In general, Ji's $J_g$ in Eq.(26) 
is not readily separable into spin and orbital components.
New ideas have recently been investigated where one separates 
the gluon field into a ``physical" transverse part and
``pure" gauge part, with different conventions how to deal
with the gauge part 
\cite{Chen:2008ag,Wakamatsu:2010qj,Hatta:2011ku,Lorce:2012rr}.
Discussion of total orbital angular momentum 
involving gluonic degrees of freedom should 
be labelled with respect to the scheme or convention used.

To connect quark model predictions with lattice calculations and
fits to data it is necessary to use QCD evolution of the model
results from the low-energy scale where the model applies up to
the hard scale of deep inelastic scattering.
Model calculations
(and also lattice calculations without disconnected diagrams)
of $\Delta q$ are commonly understood
to refer to the scale invariant version of this quantity,
{\it e. g.}\
the chiral/JET or AB scheme quark spin contributions in Eq.~(22).
One chooses a model ansatz for the gluon polarization and
total angular momentum, typically $\Delta g = J_g = 0$ at
the model input scale.
For illustration, Fig.~13 shows the evolution of total 
and orbital angular momentum contributions 
in the Cloudy Bag from the model scale up to $Q^2=4$~GeV$^2$.
Various phenomenological investigations
\cite{Steffens:1995at, Mattingly:1993ej}
found that 
by going beyond leading order QCD 
(and including pions in the nucleon wavefunction),
the optimal fit 
to high-energy scattering data
involved taking the running coupling
$\alpha_s$ about 0.6--0.8 at the low energy input scale.
For this range of $\alpha_s$
the scale dependence of $\Delta \Sigma$ (in full QCD)
through Eq.(15) converges well.
(Going to higher orders in the model fits to data and putting in 
 pions raises the model input scale $\mu_0$ needed
 for the calculations.)
Table V compares the results of lattice calculations 
\cite{Hagler:2007xi}
for up- and down-quark spin and total angular momentum 
with the Cloudy Bag model results and
the values extracted from QCD fits 
to hard exclusive reaction data, GPDs \cite{Goloskokov:2008ib},
and transverse single spin asymmetries, TMDs \cite{Bacchetta:2011gx}.
The lattice calculation involves connected diagrams only 
(no axial-anomaly contribution) plus chiral extrapolation. 
The QCD fit numbers are central values modulo (possibly large) systematic errors from the model functional forms of distributions used in the fits.
There is good convergence of the different theoretical values with
``data".
Here, one has 
$L_u \sim -L_d \sim 15 \%$
at the scale of typical deep inelastic measurements.

\begin{figure}[tbp]
\centering
\includegraphics[width=0.45\textwidth,clip]{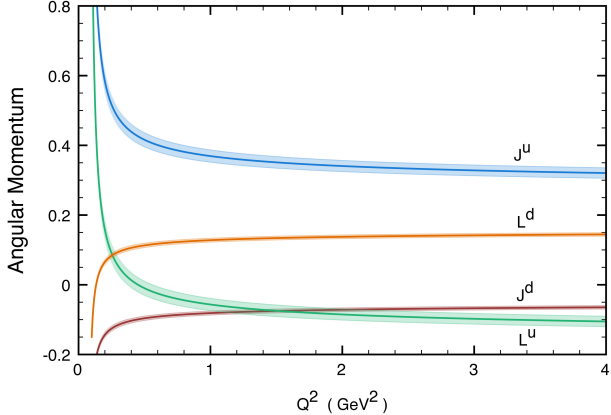}
\caption{Calculation of the NLO QCD evolution of
$J_u$, $L_d$, $J_d$, $L_u$
in the Cloudy Bag with model input scale $Q_0 = 0.4\,{\rm GeV}$.
Figure from \textcite{Thomas:2010zzc}, copyright 2010 WSPC.
}
\label{fig:ds_dsbar}
\end{figure}

In an alternative approach, the proton spin puzzle has also been
addressed in the Skyrme model,
where baryons emerge as topological solitons in the meson fields
at large number of colors $N_c$,
and in the chiral quark soliton model (ChQSM),
where explicit quark degrees of freedom are also present in the model.
The nucleon's axial charges in these models are sensitive to which
mesons are included in the model and the relative contribution of a
quark source and pure meson component.
In an early calculation \textcite{Brodsky:1988ip} found that 
$g_A^{(0)}$ vanishes in a particular version of the Skyrme model
with just pseudoscalar mesons.
Finite values of $\Delta s \sim -0.08$, 
close to the value obtained from inclusive $g_1$ measurements 
with good SU(3) assumed for $g_A^{(8)}$ 
are found in the ChQSM \cite{Wakamatsu:2006ba}.

\begin{table}[htp]
\caption{\small
Model, lattice and fit extractions of angular momentum contributions
in the proton, quoted for 4 GeV$^2$ (except GPD at 2 GeV$^2$).
}
\begin{center}
\begin{tabular}{|c|r|r|r|r|} \hline
            & Cloudy Bag         &  Lattice            & GPD & TMD \\ \hline
$\Delta u$  &  $0.85 \pm 0.06$    &  $0.82 \pm 0.07$  &     &     \\ \hline
$\Delta d$  &  $-0.42 \pm 0.06$   &  $-0.41 \pm 0.07$  &     &     \\ \hline
$J_u$       &  0.30            &  $0.24 \pm 0.05$    & 0.24    & 0.24 \\ \hline
$J_d$       &  $-0.04$           &  $0.00 \pm 0.05$    & 0.02    & 0.02 
\\
\hline
\end{tabular}
\end{center}
\label{tab1}
\end{table}

\section{Transverse Nucleon Structure and Orbital Angular Momentum}

Confinement induces transverse hadronic scales in the nucleon 
with accompanying finite quark orbital angular momentum and 
finite spin-orbit couplings which can be probed in experiments.
The search for orbital angular momentum has motivated new theoretical and experimental investigations 
of the three dimensional structure of the nucleon.
Key observables in deeply virtual Compton scattering (DVCS) 
and 
transverse single spin asymmetries in lepton-nucleon and 
proton-proton scattering,
and also the large $x$ limit of 
the down quark helicity distribution
are sensitive to orbital angular momentum in the nucleon.

Hard exclusive reactions such as DVCS
are
described theoretically using the formalism 
of generalized parton distributions (GPDs) and
probe the three-dimensional spatial structure of the nucleon,
as reviewed in \textcite{Ji:1998pc}, 
\textcite{Goeke:2001tz} and \textcite{Diehl:2003ny}.
Ji has derived a sum-rule connecting the forward limit of 
GPDs to information about the quark and gluon total angular 
momentum in the proton~\cite{Ji:1996ek}.
Considerable experimental and theoretical effort has and
continues to be invested aimed at accessing this information.

Studies of single spin asymmetries for semi-inclusive meson production in high-energy lepton-nucleon and proton-proton 
collisions are sensitive to possible spin-orbit coupling 
both in the nucleon 
and in the final-state hadronization process;
for a recent review see 
\textcite{Barone:2010zz}.
One studies correlations between the 
transverse momentum
(orbital motion) of partons,
their spin and the spin polarization of the nucleon.
The theoretical tools are transverse momentum dependent 
(TMD) distributions and fragmentation functions. 
TMDs probe the three-dimensional transverse momentum structure of the
nucleon and are associated, in part, with finite orbital angular momentum.

Experimental studies of three-dimensional nucleon structure have been pioneered at HERMES and JLab for GPDs and at 
COMPASS,
HERMES and RHIC for TMDs in single spin asymmetry measurements.
There has also been considerable theoretical effort aimed at model
and lattice calculations of these observables.

In the rest of this Section we present the theory and present 
status of these new GPDs and TMD distributions plus spin-orbit coupling in fragmentation
and the prospects for future experiments 
including key observables that will be studied.
The aim for experiments should be to focus on observables that
have the cleanest theoretical interpretation with minimal model dependence.

Quark orbital angular momentum in the nucleon may also be manifest 
in future measurements of the large $x$ behavior of the polarized down-quark distribution $\Delta d/d$ and 
in the ratio of 
the proton's spin-flip Pauli form-factor to the Dirac form-factor 
at large $Q^2$.
These observables can be studied with the 12 GeV upgrade of JLab.
Valence Fock states with non-zero orbital angular momentum induce
a logarithmic correction to the QCD counting rules predictions 
for these observables. 
Perturbative QCD calculations which take into account orbital 
angular momentum 
give
\begin{equation}
F_2/F_1 \sim (\log^2  Q^2 / \Lambda^2)/Q^2
\end{equation}
for the ratio of Pauli to Dirac form-factors at large $Q^2$
\cite{Belitsky:2002kj}.
Form-factor measurements at JLab \cite{Jones:1999rz,Gayou:2001qd}
are consistent with 
this behavior and also with
$F_2/F_1 \sim 1/\sqrt{Q^2}$ for $Q^2$ between 4 and 6 GeV$^2$,
in contrast to the counting rules prediction without
orbital angular momentum $F_2/F_1 \sim 1/Q^2$ \cite{Lepage:1980fj}.
One also finds a logarithmic correction to the leading 
large-$x$ behavior of 
the negative-helicity spin-dependent quark distributions 
$
\sim (1-x)^5 \log^2 (1-x)
$
\cite{Avakian:2007xa}. 
An interesting prediction here is that $\Delta d / d$ should 
cross zero and become positive at a value $x \sim 0.75$
when this term is included, 
in contrast to the model expectation that crossing occurs at 
$x \sim 0.5$ when this orbital angular momentum effect is neglected.
An accurate measurement of $\Delta d / d$ at $x$ close 
to unity would be very interesting 
if this quantity can be extracted free of uncertainties 
from nuclear effects~\cite{Kulagin:2007ph,Kulagin:2008fm}
in the neutron structure functions measured from deuteron or $^3$He targets.

\subsection{\label{sec:GPDs}Generalized parton distributions}

Observables in deeply virtual Compton scattering and deeply 
virtual meson production are sensitive to information about 
total angular momentum in the nucleon.
In these hard exclusive reactions a deeply virtual photon impacts
on a nucleon target and a real photon or a meson is liberated from the struck nucleon into the final state, leaving the target nucleon intact.
These processes can be described using the formalism of generalized parton distributions (GPDs),
involving the Fourier transforms of off-diagonal nucleon matrix elements
~\cite{Mueller:1998fv,Ji:1996nm,Ji:1996ek,Radyushkin:1997ki}.

The important kinematic variables are the virtuality of the hard photon $Q^2$,
the momenta $p - \Delta/2$ of the incident proton and
$p + \Delta /2$ of the outgoing proton,
the invariant four-momentum transferred to the target $t = \Delta^2$,
the average nucleon momentum $P$,
the generalized Bjorken variable $k^+ = x P^+$
and the light-cone momentum transferred to the target proton
$\xi = - \Delta^+/2p^+$.
In the Bjorken limit, $\xi$ is related to Bjorken $x_B$ via 
$\xi = x_B / (2-x_B) $. 
The generalized parton distributions are defined as
the light-cone Fourier transform of the point-split matrix 
element\footnote{
We work in the light-cone gauge $A_+=0$
(so the path-ordered gauge-link needed for gauge-invariance in
 the  correlation function becomes trivial and set equal to one).
}
\begin{eqnarray}
& &
{P_+ \over 2 \pi}
\int d y^- e^{-ixP^+y^-}
\langle p' | {\bar \psi}_{\alpha}(y) \psi_{\beta}(0) |p \rangle
_{y^+ = y_{\perp} = 0}
\nonumber \\
& &
=
{1 \over 4} \gamma^-_{\alpha \beta}
\biggl[ H(x,\xi,t) {\bar u}(p') \gamma^+ u(p)
\nonumber \\
& &
\ \ \ \ \ \ \ \ \ \ \ \ \ \ \ \ \ \ \ \ \ \
+
       E(x,\xi,t) {\bar u}(p') \sigma^{+ \mu} {\Delta_{\mu} \over 2M}
       u(p) \biggr]
\nonumber \\
& &
\ \ \
+ {1 \over 4} ( \gamma_5 \gamma^-)_{\alpha \beta}
\biggl[ {\widetilde H} (x,\xi,t) {\bar u}(p') \gamma^+ \gamma_5 u(p)
\nonumber \\
& &
\ \ \ \ \ \ \ \ \ \ \ \ \ \ \ \ \ \ \ \ \ \ \ \ \
+
       {\widetilde E} (x,\xi,t)
       {\bar u}(p') \gamma_5 {\Delta^+ \over 2M} u(p) \biggr] .
\nonumber \\
\label{eqk152}
\end{eqnarray}

The physical interpretation of the generalized parton distributions
(before worrying about possible renormalization effects and higher
 order corrections) is the following.
Expanding out the quark field operators in Eq.~(\ref{eqk152})
in terms of light-cone quantized creation and annihilation operators
one finds that for $x > \xi$ ($x < \xi$) the GPD is
the amplitude to take a quark (anti-quark) of momentum
$k - \Delta / 2$ out of the proton and reinsert a quark
(anti-quark) of momentum $k+\Delta/2$ into the proton some
distance along the light-cone to reform the recoiling proton.
In this region the GPD is a simple generalization of the usual
parton distributions studied in inclusive and semi-inclusive scattering
which are formally defined via light-cone correlation functions
-- see {\it e.g.}\ \textcite{Bass:2004xa}.
In the remaining region $-\xi < x < \xi$ the GPD involves taking out (or inserting) a $q {\bar q}$ pair with momentum $k - \Delta / 2$ and
$-k - \Delta / 2$ (or $k + \Delta /2$ and $-k + \Delta / 2$) respectively.
Note that the GPDs are interpreted as probability amplitudes rather than densities.
The non-forward matrix elements give access to transverse degrees of freedom in the nucleon.

In the forward limit the GPDs $H$ and ${\tilde H}$ are related 
to the parton distributions studied in deep inelastic scattering
\begin{eqnarray}
H (x,\xi,t)|_{\xi=t=0} &=& q (x)
\nonumber \\
{\widetilde H} (x,\xi,t)|_{\xi=t=0} &=& \Delta q(x)
\label{eqk153}
\end{eqnarray}
whereas the GPDs $E$ and ${\tilde E}$ have no such analogue.
Integrating over $x$ the first moments of the GPDs are related
to the nucleon form-factors
\begin{eqnarray}
\int_{-1}^{+1} dx H (x,\xi, t) &=& F_1 (t)
\nonumber \\
\int_{-1}^{+1} dx E (x,\xi, t) &=& F_2 (t)
\nonumber \\
\int_{-1}^{+1} dx {\widetilde H} (x,\xi, t) &=& G_A (t)
\nonumber \\
\int_{-1}^{+1} dx {\widetilde E} (x,\xi, t) &=& G_P (t)
.
\label{eqk154}
\end{eqnarray}
Here
$F_1$ and $F_2$ are the Dirac and Pauli form-factors of the nucleon,
and
$G_A$ and $G_P$ are the axial and induced-pseudoscalar form-factors
respectively.
(The dependence on $\xi$ drops out after integration over $x$.)

GPDs contain vital information about quark total angular momentum
in the nucleon.
Ji's sum-rule \cite{Ji:1996ek} 
relates $J_q$ to the forward limit of 
the second moment in $x$ of the spin-independent quark GPDs
\begin{equation}
J_q = {1 \over 2}
\int_{-1}^{+1} dx x
\biggl[ H^q (x,\xi,t=0) + E^q (x,\xi,t=0) \biggr].
\label{eq:Ji-SR}
\end{equation}
The gluon ``total angular momentum'' could then be obtained
through the equation
\begin{equation}
\sum_q J_q + J_g = {1 \over 2} .
\label{eq:Ji-SR-glue}
\end{equation}
In principle, it could be extracted from precision measurements of 
the $Q^2$ dependence of DVCS at next-to-leading-order accuracy 
where the quark GPDs mix with glue under QCD evolution
or via
$
J_g = {1 \over 2} \int_{-1}^{+1} dx x
\{ H^g(x,\xi,t=0) + E^g(x,\xi,t=0) \}
$
if the gluon GPD can be accurately measured in more direct experiments.
In these equations $J_q$ and $J_g$ are defined through the 
proton matrix elements of the angular momentum operators in Eq.(26).
If information about 
$J_q$ can be extracted from experiments,
then the corresponding
quark 
``orbital angular momentum" 
can be deduced by subtracting the value of the quark spin content $\Delta q$ extracted from deep inelastic scattering and polarized proton-proton collisions.
\footnote{We note recent discussion of a J=0 fixed pole contribution to DVCS \cite{Brodsky:2008qu,Brodsky:2009bp},
which corresponds to a $x \delta(x)$ term in the GPD $H$ and 
affects the $1/x$ moment of this GPD though not the sum-rules in 
Eqs.(31) and (32).
The same fixed pole also contributes to the Schwinger term 
sum-rule for the $1/x$ moment of the longitudinal structure function $F_L$ \cite{Broadhurst:1973fr}.
}

Experimental attempts to access $J_q$ via Eq.(32) require 
accurate determination of the two unpolarized GPDs $H$ and $E$.
Measurements from a proton target are more sensitive to $J_u$,
the total angular momentum carried by up quarks, 
while the neutron (via a deuteron or $^3$He target) 
is most sensitive to $J_d$.
The experiments require high luminosity to measure the small
exclusive cross-section, 
plus measurements over a wide range of kinematics in $Q^2$, 
$x$ and $t$ 
(since sum-rule tests and evaluations depend on making reliable extrapolations into unmeasured kinematics). 
In particular, one has to extrapolate the GPDs to $t=0$.
One also needs reliable theoretical technology 
to extract the GPDs from the measured cross-sections.

GPDs appear in the amplitudes for DVCS and hard exclusive meson
production as convolutions with the hard scattering coefficient 
and only these so-called Compton form factors (CFF) are experimentally accessible.
Measuring photon and also meson production
in the final state gives access to different flavor combinations of GPDs, like in semi-inclusive 
DIS.
However, meson production is more sensitive to QCD radiative
corrections and power corrections in $1/Q$, and 
reliable theoretical description requires larger values of $Q^2$ 
compared to DVCS.
Channels particularly sensitive to gluons in the proton are 
hard exclusive vector meson production where  both quark and 
gluon GPDs appear at lowest order in the strong coupling constant. 
There is a challenging program to disentangle the GPDs from the formalism and to undo the convolution integrals which relate the 
GPDs to measured cross-sections.
In practice, the approach used is to constrain models of GPDs 
against experimental data in measured kinematics.
These models are then integrated to obtain the Ji moments of $J_u$ and $J_d$, 
which may then be compared to the predictions of QCD inspired model
plus lattice calculations -- see Table V.

In the rest of this discussion we focus on deeply virtual Compton
scattering.

\subsubsection{Deeply virtual Compton scattering}

Measurements of hard exclusive processes are much more challenging than traditional inclusive and semi-inclusive scattering experiments. 
These exclusive processes require a difficult full reconstruction of final state particles and their cross-sections are usually small, demanding high luminosity machines.

\begin{figure}[tbp]
\includegraphics[width=0.8\hsize]{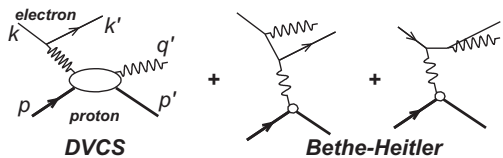}
\caption[Delta]
{\small
The leading DVCS and Bethe-Heitler processes.}
\label{fig:DVCS-BH-diagr}
\end{figure}

DVCS experiments have to be careful to choose the kinematics
so as not to be saturated by a large Bethe-Heitler (BH)
background where the emitted real photon
is radiated from the incident lepton rather than from the proton 
target
 -- see Fig.~14.

Most of the DVCS program so far has focused on the DVCS-BH
interference term.
Use of different combinations of beam and target polarization
plus changing the electric charge of the incident lepton beam
gives maximum access to most combinations of DVCS observables.
Measurement of the DVCS-BH interference term 
-- see Eq.(34) below --
allows one to measure not only the size of the DVCS amplitude 
but also its phase -- that is, it gives separate information 
about the real and imaginary parts of the Compton form factors.

Pioneering measurements of DVCS have been performed at 
DESY (HERMES, H1 and ZEUS) and JLab 
(Hall A and Hall B), 
which complement each other 
in the covered kinematic phase space and the extracted observables.

The experiments use different measurement techniques to access exclusive reactions.
The HERA collider experiments at DESY, H1 and ZEUS, as well as 
CLAS (Hall B at JLab) have the advantage of nearly hermetic spectrometers, 
whereas the fixed target experiments HERMES and JLab Hall A 
had to deal with the restrictions caused by incomplete event reconstruction due to their forward spectrometers.
Hall A and HERMES successfully employed the so-called 
missing mass 
technique together with careful background subtraction 
~\cite{Camacho:2006hx,Airapetian:2008aa}.
For Hall A 
the low beam energy and high resolution spectrometer 
allowed one to resolve pure elastic scattering 
from associated production with an excited nucleon in the final state.
The latter contribution was treated as part of the signal 
in HERMES results.
Very recently, beam-spin asymmetries for a pure DVCS sample 
have also been reported by HERMES~\cite{Airapetian:2012pg}.  
In the fixed target experiments the spin-dependent DVCS 
cross-sections have been explored using longitudinally 
polarized lepton beams with longitudinally (JLab and HERMES) and 
transversely (HERMES) polarized targets.
HERMES also took advantage of the available different beam charges.

JLab experiments focus on kinematics dominated by valence quarks.
Data from Hall A suggest leading twist-2 dominance of DVCS even at the relatively low $Q^2$ of 1.5--2.3 GeV$^2$ \cite{Camacho:2006hx}.

\begin{figure}[tbp]
\begin{center}
\includegraphics[width=0.40\textwidth]{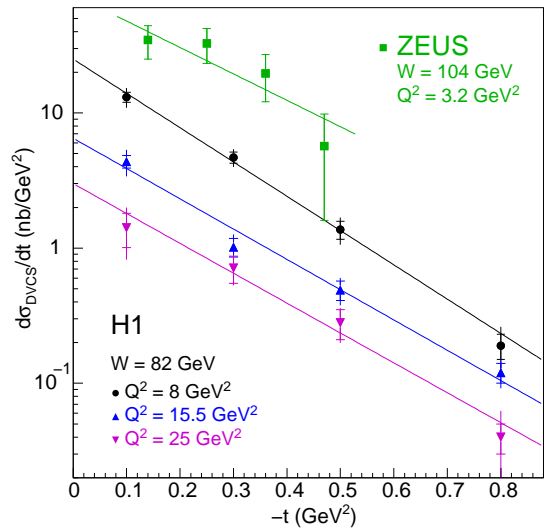}
\caption{\small 
The $t$ 
dependence of the DVCS cross-section for several values of $Q^2$ as measured by H1 and ZEUS.
The curves are results of fits of the form $e^{-b|t|}$ with $b$ being related to the 
transverse extension of partons in the proton at a given $x$ and $Q^2$ (see text).
The inner error bar represents the statistical uncertainty; the full bar the quadratic sum of statistical and systematic uncertainties.
}
\label{fig:DVCS-cs-dt}
\end{center}
\end{figure}

The HERA collider experiments H1 and ZEUS measured the 
DVCS cross-section 
close to the forward direction with $\xi < 10^{-2}$,
integrated over its azimuthal dependence, in an $x_B$ 
range where two-gluon exchange plays a major role in addition 
to the leading order quark-photon scattering process.
Figure~\ref{fig:DVCS-cs-dt} shows the cross-section differential 
in $t$ for different ranges in $Q^2$ measured 
by H1~\cite{Aaron:2007ab} and ZEUS~\cite{Chekanov:2008vy}.
The data are well described by the exponential behavior 
$d\sigma/dt \propto e^{-b|t|}$. 
The distribution of partons in the transverse plane is then obtained from this dependence by a Fourier transform
with respect to $\Delta_T$ 
(the transverse momentum shift in Eq.~(29))
$F(b,x,Q^2) \propto 
\int d^2\Delta_T \exp^{-ib\Delta_T} \sqrt{d\sigma/dt}$
~\cite{Burkardt:2002hr,Diehl:2002he}.
The impact parameter provides an estimate of the transverse extension of the partons probed during the hard process. 
While DVCS data provide information about the transverse distribution of quarks in the proton,
data on exclusive heavy vector meson production 
($J/\Psi$ or $\Upsilon$) 
describe the transverse distribution of glue at specific values of $x$.

The full DVCS cross-section reads~\cite{Diehl:2005pc}
\begin{alignat}{3}
& d\sigma(\ell p\to \ell \gamma p) 
\,\sim\, 
&& && 
\nonumber \\
&
\ \ \ \ \ \ \ {d\sigma^{\mathit{BH}}_{UU}}
&& + e_{\ell} \, {d\sigma^I_{UU}}
&& + {d\sigma^{DVCS}_{UU}}
\nonumber \\
&\quad + P_\ell S_L\, {d\sigma^{\mathit{BH}}_{LL}}
&& + e_\ell P_\ell S_L\, {d\sigma^I_{LL}}
&& + P_\ell S_L\, {d\sigma^{DVCS}_{LL}}
\nonumber \\
&\quad + P_\ell S_T\, {d\sigma^{\mathit{BH}}_{LT}}
&& + e_\ell P_\ell S_T\, {d\sigma^I_{LT}}
&& + P_\ell S_T\, {d\sigma^{DVCS}_{LT}}
\nonumber \\
&
&& + e_\ell P_\ell \, {d\sigma^I_{LU}}
&& + P_\ell \, {d\sigma^{DVCS}_{LU}}
\nonumber \\
&
&& + e_\ell S_L\, {d\sigma^I_{UL}}
&& + S_L\, {d\sigma^{DVCS}_{UL}}
\nonumber \\
&
&& + e_\ell S_T\, {d\sigma^I_{UT}}
&& + S_T\, {d\sigma^{DVCS}_{UT}}  
.
\nonumber \\
\label{eqn:full-dvcs}
\end{alignat}
Here the first subscript $U,L$ on $d\sigma$ indicates an unpolarized 
or longitudinally polarized lepton beam and the second subscript 
$U,L,T$ denotes an unpolarized, longitudinally or transversely polarized proton target; $P_\ell$ is the lepton beam polarization; 
$S_L$ and $S_T$ denote longitudinal and transverse proton polarization.  
Of particular interest is also the dependence on the sign of the 
charge of the beam lepton $e_\ell$, which 
allows one to disentangle contributions 
from the pure interference term and the DVCS term as 
pioneered in \textcite{Airapetian:2006zr,Airapetian:2008aa}.
The various cross-section terms depend on the azimuthal angle $\phi$ between the lepton scattering plane and the photon production plane,
and, in case of a transversely polarized proton target, also on the
azimuthal angle $\phi_S$ between the lepton plane and the transverse
target spin vector.  
Equation (\ref{eqn:full-dvcs}) 
indicates the large variety of observables accessible with polarized beams and/or targets.

As an example for the azimuthal dependence of the cross-section 
we give the expression for the interference term for the case 
of an unpolarized target and polarized beam ~\cite{Belitsky:2001ns}
\begin{eqnarray}
 I & & \propto - e_\ell
\left(\sum_{n=0}^3 c_{n}^{I}\cos(n\phi) + \lambda \sum_{n=1}^2
s_{n}^{I}\sin(n\phi)\right).
\label{eqn:unpol-dvcs}
\end{eqnarray}
The proportionality involves a kinematic factor and the lepton propagators of the BH process; 
$\lambda$ is the helicity of the incoming lepton.
The Fourier coefficients $c_{n}^{I}$ provide an experimental constraint on the real part of the Compton form factor and 
$s_{n}^{I}$ on the imaginary part.
Their relation to linear combinations of Compton form factors and 
hence to the respective GPDs is listed in Table~\ref{tab:gpds-dvcs}. 
A specific Fourier coefficient can be accessed experimentally by weighting the cross-section with the respective azimuthal modulation.

\begin{table}
\caption{\small
\label{tab:gpds-dvcs} 
Linear combinations of Compton form factors (CFF)
    in the DVCS-BH interference terms.  
Here, $F_1$ and $F_2$ are the electromagnetic form factors.
Subleading terms not shown are suppressed in a wide range of kinematics.
}
\begin{center}
\begin{tabular}{ll} \hline
 Target polarization & CFF combination \\ \hline
 unpolarized / charge & $F_1 H + \xi (F_1+F_2) \widetilde{H}
                     - \frac{t}{4m^2} F_2\, E$ \\[0.3em]
 longitudinal & $F_1 \widetilde{H} + \xi (F_1+F_2) H
                    - \ldots$ \\[0.3em]
 transverse $\propto \sin(\phi-\phi_S)$
              & $F_2 H - F_1\,  E
                      + \ldots$ \\[0.3em]
 transverse $\propto \cos(\phi-\phi_S)$
              & $F_2 \widetilde{H} - F_1\
                      \xi\widetilde{E} + \ldots$ \\ \hline
\end{tabular}
\end{center}
\end{table}

The DVCS-BH interference term was extracted by varying the 
electric charge of the incident lepton (HERMES) and 
studying polarization observables, varying the beam or
target helicity (JLab and HERMES).
JLab experiments have focused on 
studying their accessible observables fully differentially.
HERMES explored the advantages of using simultaneously polarization 
and charge observables to cleanly isolate the interference term and obtained
the most complete set of DVCS observables measured so far providing access to all interference terms listed in Eq.~(\ref{eqn:full-dvcs}).

\begin{figure}[tbp]
\begin{center}
\hspace{-0.5cm}
\includegraphics[width=0.47\textwidth]{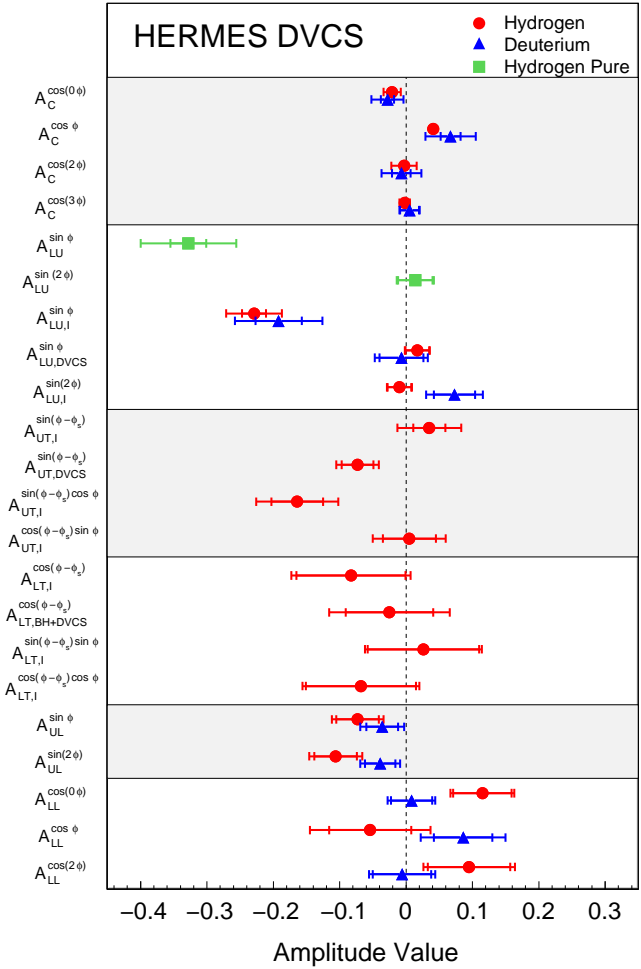}
\caption{\small 
Overview of all DVCS azimuthal asymmetry amplitudes measured at HERMES with proton and
deuterium targets, given at the average kinematics.
The inner error bar represents the statistical uncertainty; the full bar the quadratic sum of statistical and systematic uncertainties.}
\label{fig:hermes-dvcs-overview}
\end{center}
\end{figure}

Figure~\ref{fig:hermes-dvcs-overview} shows a summary of the 
HERMES DVCS measurements with polarized proton and deuterium targets
at their average kinematics
~\cite{Airapetian:2008aa,Airapetian:2009aa,Airapetian:2009bm,Airapetian:2010ab,Airapetian:2010aa,Airapetian:2011uq,Airapetian:2012mq,Airapetian:2012pg}.  
Here, $A_C$ is the charge asymmetry and $A_{XY}$ are the 
polarization-dependent asymmetries with $X$ and $Y$ indicating the beam and target polarization, respectively, which could be longitudinal ($L$) or
transverse ($T$). 
The subscript $I$ indicates an extraction of the pure interference term.
The measured asymmetries are subject to a harmonic expansion with respect to the azimuthal angle(s) as given by the superscript of
$A_{XY}$ in the Figure.
These data denoted by squares in Fig.~16
show results extracted from a DVCS sample with 
kinematically complete event reconstruction~\cite{Airapetian:2012pg}. 
The dependence on the kinematic variables $t$, $Q^2$, and $x_B$ was explored for each observable.

\begin{figure}[tbp]
\begin{center}
\includegraphics[width=0.47\textwidth]{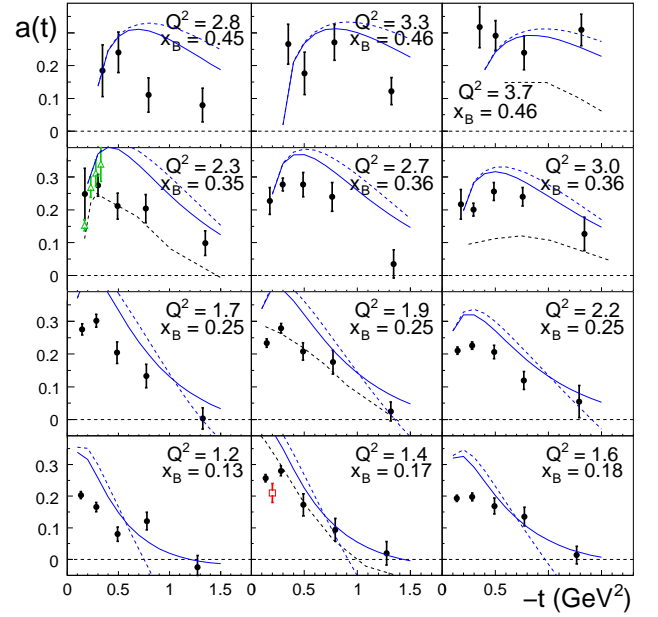}
\caption{\small 
The leading beam-spin asymmetry amplitude 
$a(t) = A_{LU}^{\sin\phi}$ differential in 
$t$, $x$ and $Q^2$ as measured by CLAS, from \textcite{Girod:2007aa}.
An earlier CLAS measurement~\cite{Stepanyan:2001sm} is indicated by the 
square.
The open triangles represent the cross-section data 
from Hall A~\cite{Camacho:2006hx}.
Error bars are statistical errors only.}
\label{fig:clas-dvcs}
\end{center}
\end{figure}

An example of the high statistics data from JLab is shown in 
Fig.~\ref{fig:clas-dvcs} 
for the beam-spin asymmetry 
$A_{LU}$ measured fully differentially by CLAS~\cite{Girod:2007aa}.
The presented data contain an admixture of 
the $A_{LU,I}^{\sin\phi}$ and  
$A_{LU,DVCS}^{\sin\phi}$ contributions
from the interference and pure DVCS terms, which cannot be separated here.
CLAS also provides measurements of 
$A_{UL}^{\sin\phi}$ and $A_{UL}^{\sin2\phi}$~\cite{Chen:2006na}.

\subsubsection{The quest for orbital angular momentum and GPD parametrizations}

Of the two GPDs $H$ and $E$ entering Ji's sum-rule, Eq.(32),
measurements with unpolarized targets but longitudinally polarized beams and also beam-charge asymmetries are mainly sensitive to $H$. 
As indicated in Table~\ref{tab:gpds-dvcs}, 
transverse target polarization provides kinematics-wise unsuppressed access to $E$. 
The GPD $E$ is essentially unknown. 
In contrast to $H$, it is not related to a deep inelastic parton distribution in the forward limit;
$E$ describes helicity flip at the proton vertex and 
requires finite orbital angular momentum in the nucleon. 
Contributions from $E$ to most DVCS observables are damped 
by kinematic factors $\sim |t|/M^2_p$, with the average $|t|$
value generally much smaller than 1 GeV$^2$ in the experiments.
To access $E$ requires DVCS and/or vector meson production 
asymmetry measurements with transversely polarized nucleon targets.
It may also be accessed through the beam polarization dependence of
DVCS with a neutron target because of the different 
size of the form-factors for the neutron~\cite{Belitsky:2001ns}.
Measurements have been performed already for all 
channels~\cite{Airapetian:2008aa,Airapetian:2009ad,Mazouz:2007aa,Adolph:2012ht}. 
Despite the lack of precision for these observables, 
attempts to extract information about quark total angular momentum
have been performed
by fitting theoretical models of GPDs 
to the DVCS measurements \cite{Airapetian:2008aa,Mazouz:2007aa}.
Although this analysis is very model dependent,
the results agree (surprisingly) well with model and 
lattice expectations, {\it e.g.} the calculations reported in 
Table V.
For example, 
within the model of \textcite{Vanderhaeghen:1999xj}
JLab Hall A DVCS measurements from the neutron were interpreted 
to give
$J_d + J_u/5.0 = 0.18 \pm 0.14 ({\rm expt.})$
\cite{Mazouz:2007aa}
whereas
HERMES results from the proton gave
$J_u + J_d/2.8 = 0.49 \pm 0.17 ({\rm expt.})$
\cite{Airapetian:2008aa} in the same model.

To go further and perform global fits of GPDs to hard exclusive observables one faces several challenging theoretical issues.
Parameterizations of GPDs have to deal with two longitudinal variables instead of one plus the $t$ dependence of DVCS.
It is also not yet known whether relatively simple and smooth functions like those used in QCD fits to deep inelastic data 
are sufficient to describe GPDs. 
A reliable parameterization of GPDs might therefore require a 
larger number of moments than employed in usual QCD parton descriptions. 
In addition, the dependence of the functions on the variable $x$ 
is not directly accessible, as $x$ represents a mute variable 
which is integrated over. 
In the interpretation of DVCS observables one has to deal with complex amplitudes;
the GPDs are embedded in the Compton form factors which relate to the measured cross-sections.
Despite these complications and the early stage of global fitting 
for GPDs, many results have been obtained in recent years fitting 
to different hard exclusive scattering data.
Interested readers are referred to the original literature in
\textcite{Vanderhaeghen:1999xj},
\textcite{Goloskokov:2007nt},
\textcite{Guidal:2010de}
\textcite{Kumericki:2009uq} and
\textcite{Goldstein:2010gu}.
This phenomenology is complemented by progress in lattice 
QCD calculations of 
GPD moments~\cite{Hagler:2007xi,Brommel:2007sb,Gockeler:2006zu}.

\subsection{\label{sec:TMDs}
Transversity,
transverse-momentum-dependent distributions and
fragmentation functions}

Striking single spin asymmetries associated with spin-momentum correlations 
(expected with parton orbital angular momentum) 
were first observed in the 1970s. 
Using a 12 GeV polarized proton beam from the Argonne National Laboratory Zero Gradient Synchrotron on a fixed target, up to 40\% more positive pions were produced left of the beam 
when the beam was polarized up, 
and up to 20\% more negative pions were produced 
to the right of the beam~\cite{Klem:1976ui}. 
These measurements were confirmed by similar 
experiments~\cite{Dragoset:1978gg,Antille:1980th,Saroff:1989gn,Apokin:1990ik}, 
but it was not until the 1990s that a theoretical framework was developed to attempt interpreting them.

Single-spin asymmetries have now also been observed in 
proton-proton collisions at RHIC, 
where they reach up to $\sim 40\%$, and
in lepton-nucleon collisions at COMPASS, HERMES and JLab, 
where they are typically $5-10\%$.
Single-spin asymmetries for hadron production from transversely polarized targets tell us about spin-orbit coupling in the nucleon and/or in the fragmentation process.
Transverse-momentum-dependent distributions 
simultaneously describe the dependence on longitudinal momentum fraction of the parton within the parent hadron as well as the parton's transverse momentum.  
Similarly, transverse-momentum-dependent fragmentation functions describe the dependence on longitudinal momentum fraction of the produced hadron with respect to the scattering parton as well as 
the hadronic transverse momentum with respect to the jet axis.

\begin{table}[htp]
\caption{\small
Leading-twist transverse-momentum dependent parton distributions.
$U$, $L$, and $T$ stand for unpolarized, longitudinally polarized, 
and transversely polarized nucleons (rows) and quarks (columns)
respectively.}
\begin{center}
\begin{tabular}{|c|c|c|c|} \hline
\backslashbox{N}{q}
           & U & L & T \\ \hline
           {U} & ${ \bf f_1}$   & & $ h_{1}^\perp$ \\
           \hline
           {L} & &${ \bf g_1}$ &    $ h_{1L}^\perp$ \\
           \hline
           {T} & $ f_{1T}^\perp $ &  $ g_{1T}^\perp $ &  {$\bf h_1$} \,
           ${ h_{1T}^\perp }$ \\
           \hline
           \end{tabular}
\end{center}
\label{tab1}
\end{table}

We next introduce these distributions and fragmentation functions,
and discuss their phenomenology.
In QCD there are eight leading-twist quark TMDs.
These are listed in Table~\ref{tab1} and discussed in
\textcite{Mulders:1995dh} and \textcite{Bacchetta:2006tn}. 
The three distributions highlighted in boldface survive integration 
over transverse momentum $k_t$. 
These yield the unpolarized parton distribution, $f_1 (x,k_t)$,
the spin-dependent parton distribution
$g_1(x,k_t)$ and the transversity distribution $h_1(x,k_t)$.
The other five distributions do not survive integration over $k_t$.
They describe correlations between the quark transverse momentum 
with the spin of the quark and/or the spin of the parent nucleon,
{\it viz.} spin-orbit correlations.
The three TMDs denoted by $h$
describe the distribution of transversely polarized partons.
They are chiral-odd distributions and
appear only in observables involving two chiral-odd partners,
such as Drell-Yan processes (two chiral-odd parton distributions) 
or SIDIS (chiral-odd parton distribution and the
Collins fragmentation function discussed below).
The three distributions $f_{1T}^\perp $ (the Sivers distribution), 
$h_{1}^\perp $ (the Boer-Mulders distribution) and 
$ h_{1T}^\perp $ (pretzelosity) 
require orbital angular momentum in the nucleon since they involve 
a transition between initial and final nucleon states 
whose orbital angular momentum 
differ by $\Delta L_z^q = \pm 1$ (Sivers and Boer-Mulders)
or $\Delta L_z^q = \pm 2$ (pretzelosity).
The ``worm-gear" functions 
$h_{1 L}^{\perp}$ and $g_{1 T}^{\perp}$ 
link two perpendicular spin directions 
and are also connected to quark orbital motion inside nucleons.

Transverse momentum distributions have been studied most in 
semi-inclusive DIS experiments
where 
they appear in combination with 
the usual unpolarized fragmentation function $D(z,p_t)$ 
or, 
in case of the chiral-odd TMD distributions, with a 
chiral-odd Collins fragmentation function $H_1^\perp(z,p_t)$
discussed in Section VII.B.2.
One measures the azimuthal distribution of the produced 
final-state hadron with respect to the virtual-photon axis.
Each species of TMD comes with a different angular modulation 
in the semi-inclusive cross-section
allowing it to be projected out to yield information about 
the different spin-momentum correlations~\cite{Bacchetta:2004jz}.
All these leading-twist TMDs have been measured in semi-inclusive 
DIS over the last decade.
However, several have been the focus of more intense studies and 
we focus on those here.

The different modulation combinations are listed in Table VIII
together with present experimental measurements.
Results quoted at $\sqrt{s}=18$~GeV are from COMPASS,
7.4~GeV from HERMES and 3.5~GeV from JLab.
Here $\phi$ is the angle between the lepton direction and 
the plane spanned by the exchanged photon and tagged final-state
hadron, {\it e.g.}\ a high-energy meson;
$\phi_S$ is the angle between the lepton direction and 
the transverse nucleon target spin.
The convolution is taken over the involved transverse momenta of 
the quark and the hadron produced in the fragmentation process.

\begin{widetext}

\begin{table}[t!]
\caption{\label{tab:TMD+FF-meas}
Experimental access to the leading twist TMD distributions 
in  SIDIS with unpolarized (U), longitudinally (L) or transversely polarized (T) beam (modulation first subscript) 
and/or target (modulation second subscript). 
}
\begin{center}
{\begin{tabular}{lccccl}
\hline
 Modulation & Combination & $\sqrt{s}$ &  Target  &  Observed & Measurement \\
 & Distribution name & GeV & type & hadron types & \\
\hline
 $\sin (\phi+\phi_S)_{UT}$ & $h_1 \otimes H_1^{\perp}$ 
    & 18 & d & $h^\pm,\pi^\pm, K^\pm, K^0$  & \textcite{Ageev:2006da,Alekseev:2008dn}\\
    &Transversity & & p & $h^\pm$ & \textcite{Alekseev:2010rw,Adolph:2012sn}\\
  &      & & p & $\pi^\pm, K^\pm$  & prelim.~\textcite{Pesaro:2011zz} \\
   &  & 7.4 & p & $\pi^\pm, \pi^0, K^\pm$ & \textcite{Airapetian:2004tw,Airapetian:2010ds} \\
   &  & 3.5 & n & $\pi^\pm$ & \textcite{Qian:2011py} \\
\hline
 $\sin (\phi-\phi_S)_{UT}$ & $f_{1T}^{\perp} \otimes D$ 
     & 18 & d & $h^\pm,\pi^\pm, K^\pm, K^0$  & \textcite{Ageev:2006da,Alekseev:2008dn}\\
 & Sivers & & p & $h^\pm$ & 
\textcite{Alekseev:2010rw,Adolph:2012sp}\\
   &       & & p & $\pi^\pm, K^\pm$  & prelim.~\textcite{Pesaro:2011zz} \\
 &    & 7.4 & p & $\pi^\pm, \pi^0, K^\pm$ & \textcite{Airapetian:2004tw,Airapetian:2009ae} \\
 &    & 3.5 & n & $\pi^\pm$ & \textcite{Qian:2011py} \\
\hline
 $\cos (2 \phi)_{UU}$ &  $h_1^{\perp} \otimes H_1^{\perp}$  
     & 18 & d & $h^\pm$ & prelim.~\textcite{Sbrizzai:2010fi}\\
 & Boer-Mulders & 7.4 & p & $\pi^\pm, K^\pm$ & \textcite{Airapetian:2012yg} \\
 &    & 3.5 & n & $\pi^+$ & \textcite{Osipenko:2008rv} \\
\hline
 $\sin (3\phi - \phi_S)_{UT}$ &  $h_{1 T}^{\perp} \otimes H_1^{\perp}$  
    & 18 & d & $h^\pm$ & prelim.~\textcite{Kotzinian:2007uv}\\
 & Pretzelosity & 18 & p & $h^{\pm}$ & prelim.~\textcite{Parsamyan:2010se} \\
& & 7.4 & p & $\pi^\pm, K^\pm$ & prelim.~\textcite{Pappalardo:2010zza} \\
\hline
 $\sin (2 \phi)_{UL}$ &  $h_{1 L}^{\perp} \otimes H_1^{\perp}$  
    & 18 & d & $h^{\pm}$ & \textcite{Alekseev:2010dm} \\
 & Worm-gear 1 & 7.4 & p & $\pi^\pm, \pi^0$ & \textcite{Airapetian:1999tv,Airapetian:2001eg} \\
 & &  & d & $\pi^\pm, \pi^0, K^+ $& \textcite{Airapetian:2002mf} \\
 &     & 3.5 & n & $\pi^\pm, \pi^0 $ & \textcite{Avakian:2010ae} \\
\hline
 $\cos (\phi - \phi_S)_{LT}$ &  $g_{1 T}^{\perp} \otimes D$  
     & 18 & d & $h^\pm$ & prelim.~\textcite{Kotzinian:2007uv}\\
 & Worm-gear 2 & 18 
& p & $h^{\pm}$ & prelim.~\textcite{Parsamyan:2010se} \\
& & 7.4 & p & $\pi^\pm, \pi^0, K^\pm$ & prelim.~\textcite{Pappalardo:2011cu} \\
 &     & 3.5 & n & $\pi^\pm$ & \textcite{Huang:2011bc} \\
\hline
\end{tabular}}
\end{center}
\end{table}
\end{widetext}

When one projects out the terms with different azimuthal 
angular dependence
summarized in Table VIII, 
the
COMPASS, HERMES and JLab data
suggest that the Sivers, Collins and Boer-Mulders effects 
are all present in the proton target data -- see below.
JLab data from CLAS 
reveal a clear signal for the worm-gear-1 distribution; 
there is some hint for a non-zero worm-gear-2 distribution
(with low significance)
and
(so far) no significant signal for pretzolosity in the proton.
For the deuteron target, 
there is evidence 
for a Boer-Mulders effect from COMPASS and HERMES.
The Sivers, Collins, worm-gear and pretzelosity effects are all
consistent with zero in the deuteron target data. 
The Collins and Sivers effects observed in the proton data 
therefore contain a predominant isovector contribution.

We next focus on the Sivers, Boer-Mulders and Collins effects.

\subsubsection{The Sivers and Boer-Mulders TMD distributions}

The Sivers distribution was first proposed in 
\textcite{Sivers:1989cc}
in an attempt to explain the large transverse 
single spin asymmetries observed in the 1970s and 1980s.
It describes the correlation 
between the transverse momentum $k_t$ of the struck quark
and 
the spin $S$ and momentum $p$ of its parent nucleon
\begin{equation}
f_{q/p^{\uparrow}}(x, k_t)
=
f_1^q (x, k_t^2)
-
f_{1t}^{\perp q}(x, k_t) 
\frac{{\bf S \cdot (k_t \times {\hat p})}}{M} .
\end{equation}
The $k_t$ dependence means that the Sivers distribution 
is sensitive to non-zero parton orbital angular momentum 
in the nucleon, 
though the mapping from Sivers observables to quark 
(and gluon) orbital angular momentum is (so far) model dependent
with present theoretical technology.

The Sivers distribution has the interesting property that it is odd under time reversal.  
Due to this feature, such a correlation was believed to be forbidden for more than a decade. 
Then
\textcite{Brodsky:2002cx,Brodsky:2002rv} 
showed that, with initial- or final-state interactions,
the Sivers effect could be non-zero in QCD processes.
Final-state interactions in SIDIS can generate the azimuthal 
asymmetry before the quark fragments into hadrons.
Shortly afterwards, 
\textcite{Collins:2002kn}
realized that initial-state color interactions in 
the case of Drell-Yan and final-state interactions 
in the case of SIDIS would lead 
to a process-dependent sign difference in the Sivers distribution.
SIDIS measurements~\cite{Airapetian:2004tw,Airapetian:2009ae,Qian:2011py,Adolph:2012sp,Alekseev:2010rw,Pesaro:2011zz}, 
suggest sizable asymmetries at the level of about $5-10\%$ 
for a proton and a neutron target, 
while Drell-Yan measurements are planned for the future.

\begin{figure}[tbp]
\begin{center}
\includegraphics[width=0.47\textwidth]{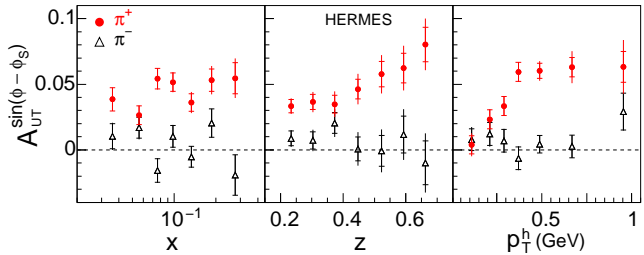}
\caption{\small 
Sivers amplitudes for charged pions measured by HERMES with a 
proton target; from ~\textcite{Airapetian:2009ae}.
The Sivers amplitudes for $K^+$ (not shown here) 
appear to be nearly twice as large as those for $\pi^+$.
The inner error bar represents the statistical uncertainty; the full bar the quadratic sum of statistical and systematic uncertainties.
}
\label{fig:sivers-hermes}
\end{center}
\end{figure}
\begin{figure}[tbp]
\begin{center}
\includegraphics[width=0.47\textwidth]{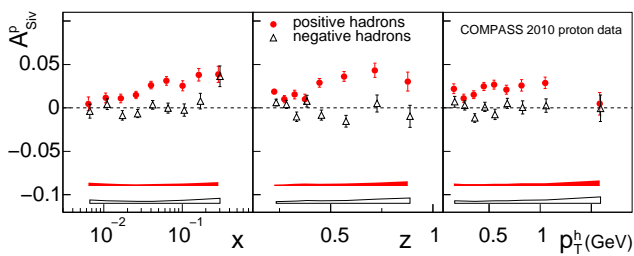}
\caption{\small 
Sivers amplitudes for unidentified charged hadrons 
measured by COMPASS with a proton target~\cite{Adolph:2012sp}.
The hadron yield is dominated by pions. 
The bands indicate the systematic uncertainties.
}
\label{fig:sivers-compass}
\end{center}
\end{figure}

A qualitative picture of 
the Sivers distribution can already be deduced from SIDIS measurements.
The non-zero amplitudes shown in Figs.~\ref{fig:sivers-hermes} and 
~\ref{fig:sivers-compass} were obtained with a proton target.
For HERMES the amplitude includes a kinematic factor depending on 
the ratio of transverse-to-longitudinal photon flux, which in the COMPASS data is divided out.
Since scattering off $u$ quarks dominates these data 
due to the quark charge factor,
the positive Sivers amplitudes for $\pi^+$ 
(and $h^+$, which is dominated by the pion yield),
suggest a large and negative Sivers function for up quarks.
The vanishing amplitudes for $\pi^-$ ($h^-$) require cancellation effects, {\it e.g.}\ from a
$d$ quark Sivers distribution opposite in sign to the $u$ quark Sivers distribution.
These cancellation effects between Sivers distributions 
for up and down quarks are supported
by the vanishing Sivers amplitudes extracted from deuteron
data by the 
COMPASS Collaboration~\cite{Ageev:2006da,Alekseev:2008dn}.
An interesting facet of the HERMES data is the magnitude of the 
$K^+$ amplitude, 
which is nearly twice as large as that of 
$\pi^+$~\cite{Airapetian:2009ae}.
Again, on the basis of $u$ quark dominance,
one might naively expect that 
the $\pi^+$ and $K^+$ amplitudes should be similar.
Their difference in size may thus point to a significant role of other quark flavors,
{\it e.g.}\ sea  quarks.
A sizable Sivers amplitude for $\pi^+$ was also recently reported 
by JLab Hall A~\cite{Qian:2011py} for measurements with a $^3$He (neutron) target.  
In that data a negative Sivers amplitude for $\pi^+$ was found 
which independently supports a 
$d$-quark Sivers distribution opposite in sign to the $u$-quark one.

The Boer-Mulders distribution \cite{Boer:1997nt}
describes the correlation between transversely polarized 
quarks in an unpolarized nucleon and the quarks' transverse momentum,
${\bf s_q \cdot (k_t \times \hat{p})}$,
where $s_q$ denotes the spin of the quark
and might hence yield unexpected spin effects even in an unpolarized nucleon.  
It is similar to the Sivers distribution in that it is $T$-odd.
However, it is also chiral-odd and 
hence must be probed in conjunction with a second chiral-odd function.  
For Drell-Yan production, 
the second function is a
Boer-Mulders distribution in the second incident hadron.
For SIDIS, 
the Collins fragmentation function described below is involved.
Like the $T$-odd Sivers distribution, 
the Boer-Mulders distribution is 
also expected to change sign between Drell-Yan production and SIDIS.
Future experimental effort is planned to test this QCD prediction.

Azimuthal distributions sensitive to the Boer-Mulders distribution 
were originally measured in Drell-Yan experiments~\cite{Falciano:1986wk,Guanziroli:1987rp,Conway:1989fs,Zhu:2006gx,Zhu:2008sj}.
The SeaQuest fixed-target Drell-Yan experiment currently underway at Fermilab ~\cite{Reimer:2007zza}
expects to be sensitive to the Boer-Mulders distribution at high $x$.
In SIDIS the distinctive pattern of Boer-Mulders modulations for oppositely charged pions and for pions and kaons was recently reported by HERMES~\cite{Airapetian:2012yg}.
The amplitudes for kaons are larger in magnitude than the 
amplitudes for pions.
The amplitudes for the negative pions 
have the opposite sign to the amplitudes for negative kaons.
This hints at a significant contribution from sea quarks, 
in particular from strange quarks.
Measurements of the Boer-Mulders amplitudes were also reported 
by COMPASS for unidentified hadrons~\cite{Sbrizzai:2010fi}
and by CLAS for pions~\cite{Osipenko:2008rv}.
The interpretation of the SIDIS amplitudes for the Boer-Mulders 
distribution, 
is, however, complicated by contributions 
from the twist-4
Cahn effect~\cite{Cahn:1989yf,Cahn:1978se}
which have been estimated 
to be sizable even at COMPASS kinematics~\cite{Anselmino:2006rv}.
The Cahn effect accounts 
for the parton intrinsic transverse momenta in the target 
nucleon and the fact that produced hadrons might acquire 
transverse momenta during the fragmentation process.
Theoretical estimates of the Boer-Mulders effect are still 
plagued by large uncertainties, 
mainly related to the insufficient knowledge of 
the transverse-momentum dependence of 
the unpolarized distribution
$f_1(x,k_t)$ and fragmentation function $D (z,p_t)$.

\subsubsection{The Collins TMD fragmentation function}

The Collins TMD fragmentation function describes a
spin-momentum correlation in the hadronization process,
${\bf \; {s}_q \cdot  ( {k}_q  \times {p}_t )}$,
with a hadron produced in fragmentation having some transverse momentum $p_t$ with respect to the momentum direction $k$ of a 
transversely polarized fragmenting quark with spin $s_q$
~\cite{Collins:1992kk,Collins:1993kq}.
The Collins fragmentation function
has been investigated in semi-inclusive
lepton-nucleon scattering and $e^+e^-$ annihilation.
The magnitude of the effect is approximately 5--10\%, like that
found for the Sivers asymmetries.

\begin{figure}[t]
\begin{center}
\includegraphics[width=0.47\textwidth]{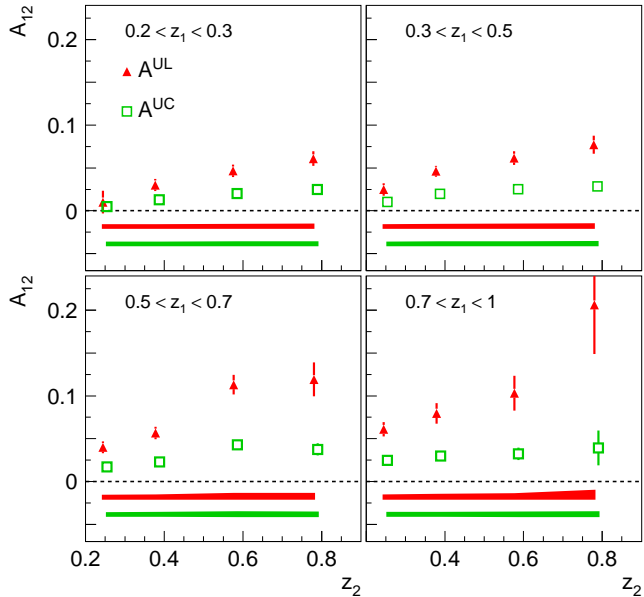}
\caption{\small 
Collins asymmetry for the double ratios of like-sign (L), unlike-sign (U) and any charged (C) pion pairs
from Belle~\cite{Seidl:2008xc}.
$A^{UL}$ and $A^{UC}$ are sensitive to different combinations of the favored and unfavored Collins
fragmentation functions.
The bands indicate the systematic uncertainties.
}
\label{fig:Belle-CollinsFF}
\end{center}
\end{figure}

For $e^+e^-$ annihilation 
the chiral-odd Collins fragmentation function enters with
a second Collins function in the opposing jet.  
The Collins function has been measured to be non-zero for the production of charged pions 
in $e^+e^-$ annihilation at 
Belle~\cite{Abe:2005zx,Seidl:2008xc},
as shown in Fig.~\ref{fig:Belle-CollinsFF}, 
and in recent preliminary data from BABAR~\cite{Garzia:2012fj}.

\begin{figure}[t]
\begin{center}
\includegraphics[width=0.47\textwidth]{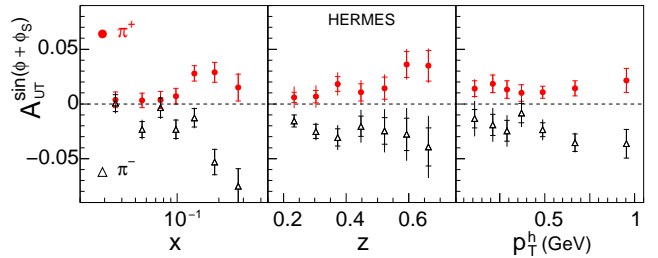}
\caption{\small 
Collins amplitudes for charged pions measured by HERMES 
with a proton target; from ~\textcite{Airapetian:2010ds}.
The inner error bar represents the statistical uncertainty; the full bar the quadratic sum of statistical and systematic uncertainties.
}
\label{fig:collins-hermes}
\end{center}
\end{figure}

\begin{figure}[th]
\begin{center}
\includegraphics[width=0.47\textwidth]{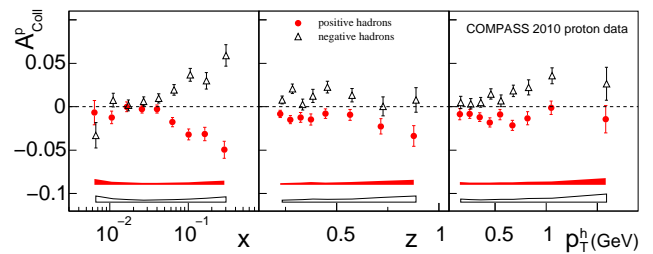}
\caption{\small 
Collins amplitudes for unidentified charged hadrons measured by COMPASS with a proton target~\cite{Adolph:2012sn}.
The hadron yield is dominated by pions. 
Note that a different definition of the Collins angle results in 
amplitudes with the opposite sign compared to other measurements.
The bands indicate the systematic uncertainties.}
\label{fig:collins-compass}
\end{center}
\end{figure}

In SIDIS the second chiral-odd function is the transversity   distribution introduced in Section II and discussed further below
(or the Boer-Mulders distribution).  
The HERMES~\cite{Airapetian:2004tw,Airapetian:2010ds}, 
COMPASS~\cite{Ageev:2006da,Alekseev:2008dn,Alekseev:2010rw,Adolph:2012sn,Pesaro:2011zz} 
and 
JLab Hall A~\cite{Qian:2011py} experiments 
have performed 
SIDIS measurements of the Collins effect.
The measurements for a proton target 
are shown in Figs.~\ref{fig:collins-hermes} and 
~\ref{fig:collins-compass} for HERMES and COMPASS respectively.
(Note that COMPASS uses a definition of the Collins angle which 
 results in Collins amplitudes with opposite sign
 to the ``Trento convention" of \textcite{Bacchetta:2004jz}
 used by HERMES, JLab and commonly in theoretical papers).   
There is excellent agreement between the measurements in similar
kinematics.
One finds the striking observation that the
Collins amplitude for $\pi^-$ is of similar size to $\pi^+$ 
production but comes with opposite sign.
This hints at an unfavored Collins function of 
similar size and opposite sign 
than the favored one, a situation very different 
from that observed with unpolarized fragmentation functions.


\subsubsection{Probing transversity}

The transversity distribution introduced in Section II describes 
the transverse polarization of quarks within a 
transversely polarized nucleon.  
Along with the unpolarized and helicity distributions, 
it survives integration over 
partonic transverse momentum and is thus a collinear distribution.

The first moment of the transversity distribution is proportional
to the nucleon's $C$-odd tensor charge,
{\it viz.}
$\delta q = \int_0^1 dx h_1^q(x)$ with
\begin{equation}
\langle p, s | \ {\bar q} i \sigma_{\mu \nu} \gamma_5 q \ | p,s
\rangle
=
(1 / M) ( s_{\mu} p_{\nu} - s_{\nu} p_{\mu} ) \delta q.
\end{equation}
The difference between the transversity and helicity spin
distributions
reflects the relativistic character of quark motion in the nucleon.
In Bag models this effect is manifest as follows.
The lower component of the Dirac spinor enters the relativistic
spin depolarization factor with the opposite sign to $\Delta q$
because of the extra factor of $\gamma_{\mu}$ in the tensor charge
\cite{Jaffe:1991ra}.
The relativistic Bag depolarization factor mentioned in Section VI
becomes 0.83 for transversity in contrast to 0.65 for helicity and
the nucleon's axial-charges.
In leading order QCD the transversity distributions are bound
by Soffer's inequality
$|h_1^q (x,Q^2)| 
 \leq \frac{1}{2} [\{q + {\bar q} \}(x,Q^2) + \Delta q(x,Q^2)]$,
\textcite{Soffer:1994ww}.
QCD motivated fits to transversity observables are reported in
\textcite{Anselmino:2008jk}, which also reviews the comparison
to model predictions.

Transversity is measured through the Collins effect and also in
in dihadron production,
where the chiral-odd partner of 
$h_1^q$ is given by the dihadron
fragmentation function $H_1^{\open q}$~\cite{Collins:1993kq,Bianconi:1999cd,Bacchetta:2011ip}.
This describes how the transverse spin of the fragmenting quark is transferred to the relative orbital angular momentum of the hadron pair.
Consequently, this mechanism does not require transverse momentum of the produced hadron pair.
Standard collinear factorization applies allowing one to 
study the transversity distribution without having to worry
about solving convolution integrals over 
transverse momentum or issues of TMD factorization and evolution.

Pioneering measurements of two-pion production in 
polarized semi-inclusive DIS by HERMES~\cite{Airapetian:2008sk} and COMPASS~\cite{Adolph:2012nw} reveal a sizable effect
and have already been
employed for an extraction of transversity~\cite{Courtoy:2012in}. 
First measurements of azimuthal correlations of two pion pairs 
in back-to-back jets in $e^+e^-$ annihilation 
related to the dihadron fragmentation function 
have just become available from Belle~\cite{Vossen:2011fk} 
and a first extraction of the dihadron fragmentation function 
from these data was performed in \textcite{Courtoy:2012ry}.

\subsubsection{Current status and recent progress with TMD
distributions}

There has been considerable progress in the understanding of intrinsic transverse momentum and spin-momentum correlations in QCD over the past decade,
motivated by the theoretical breakthroughs regarding $T$-odd TMD distributions~\cite{Brodsky:2002cx,Brodsky:2002rv,Collins:2002kn} 
and by a vast program of theoretical and experimental activity.

A recent monograph, \textcite{Collins:2011zzd} gives definitions of TMD distributions which allow QCD evolution to be applied 
rigorously for the first time with separately identifiable TMD distributions and fragmentation functions. 
Building upon this progress, the evolution of previously unevolved models and fits has now been published for unpolarized 
TMD distributions and fragmentation functions~\cite{Aybat:2011zv} 
and
the Sivers distribution~\cite{Aybat:2011ge}.
QCD evolution is just starting to be applied to phenomenological studies~\cite{Aybat:2011ta}, which will be a major step forward in interpreting and comparing results from different experiments.  
The new definitions of TMD distributions also recently made
possible a determination of the hard parts for SIDIS and 
Drell-Yan at 
next-to-leading order~\cite{Aybat:2011vb}, 
which should lead to improved phenomenology.

Much effort has been dedicated to phenomenological extractions of 
TMDs and parameterizations of the Sivers distribution from SIDIS 
data, 
see e.g.\ \textcite{Anselmino:2008sga} and \textcite{Anselmino:2011gs}, 
with QCD evolution 
now starting to be considered~\cite{Anselmino:2012aa}.  
One parameterization of the Sivers function includes 
both semi-inclusive deep inelastic and proton-proton 
data~\cite{Kang:2012xf},
modulo
issues related to factorization breaking~\cite{Rogers:2010dm} 
discussed below.
Fits to the Collins TMD fragmentation function
have been performed using both $e^+e^-$ and SIDIS data as 
input~\cite{Efremov:2006qm,Anselmino:2007fs}.
The Boer-Mulders distribution has been extracted based on 
Drell-Yan~\cite{Zhang:2008nu,Lu:2009ip} 
as well as SIDIS data~\cite{Barone:2009hw}.

These first phenomenological fits to TMD observables have been
performed using a simple Gaussian ansatz 
for the transverse momentum 
dependence of quarks in the nucleon and fragmentation functions.
For example, the Sivers function in Eq.(36) was parametrized in 
the fits by the product of the unpolarized distribution 
$f_{q/p^{\uparrow}}(x, k_t)$ 
with 
an $x$ dependent factor and an $x$-independent Gaussian 
$\sim {k_t \over M_1} e^{- k_t^2 / M_1^2 }$ containing
all the $k_t$ dependence.
While the Gaussian ansatz is unstable with respect to QCD evolution
with increasing $Q^2$, 
the method does provide a reasonable fit to present data
with values
$
\langle k_t^2 \rangle = 0.25 \ \GeV^2$ and 
$
\langle p_t^2 \rangle = 0.20 \ \GeV^2
$
taken from fits to the Cahn effect in unpolarized scattering
\cite{Anselmino:2005nn}.
A longer term goal for TMD experiments is 
to observe deviation from Gaussian behavior for transverse momentum
dependence.
With extensive unpolarized Drell-Yan and weak boson production data available over scales from $\sim 4$~GeV$^2$ to $M_Z^2$, new fits of unpolarized TMD distributions are quite promising as a means to test the $Q^2$ evolution of TMD distributions as well as to learn more about the shape of the distributions in $k_t$.

Lattice calculations of the Sivers and Boer-Mulders distributions
have been performed
\cite{Gockeler:2006zu,Hagler:2009mb,Musch:2011er}.  
There have also been efforts in recent years to implement TMDs 
in Monte Carlo event generators
\cite{Bianconi:2011un,Hautmann:2012pf}.
Models can provide helpful insight into TMD distributions, and 
a wealth of different model calculations have been explored and published. 
We refer to \textcite{Avakian:2009jt},
\textcite{Bacchetta:2011bn},
\textcite{Lorce:2011zta}, and
\textcite{Pasquini:2011tk} 
for recent discussion of models related to TMD distributions, 
including attempts to address the relationship between 
TMD distributions and orbital angular momentum in the nucleon.

\subsubsection{Proton-proton asymmetries and TMD-factorization breaking}

Despite the fact that the large transverse single spin asymmetries observed in hadronic scattering originally inspired the development of TMDs, inclusive hadron production in $p+p$ scattering cannot be cleanly separated into Sivers, Collins, or other contributions as is possible in SIDIS.  
In recent work \textcite{Rogers:2010dm}
argue that the TMD framework is not valid in the case of hadroproduction of hadrons, as factorization does not hold.
While the short-distance (perturbative) components are still believed to factorize from the long-distance (non-perturbative) ones, the long-distance components become entangled and no longer factorize from one another into independent 
TMD distributions and/or fragmentation functions.
What is particularly interesting is that the factorization breaking effects are relevant in precisely the kinematic regime where a parton description is generally expected to apply.  
It will be exciting to see this tested experimentally in the upcoming years, exploring long-distance quantum entanglement effects in QCD.  
In the longer-term future, it may be possible to develop well-defined functions within the framework of pQCD which describe the correlations between the partons in the incoming and/or outgoing hadrons.

\begin{widetext}

\begin{figure}
\centering
\includegraphics[width=0.9\textwidth]
{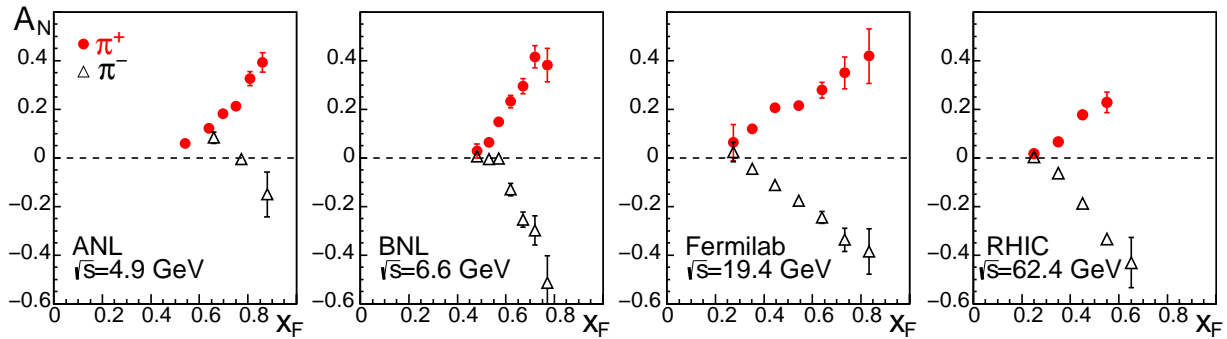}
\caption[Transverse single spin asymmetry for 
$p+p \rightarrow \pi^\pm$ at four different center-of-mass energies.]
{The transverse single spin asymmetry in forward $\pi^\pm$ production as measured in polarized proton-proton collisions across a range of center-of-mass energies.
From left to right, the data are from~\textcite{Klem:1976ui}, \textcite{Allgower:2002qi}, \textcite{Adams:1991cs}, and~\textcite{Arsene:2008aa}. 
Error bars are statistical errors only.
\label{figure:ANAcrossEnergies}}
\end{figure}

\end{widetext}

In the meantime, single spin asymmetries for forward meson production in $p+p$ collisions have been shown to remain large across a very wide range of center-of-mass energies~\cite{Allgower:2002qi,Adams:1991cs,Adams:1994yu,Arsene:2008aa,Abelev:2008af} and up to the highest measured 
$p_T$ of $\sim$5~GeV~\cite{Koster:2012zz}.
As shown in Fig.~\ref{figure:ANAcrossEnergies}, 
the transverse single spin asymmetries in charged pion production as a function of Feynman-$x$ are remarkably similar from $\sqrt{s}=$4.9~GeV all the way up to 62.4~GeV measured by the BRAHMS experiment at RHIC.

At higher energies and in particular at $p_T$ values large enough to serve as a hard scale, one can try to interpret these phenomena utilizing the tools of pQCD.  With no explicitly measured scale sensitive to the partonic transverse momentum in inclusive 
single spin asymmetries, a more appropriate framework than TMD distributions in which to interpret the asymmetries may be a collinear, twist-3 picture~\cite{Efremov:1981sh, Efremov:1984ip, Qiu:1998ia}.  
A relationship between the TMD and the collinear, twist-3 frameworks was laid out in~\textcite{Ji:2006ub}.

Surprises continue to emerge from these kinds of measurements, with large asymmetries for negative kaons as well as antiprotons from BRAHMS~\cite{Arsene:2008aa} suggesting that the pion asymmetries are not a valence quark effect as previously believed, and a recent hint from STAR~\cite{Adamczyk:2012xd} that the asymmetry for eta mesons may be larger than that of neutral pions.

\section{Future Projects}

A new program of dedicated experiments is planned 
to investigate key open questions in QCD spin physics. 
We briefly outline these experiments and their prime 
physics objectives.

Since May 2012 CEBAF is undergoing
a major upgrade that will bring 
the maximum available energy of the electron beam to 12 GeV.
The experimental equipment in all three halls will be
upgraded (Hall A and C) or completely renewed (Hall B),
in order to better match the increased energy and luminosity.
A new experimental Hall D is being built.
Commissioning of the new accelerator and of 
the experimental halls is expected for 2014.
The future physics program
focuses on dedicated studies of large $x$ phenomena, 
hard exclusive reactions and TMD effects 
in kinematics where valence quarks dominate the physics
\cite{dudek:2012vr}.

At CERN a proposal by the COMPASS Collaboration \cite{CompassII:2010} to study TMDs and GPDs in the period 2014--2017 has been approved. 
The COMPASS data will provide a link between the kinematic domains 
of HERA on one hand and of HERMES and JLAB on the other.
The program will start with the first ever polarized Drell--Yan experiment using a transversely polarized ammonia (proton) 
target and a negative pion beam. Due to the underlying 
annihilation of the 
anti up-quark from the pion and the target
up quark, the process is dominated by the up quark distribution in the valence region. 
An important goal is to check the QCD prediction of a sign change 
in the naive $T$-odd TMDs with respect to the DIS case.  
A study of GPDs in DVCS and hard exclusive meson production with a polarized muon beam will follow in 2015 using a 
liquid hydrogen target, a dedicated target recoil detector and an additional large-angle electromagnetic calorimeter. An important measurement is the beam charge-and-spin asymmetry, which uses the property of the muon beam that polarization changes sign when going from positive to negative muons. 
A first result on the correlation of transverse size and longitudinal momentum fraction might already be expected from a 2012 pilot run.
In parallel semi-inclusive DIS data will be taken on the pure hydrogen target.

There are proposals to create a polarized fixed-target Drell-Yan program at Fermilab following the SeaQuest experiment, scheduled to complete data taking in 2014.  R\&D has begun for a suitable polarized target, and a formal proposal to polarize the 120 GeV proton beam in the Main Injector has been submitted to Fermilab management~\cite{SPINFermilab:2011aa}.
One of the primary physics motivations for such a program would be to explore in detail the QCD spin-momentum correlations described by TMD distributions such as the Sivers distribution, in particular the role of color flow in Drell-Yan versus semi-inclusive DIS interactions.

A variety of possibilities for the medium-term future of RHIC is currently under discussion.  R\&D is ongoing for a polarized $^3$He source for RHIC~\cite{Zelenski:2008zz}, 
which would allow the neutron spin structure to be studied in collider kinematics for the first time.  
There are also proposals to significantly extend the detector capabilities at RHIC; see e.g.\ \textcite{Aidala:2012nz}.  Of particular interest to nucleon structure studies are potential upgraded forward spectrometers capable of reconstructing jets, with hadronic particle identification and Drell-Yan measurement capabilities up to pseudorapidities of $\sim 4$.  
The ability to perform full jet reconstruction in the forward region where large transverse single spin asymmetries are observed, and in addition to measure and identify hadrons within the jet, would allow separation of effects due to distribution versus fragmentation functions and shed great light on the origin of these significant 
spin-momentum correlations.
An integrated design process for new detectors is underway such that they would be able to take full advantage of electron-proton and electron-ion collisions in the longer-term future should an electron beam be added to RHIC.

Ideas for future polarization measurements are also discussed and investigated in more detail at 
FAIR (Germany), J-PARC (Japan) and NICA (Russia).

A possible Electron-Ion Collider (EIC) 
is being discussed in connection with the future of RHIC and JLab.
The goal is to achieve highly polarized (greater than $70\%$) 
electron and 
light-nucleus beams with center-of-mass energies ranging from 
about 20--150 GeV at maximum collision luminosities typically 
$\sim 10^{34}$ cm$^{-2}$ s$^{-1}$.
Significant R\&D is ongoing to realize the technical challenges 
for reaching this luminosity frontier for colliders and 
for achieving and maintaining polarization of light nuclei 
(D and $^3$He) in a storage ring.
An EIC with the above performance would offer unique access 
to the small-$x$ region where gluons dominate as well as to the intermediate and high $x$ regions at unprecedented high $Q^2$.  
One could then study gluon polarization down to $x$ values of 
about $10^{-4}$~\cite{Aschenauer:2012ve}.
The high luminosity would enable us to measure and map GPDs 
over a broad range of the kinematic variables and study the 
QCD evolution of the DVCS process plus TMD distributions in
kinematics where sea/glue effects are expected to be important. 
In addition to being the first $ep$ collider exploring the 
structure of polarized protons,
an EIC would also be the first electron-nucleus collider allowing
precision studies of the gluon and sea quark structure of nuclei.
Unpolarized ion beams from deuterium to the heaviest nuclei 
-- uranium or lead -- would also be accelerated.
Knowledge about the spatial distribution of quarks and gluons in nuclei
is needed
for example in the interpretation of 
heavy-ion collision data and the search for quark-gluon plasma.
A comprehensive review of EIC physics opportunities is given in 
\textcite{Boer:2011fh}.

\section{Conclusions and Outlook}

The challenge to understand the internal spin structure of the 
nucleon has inspired a global program of enormous 
experimental and theoretical work in QCD during 
the last 25 years.

For longitudinal spin structure, there is a good convergence of
spin measurements from CERN, DESY, JLab, RHIC and SLAC  
taking into account the $Q^2$ dependence of 
the data and kinematics of the different experiments.
There is also good convergence of theoretical understanding
with the data, including QCD inspired models of the nucleon
and lattice calculations with disconnected diagrams included.
Semi-inclusive measurements in polarized lepton-nucleon and 
proton-proton collisions 
have yielded much information about the size of the separate
valence, sea and gluon spin contributions to the nucleon's spin.
The small value of 
the nucleon's flavor-singlet axial-charge, about 0.35,
extracted from polarized deep inelastic scattering seems to 
be a valence quark effect.
No significant sea-quark polarization is observed in 
semi-inclusive deep inelastic scattering experiments;
the sum of 
valence spin contributions is in close agreement with
the measured total spin contribution $g_A^{(0)}|_{\rm pDIS}$.
While gluon polarization $\Delta g$ at the scale of the 
experiments may be as much as 50\% of the nucleon's spin
at the scale of the experiments,
the QCD anomaly correction 
$-3 \frac{\alpha_s}{2 \pi} \Delta g$ 
is too small to resolve 
the difference between
$g_A^{(0)}|_{\rm pDIS}$ and 
the early quark models predictions, about 0.6.
Prime theory candidates to explain the small 
``quark spin content"
include transfer of valence quark spin 
to quark orbital angular momentum 
through the pion cloud and a possible topological effect 
whereby some fraction of the valence quarks' 
``spin" resides at Bjorken $x=0$, 
where it is missed by polarized deep inelastic scattering 
experiments.
The proton spin puzzle seems to be telling us about 
the interplay of valence quarks with 
chiral dynamics and the complex vacuum structure of QCD.
Ongoing and planned experimental activity will improve
the precision 
on the size of gluon and strangeness polarization in the nucleon.

Finite orbital angular momentum of the valence quarks is expected,
induced also by confinement which introduces a transverse scale in
the physics.
Quark orbital angular momentum through spin-orbit coupling is a 
prime candidate to explain the large transverse single spin 
asymmetries observed in proton-proton collisions and 
lepton-nucleon scattering.
The desire to understand and measure QCD orbital angular momentum
effects in the nucleon has spawned a new program to explore and
map the three-dimensional structure of the nucleon
--
both in spatial co-ordinates
(generalized parton distributions)
and transverse momentum dependence.

Studies of transverse nucleon structure will drive the experimental
program in the near future, 
with dedicated running or approved
programs at COMPASS, the 12 GeV upgrade of JLab, FNAL and RHIC. 
These experiments will test our understanding of
initial and final state interactions in QCD 
(through comparison of Sivers and Boer-Mulders observables 
 in Drell-Yan and semi-inclusive deep inelastic scattering).
Precise measurements of
GPDs and TMDs will test QCD evolution
in a regime where transverse structure becomes important.
The aim for precise information about quark (and gluon) total and orbital angular momentum in the nucleon is also a driving force for much theoretical work.
Highlights include models of transverse spin phenomena, lattice calculations and development of QCD fitting technology to extract 
GPDs and TMDs from the newly measurable observables.

\section*{Acknowledgments}

The research of SDB is supported by the Austrian Science Fund, 
FWF, through grants P20436 and P23753.

We thank 
M. Diehl, R. Fatemi, A. Korzenev, S. Kuhn, W. Melnitchouk, 
B. Pasquini, T.C. Rogers, M. Stratmann, S. Taneja, 
A. W. Thomas, F. Videbaek, R. Windmolders, A. Zelenski and
E. Zemlyanichkina for helpful discussions.

\newpage

\bibliography{rmp}

\end{document}